\documentclass[dvipsnames]{article}

\usepackage[margin=1in]{geometry}

\usepackage{graphicx,amsmath,upgreek,bm}
\usepackage{colortbl}
\usepackage{wrapfig}
\usepackage[rgb, table]{xcolor}
\usepackage{booktabs}
\usepackage{makecell}
\usepackage{pifont}
\usepackage{todonotes}
\usepackage[toc,page]{appendix}
\usepackage{multirow}
\usepackage{booktabs}
\usepackage{algorithm}
\usepackage{enumitem}
\usepackage{appendix}
\usepackage{color}
\usepackage{hyperref}
\hypersetup{
   colorlinks=true,
   linkcolor=blue,
   filecolor=magenta,
   urlcolor=blue,
   citecolor=blue,
	}
\usepackage[
  separate-uncertainty = true,
  multi-part-units = repeat
]{siunitx}
\usepackage{caption} 
\usepackage{subcaption}
\usepackage{amsmath}
\usepackage[noend]{algpseudocode}
\usepackage{stackengine}
\usepackage{multirow}
\usepackage{arydshln}
\makeatletter
  \renewcommand*\env@matrix[1][*\c@MaxMatrixCols c]{%
    \hskip -\arraycolsep
    \let\@ifnextchar\new@ifnextchar
  \array{#1}}
\makeatother
\makeatletter
\def\BState{\State\hskip-\ALG@thistlm}
\makeatother
\definecolor{light-gray}{gray}{0.9}

\graphicspath{{./figures/}{../figures}}

\title{On Transfer Learning For Chatter Detection in Turning Using Wavelet Packet Transform and Ensemble Empirical Mode Decomposition}

\author{Melih C.~Yesilli\\
				Department of Mechanical Engineering\\
				Michigan State University\\
				yesillim@egr.msu.edu
			\and
				Firas A.~Khasawneh\\
				Department of Mechanical Engineering\\
				Michigan State University\\
				khasawn3@egr.msu.edu
			\and
				Andreas Otto\\
				Institute of Physics\\
				Chemnitz University of Technology\\
				andreas.otto@physik.tu-chemnitz.de
				}
				
\date{}

\begin{document}
\maketitle

\begin{abstract}
The increasing availability of sensor data at machine tools makes automatic chatter detection algorithms a trending topic in metal cutting. 
Two prominent and advanced methods for feature extraction via signal decomposition are Wavelet Packet Transform (WPT) and Ensemble Empirical Mode Decomposition (EEMD).
We apply these two methods to time series acquired from an acceleration sensor at the tool holder of a lathe. 
Different turning experiments with varying dynamic behavior of the machine tool structure were performed.  
We compare the performance of these two methods with Support Vector Machine (SVM), Logistic Regression, Random Forest Classification and Gradient Boosting combined with Recursive Feature Elimination (RFE). 
We also show that the common WPT-based approach of choosing wavelet packets with the highest energy ratios as representative features for chatter does not always result in packets that enclose the chatter frequency, thus reducing the classification accuracy.   
Further, we test the transfer learning capability of each of these methods by training the classifier on one of the cutting configurations and then testing it on the other cases. 
It is found that when training and testing on data from the same cutting configuration both methods yield high accuracies reaching in one of the cases as high as $94\%$ and $95\%$, respectively, for WPT and EEMD.
However, our experimental results show that EEMD can outperform WPT in transfer learning applications with accuracy of up to $95\%$. 
 
%We show that:
%- WPT is better than EEMD?
%- WPT faster than EEMD?
%- different optimal features for different processes?
\end{abstract}


%!TEX root = ../machining_ML_Wavelet_CIRP_Journal_Version.tex
%-------------------------------
%************************************
\section{Introduction}
\label{sec:intro}
%************************************
Turning, boring, milling, and drilling operations constitute a major part of manufacturing processes. 
One challenging problem that all these processes have in common is the occurrence of large amplitude, detrimental oscillations called chatter \cite{Taylor1907, 2004altintas, Quintana2011}. 
Since chatter leads to increased tool wear, poor surface finish and noise, it is extremely important to anticipate and avoid its occurrence. 
Alternatively, several chatter mitigation techniques including increasing stiffness in machine tools, and active and passive damping techniques also exist \cite{Munoa2016}. 
Efficient methods for the identification of the stability lobes that separate stable cutting and chattering motion \cite{2012altintas, 2014otto} can help keep the machine away from chatter via selecting parameters in the safe area below the stability lobes. 
However, these models often do not account for the effect of the changing dynamics or for highly complex cutting operations. 
This led to the emergence of in-situ methods for chatter detection based on instrumenting the cutting center with sensors and analyzing the resulting signals \cite{Smith1992,Altintas1992,Choi2003,Kuljanic2009,Yao2010,Elias2014}. 

\begin{figure}[H]
\centering
\includegraphics[width=0.9\textwidth,height=.48\textheight,keepaspectratio]{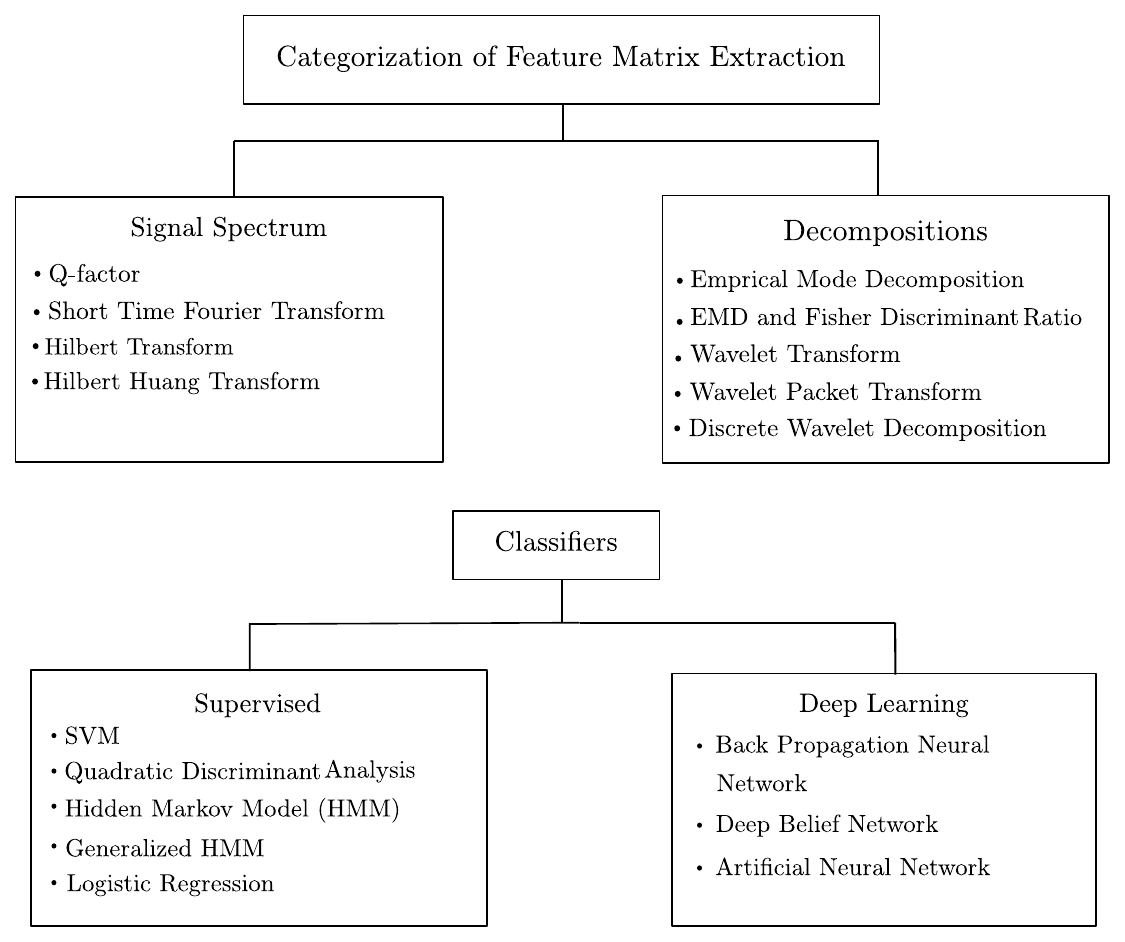}
\caption{Categorization of feature extraction methods and classifiers used for chatter detection.}
\label{fig:categ}
\end{figure}

The majority of available in-process methods for chatter identification rely on extracting certain features from the acoustic, vibration, or force signals and comparing them against some predefined markers of chatter \cite{Tlusty1983,Delio1992,Gradisek1998a,Schmitz2002,Choi2003,Bediaga2009,Sims2009,Nair2010,Dijk2010,Tsai2010,Kakinuma2011,Ma2013}. 
They can be broadly categorized into two groups as shown in Fig.~\ref{fig:categ}. 
The most prevailing methods are Wavelet Packet Transforms (WPT) and Empirical Mode Decomposition (EMD) or the Ensemble Empirical Mode Decomposition (EEMD). 
Generally, such decomposition-based methods for analyzing the cutting signal follow the same procedure. 
First, the signal is decomposed into different parts using some transformation. 
Then, the decomposed portions or packets of the signal which include the relevant information about machine tool chatter are selected to reconstruct a new signal. 
These packets are chosen by applying the Fast Fourier Transform (FFT) to the different parts or packets and choosing the ones that overlap with the known chatter frequencies of the system. 
Finally, various time and frequency domain features are computed from these packets. 
In several papers, these features are ranked and are utilized as the input for the machine learning classifiers. 
Support Vector Machine (SVM) algorithm is the most common classifier used for chatter classification \cite{Yao2010,Ji2018,Chen2018,Chen2017,Liu2011,Wang2018,Saravanamurugan2017}. Other less common classifiers include quadratic discrimination analysis \cite{Thaler2014}, Hidden Markov Model (HMM) \cite{Han2016}, generalized HMM \cite{Xie2016}, and logistic regression \cite{Ding2017} (cf. Fig.~\ref{fig:categ}). 

Wavelet packet decomposition and wavelet transform are widely adopted in machining state monitoring. Chen and Zheng \cite{Chen2017} generated feature matrices for chatter classification using wavelet packets whose frequency bands contain the chatter frequency. Yao et~al.~\cite{Yao2010} used the standard deviation and the energy of the decomposition obtained using the Discrete Wavelet Transform and the WPT for chatter detection from acceleration signals in a boring experiment. The energy of the wavelet packets were also utilized in turning experiments with comparison of different levels of WPT \cite{Qian2015,Xie2016}. Ding et~al.~\cite{Ding2017} used wavelet packet entropy as a feature for early chatter detection. In addition to WPT, EMD and EEMD are also often utilized to featurize cutting signals. Ji et~al.~\cite{Ji2018} proposed EMD to both eliminate noise from milling vibration signals and to extract features from informative Intrinsic Mode Functions (IMF). 
Chen et~al.~\cite{Chen2018} used top-ranked features extracted from the IMFs obtained from EEMD for machining state detection. Li et~al.~\cite{Li2010a} used the energy spectrum of the IMFs as features for chatter detection. The resulting features are ranked by using Fisher Discriminant Ratio (FDR) \cite{Chen2018} and, when the number of features is high, recursive feature elimination (RFE) is used to reduce the number of features \cite{Chen2017}. Although EMD/EEMD is typically applied to vibration signals, Liu et~al.~\cite{Liu2011} also used EMD to extract features from the servo motor current time series. 

In addition to WPT and EMD-based approaches, there are other methods for feature extraction from metal removal processes. For example, Thaler et~al.~\cite{Thaler2014} used Short-Time Fourier Transform to extract the frequency domain features of the feed force, acceleration, and sound pressure signals in band sawing operation. Moreover, the Q-factor and the power spectrum of the signal were used for chatter classification in milling \cite{Wang2018}. Cao et~al.~\cite{Cao2013} applied the Hilbert Huang transform to signals reconstructed using only the informative wavelet packets. Yesilli et~al. used the topological features obtained from Topological Data Analysis and to predict chatter in turning \cite{Yesilli2019} and milling \cite{Yesilli2019a} process. Also, similarity measure method, Dynamic Time Warping was used to detect chatter in turning process \cite{Yesilli2019b}.

%\textcolor{blue}{[AO: The following paragraph must be adjusted to the results that we want to highlight]}
Chatter detection strategies based on WPT or EEMD require deciding on which informative parts of the signal to use. However, since searching for the informative parts of the decomposition is a multi-step process, these approaches become impractically laborious. Although the time required to obtain the needed WPT and EEMD decompositions is relatively low, choosing the informative decompositions in WPT and EEMD is often not straightforward. This is because the featurization process involves looking into the power spectra and the energy ratio plots for each signal in order to determine the most informative parts of the decomposition.
Consequently, only a few cases are often analyzed and the chosen packets or decompositions are fixed and used for feature extraction for all the subsequent data sets. 
For example, in the WPT-based approach, the standard procedure is to pick the packets with the highest energy ratio as the most informative part of the decomposition.

Unfortunately, the resulting informative packets or decompositions may not contain chatter information especially if the system parameters shift during operation, e.g., due to the movement of the machine center which may involve changing the overhang distance of the tool and thus the flexibility of the cutting tool. 
Therefore, in these situations the classifier is required to categorize signals that may carry different characteristics and chatter features than the ones it was trained on. 
In other words, the ability of the classifier to achieve \textit{transfer learning} is tested in these situations.
However, there has not been any studies on the transfer learning capabilities of WPT and EEMD. 
Further, the common approach for picking the informative packets in WPT is to choose the packets with the highest energy. 
However, these packets do not necessarily contain the chatter frequency bands, and thus they may not be the most suitable markers for chatter detection. 

In this paper, we compare the performance and the transfer learning capabilities of two chatter detection algorithms based on WPT and EEMD on a set of turning experiments where the cutting tool is instrumented with accelerometers. 
A total of four tool stickout lengths are used, which correspond to changing the eigenfrequencies of the machine-tool structure. 
In addition, a variety of depths of cut and cutting speeds are tested. 
We establish a criteria for tagging the resulting signals into chatter-free, mild chatter, or full chatter. 
Then, we split the data for each stickout length into training and testing sets, train our two methods with the training set and use them to classify the test signal as chatter or chatter-free. 
We investigate the classification performance not only of the SVM algorithm---the most popular tool for machine learning on chatter signals, but also of the logistic regression, random forest, and gradient boosting. 
Upon obtaining a classifier for data that corresponds to data taken from a specific cutting configuration, we test the classifier on data from other cutting configurations. 
We repeat the above process ten times where each time the data is randomly split into training/testing sets and we compare the average test accuracy and standard deviation of the featurization methods. 
We then evaluate the classification results and we comment on (1) the ease of feature extraction, i.e., the effort required and the potential for automating feature extraction. (2) the classification accuracy within a fixed cutting configuration (but with varying spindle speeds and depths of cut). And (3) transfer learning capabilities, i.e., the accuracy associated with using certain feature vectors and classification algorithms to train and test on two different cutting configurations.

Based on our investigations, we believe that WPT and EEMD are more conducive to automatic feature extraction than traditional featurization methods of chatter signals. 
Further, the results based on our datasets show that classifiers based on random forest, gradient boosting, and logistic regression have higher accuracy than SVM. 
% This indicates that future works on machine learning and chatter can benefit from incorporating algorithms beyond simple SVM classification.
Our results also show that when training and testing on signals from the same cutting configuration, the WPT method gives higher classification accuracy rates than EEMD. 
However, when testing the obtained classifier on data from different cutting configurations, we show that, for our specific cutting data, the EEMD method outperforms WPT, i.e., we show that EEMD has superior transfer learning capabilities than WPT.
In addition, we discuss in Section \ref{sec:wpt} how wavelet packets with higher energy ratios do not necessarily contain chatter information. 
Specifically, we found in multiple cases that the packets ranked second or even third in terms of the energy ratio can include the chatter frequency band in their signal spectrum. 
Therefore, fixing certain informative packets or parts of the signal may not be a viable option especially when the cutting process leads to changes in the dynamic behavior of the tool-workpiece system.

The paper is organized as follows. Section \ref{sec:exp_setup} describes the experimental setup and the procedure for tagging the data as chatter versus non-chatter.
Sections \ref{sec:wpt} and \ref{sec:eemd} describe combining the WPT and the EEMD methods, respectively, with machine learning tools for chatter detection. 
Section \ref{sec:classifiers} gives reviews the classification algorithms used in this study briefly.
Section \ref{sec:results} presents the results of our investigations including comparisons of the accuracy, transfer learning capabilities, and runtime of both methods, while the discussion and our concluding remarks can be found in Section \ref{sec:conclusion}. \ref{sec:appendixA} contains supplementary classification results that are referenced from the main text of the manuscript.

%------------------
%-----------------

%!TEX root = ../machining_ML_Similarity.tex
%-------------------------------
%*******************************
\section{Experimental Setup}
\label{sec:exp_setup}
\begin{figure}[htbp]
\centering
\includegraphics[width=0.48\textwidth,height=.48\textheight,keepaspectratio]{cutting_test_fig1_zoomed_clipped_annotated}
\hspace{0.25in}
\includegraphics[width=0.40\textwidth]{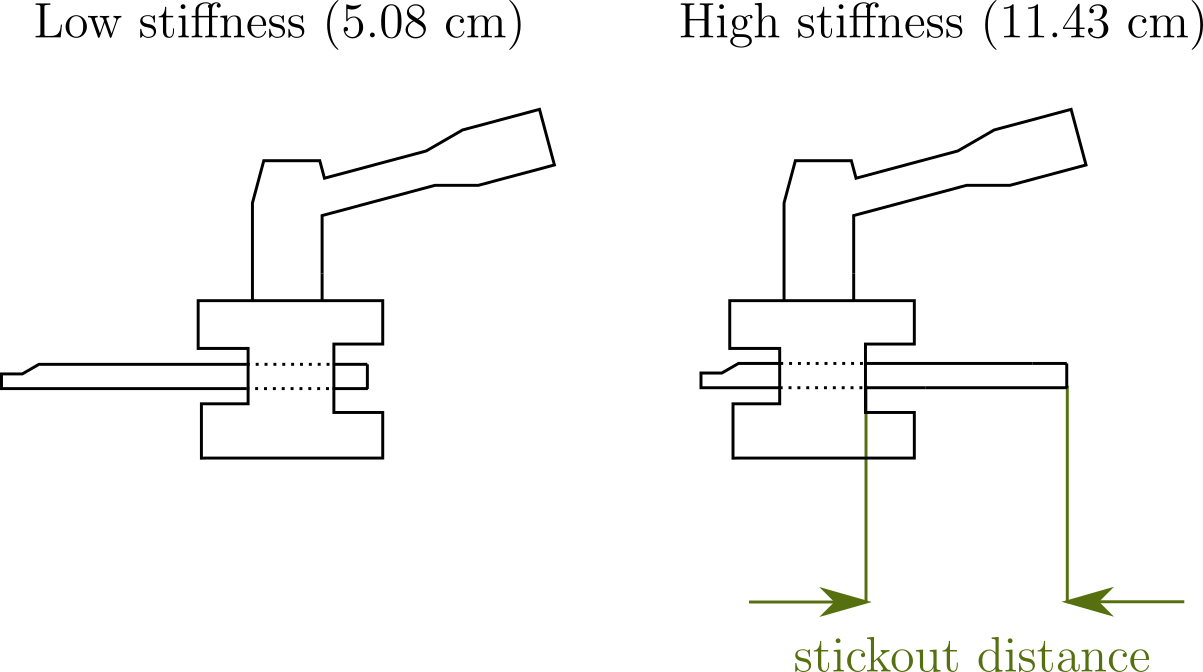}
\caption{The experimental setup showing the workpiece, the cutting tool, and the attached accelerometers (left). The visual representation of the stickout distance (right)}
\label{fig:cutting_test}
\end{figure}

% \subsection{Collecting Data}
%*******************************
Figure \ref{fig:cutting_test} shows the turning experiment that was used to collect the measurement data for training and testing of the chatter detection algorithms. It consists of a $6061$ aluminum cylindrical workpiece mounted into the chuck of the spindle of a Clausing-Gamet $33$ cm ($13$ inch) engine lathe. An S10R-SCLCR3A boring bar from the Grizzly T10439 carbide insert boring bar set with an attached $0.04$ cm ($0.015$ inch) radius Titanium nitride coated cutting insert is secured to the tool holder.

The stiffness of the rod, and therefore, the eigenfrequencies of the tool vibration, are varied by changing the overhang or stickout length of the rod. Four stickout lengths are used in the experiment:  5.08 cm ($2$ inch), 6.35 cm ($2.5$ inch), 8.89 cm ($3.5$ inch), and 11.43 cm ($4.5$ inch). In order to obtain more accurate measurements, the stickout length is measured as the distance between the flat, back surface of the tool holder and the heel of the boring rod. The visual representation of stickout distance is given in right hand side of Fig.\ref{fig:cutting_test}. This means that increasing the stickout length leads to a stiffer cutting tool and higher eigenfrequencies for lateral vibrations. Since the lateral direction is the most flexible and chatter frequencies appear in the neighborhood of dominant eigenfrequencies of the structure, the dominant chatter frequencies increase with increasing stickout length. 

The boring rod is instrumented with two PCB 352B10 miniature, lightweight, uni-axial ceramic shear accelerometers that are ninety degrees apart to measure lateral vibrations of the rod. The two accelerometers are superglued onto the rod at about 3.81 cm ($1.5$ inch) away from the cutting tool to protect them from moving parts and cutting debris. A PCB 356B11 triaxial, miniature ceramic shear accelerometer is also attached to the bottom clamp of the tool holder as shown in Fig.~\ref{fig:cutting_test}. The data from all the accelerometers are collected on the analog channels of an NI USB-6366 data acquisition box using Matlab. No in-line analog filter is used; however, the signals are oversampled at $160$kHz. Digital filtering is used before subsampling thus eliminating noise while avoiding the undesirable effects of antialising. In particular, we use a Butterworth low-pass filter with order $100$ and a cutoff frequency of $10$ kHz. The data is then downsampled to $10$ kHz without risk of causing aliasing effects. The resulting conditioned data is what we consider in Section~\ref{sec:data_labeling}. In addition, we provide both the raw and filtered data in a Mendeley repository \cite{Khasawneh2019}. Also, one can find our codes for this study in a \href{https://github.com/mcanyesilli/WPT_EEMD_ML_Machining}{GitHub repository}.

% Inputs:
% column1: time
% column2: accelerometer on the boring rod in x (x-axis is perpendicular to the rod, and it runs front and back relative to lathe operator)
% column3: accelerometer on the boring rod in y (y-axis is perpendicular to the rod, and it runs up and down)
% column4: Audio signal
% column5: laser tach signal
% column6: tri-axial accelerometer on the tool post (x-axis signal)
% column7: tri-axial accelerometer on the tool post (y-axis signal)
% column8: tri-axial accelerometer on the tool post (z-axis signal)

% \begin{figure}
% \centering
% \includegraphics[width=0.5\textwidth]{cutting_test_fig1_zoomed_clipped}
% \end{figure}

% \begin{figure}
% \centering
% \includegraphics[width=0.5\textwidth]{cutting_test_fig2_clipped}
% \end{figure}

% \begin{figure}
% \centering
% \includegraphics[width=0.5\textwidth]{cutting_test_fig1_clipped}
% \end{figure}

%!TEX root = ../machining_ML_wavelet.tex
%-------------------------------
%*******************************
\subsection{Data Labeling}
\label{sec:data_labeling}
%*******************************

\begin{figure}
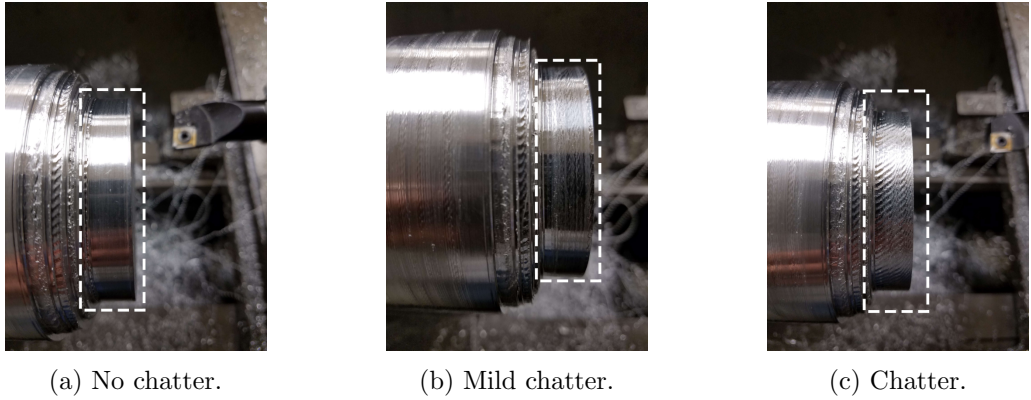

\centering
\begin{minipage}{0.3\textwidth}
\centering
\includegraphics[width=0.70\textwidth,height=.50\textheight,keepaspectratio]{2p5_stickout_320rpm_0p005doc_stable_wBorder}
\end{minipage}
\begin{minipage}{0.3\textwidth}
\centering
\includegraphics[width=0.70\textwidth,height=.50\textheight,keepaspectratio]{2p5_stickout_770rpm_0p005doc_mild_chatter_wBorder} 
\end{minipage}
\begin{minipage}{0.3\textwidth}
\centering
\includegraphics[width=0.70\textwidth,height=.50\textheight,keepaspectratio]{2p5_stickout_570rpm_0p015doc_chatter_wBorder}
\end{minipage}
\begin{minipage}[t]{0.3\textwidth}
\centering
\vspace{6pt}
(a) No chatter.
\end{minipage}
\begin{minipage}[t]{0.3\textwidth}
\centering
\vspace{6pt}
(b) Mild chatter.
\end{minipage}
\begin{minipage}[t]{0.3\textwidth}
\centering
\vspace{6pt}
(c) Chatter. 
\end{minipage}
\caption{The surface finish corresponding to (a) no-chatter case with 6.35 cm ($2.5$ inch) stickout length, $320$ rpm, and 0.127 mm ($0.005$ inch) depth of cut, 
(b) Mild chatter case with 6.35 cm ($2.5$ inch) stickout length, $770$ rpm, and 0.127 mm ($0.005$ inch) depth of cut, and 
(c) chatter case with 6.35 cm ($2.5$ inch) stickout length, $570$ rpm, and 0.127 mm ($0.005$ inch) depth of cut.}
\label{fig:surface_finish}
\end{figure}

Before tagging the signals, we analyzed the time series of the two uni-axial accelerometers on the boring rod as well as the signals from the tri-axial accelerometer on the tool post, see Fig.~\ref{fig:cutting_test}. We found that although the data of the accelerometers is mostly redundant, the $x$-vibration at the tool post, which is measured by the $x$-axis signal of the tri-axial accelerometer, had the best signal-to-noise ratio. Therefore, we performed the data tagging exclusively using the data from this channel. Another sanity check was the comparison of the tagged signals with a few photographs of the resulting machined surface taken during the experiment, as shown in Fig.~\ref{fig:surface_finish}.

%----------------------------------------------------
%----------------------------------------------------

\begin{figure}[htbp]
\centering
\begin{minipage}{0.48\textwidth}
\centering
\includegraphics[width=\textwidth,height=.50\textheight,keepaspectratio]{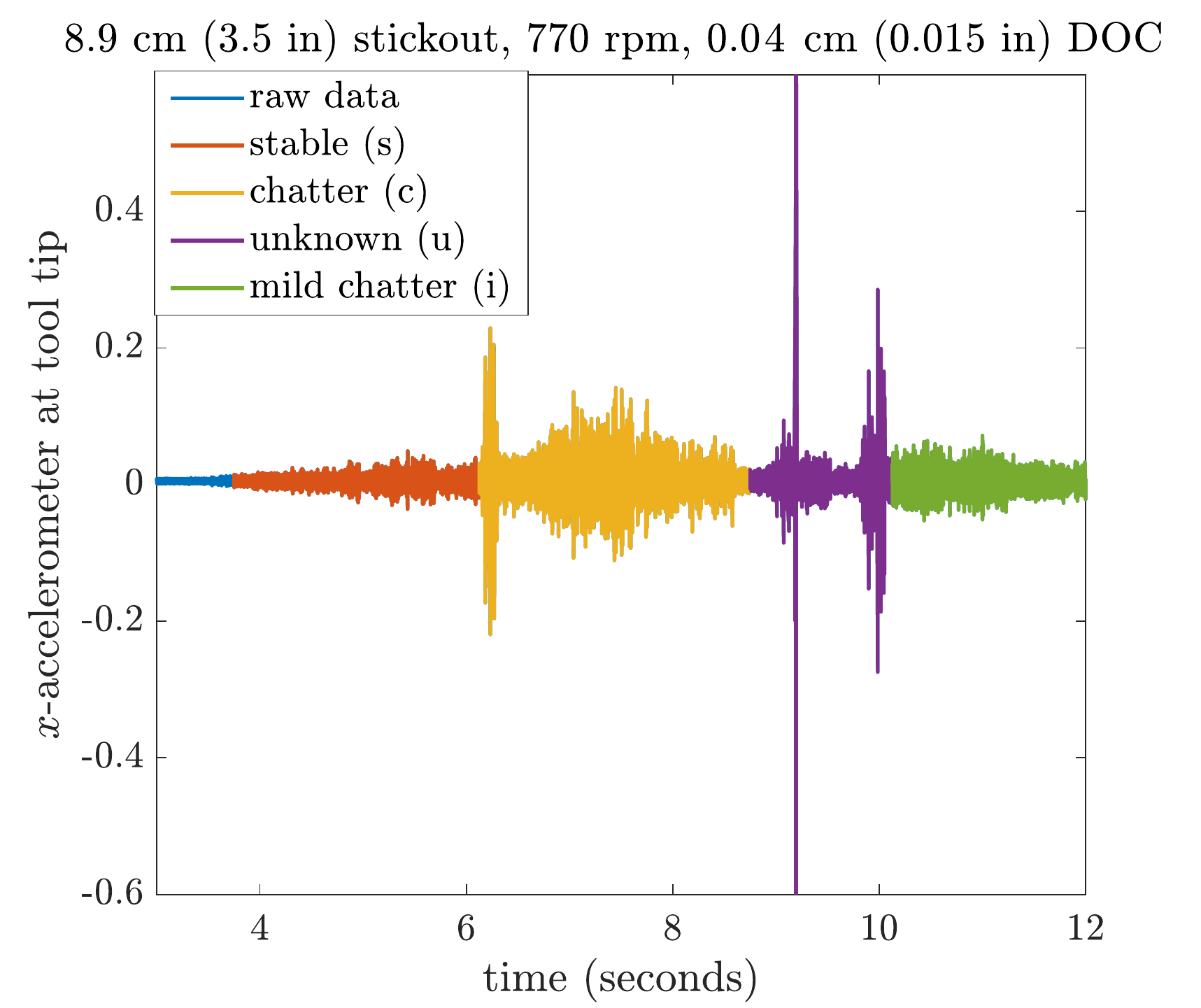}\\
\end{minipage}
\begin{minipage}{0.48\textwidth}
\centering
\includegraphics[width=\textwidth,height=.50\textheight,keepaspectratio]{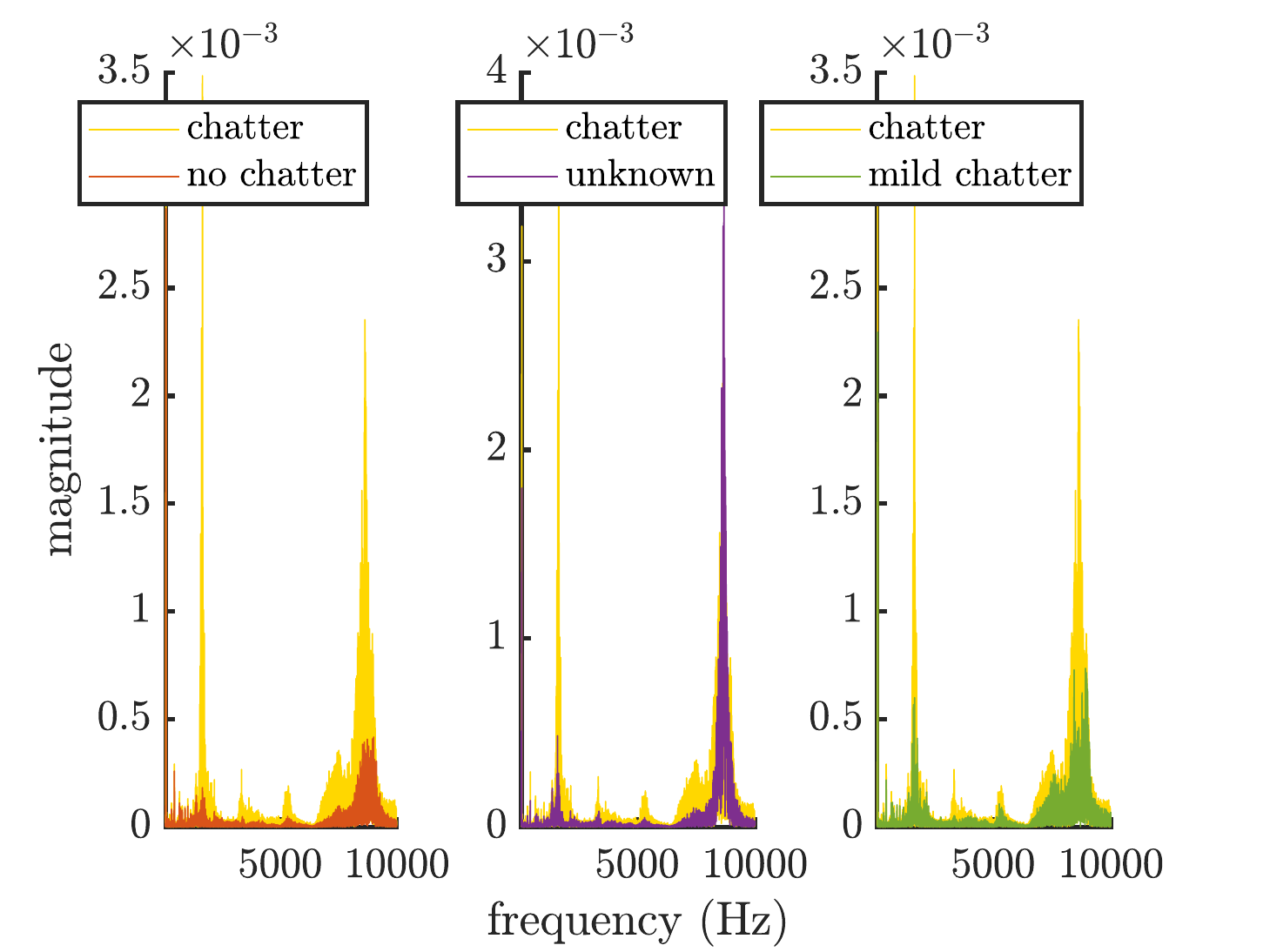}\\
\end{minipage} \\
\begin{minipage}{0.48\textwidth}
\centering
(a) time series
\end{minipage}
\begin{minipage}{0.48\textwidth}
\centering
(b) spectrum comparison
\end{minipage}
\caption{Tagging example in the (a) time domain and (b) the frequency domain for the case with $8.9$ cm (3.5 in) stickout, $770$ rpm, and $0.04$ cm (0.015 in) depth of cut.}
\label{fig:tagging_example}
\end{figure}

Each time series from every cutting test was examined and the different parts of the signals were labeled as either no chatter, mild/intermediate chatter, chatter, or unknown. Figure \ref{fig:tagging_example}a shows an example of how one time series is labeled using these categories. The separation into different parts has been done based on the characteristics of the amplitude in the time domain. In particular, parts with a low amplitude were separated from parts with a large amplitude. In addition, parts with an impact-like structure with an abrupt very strong increase of the accelerations and a relatively fast decay were also separated. Then the frequency domain characteristics were studied for a final classification of the signal. In the frequency domain only the frequency components lower than $5$ kHz were considered. Specifically, the criteria that we used for classifying the signals are:
%----------------------------------------------------
\begin{enumerate}
	\item No chatter (stable):
		\begin{enumerate}[nosep]
			\item Low amplitude in the time domain
			\item Low amplitude in the frequency domain (highest peaks at spindle rotation frequencies \cite{Insperger2008b})
		\end{enumerate}
	\item Mild or intermediate chatter:
		\begin{enumerate}[nosep]
			\item Low amplitude in the time domain
			\item Large amplitude in the frequency domain (highest peaks at chatter frequencies)
		\end{enumerate}
	\item Chatter:
		\begin{enumerate}[nosep]
			\item Large amplitude in time domain
			\item Large amplitude in the frequency domain (very high peaks at chatter frequencies which are not equal to the spindle rotation frequencies)
		\end{enumerate}
	\item Unknown:
	\begin{enumerate}[nosep]
			\item All other cases 
	\end{enumerate}
\end{enumerate}
%----------------------------------------------------
The unknown data are parts of the time series with a large amplitude in the time domain but no large peaks in the frequency domain at chatter frequencies (lower than $5$ kHz). 
Typically, this corresponds to the parts with impact-like structure, which might occur due to chip breakage or other inhomogeneities during the process. 
Also, there can be another eigenmode at 10kHz is vibrating (chattering) for unknown portion of the time series in Fig.\ref{fig:tagging_example}. 
However typical chatter frequencies in this process lie between 0-150 Hz for structural modes, 200-800 Hz for workpiece vibrations, and 1000-3000 Hz for tool vibrations. 
Therefore, it is not clear if this is chatter or something else, and therefore it was excluded from the analysis.
The time domain and the frequency domain characteristics for an example time series that includes all four classes are shown in Figure~\ref{fig:tagging_example}a and Figure~\ref{fig:tagging_example}b, respectively. The first part of the time series was not classified, because in this case the tool is still not engaged in the workpiece. 

%--------------------------------
%--------------------------------

From Figure~\ref{fig:tagging_example} it becomes also clear that a process with fixed cutting conditions is not necessarily clearly stable or unstable, which is another reason why chatter detection algorithms might be helpful for practical applications. For example, the process in Figure~\ref{fig:tagging_example} is stable at the beginning between $4$s and $6$s. Then, a strong perturbation drives the system away from the stable state to some chattering motion, which is a reasonable scenario because there can be a bistability between stable cutting and chatter \cite{Dombovari2008,2015dombovari,2017yan}. Table \ref{tab:chatter_case_number} shows breakdown and the total number of the tagged time series for each stickout length.

%----------------
\begin{table}
\centering
	\begin{tabular}{ccccc}
		\makecell{Stickout length\\ (cm (inch))} & \makecell{ Stable} & \makecell{Mild chatter} & \makecell{Chatter }  & \makecell{Total} \\
		\toprule 
		5.08 (2)	  & 17 	& 8   & 11 & 36 \\
		6.35 (2.5)	& 7 	& 4   & 3  & 14 \\
		8.89 (3.5)  & 7 	& 2   & 2  & 11 \\
		11.43 (4.5) & 13 	& 4   & 5  & 22 \\
		\bottomrule
	\end{tabular}
	\caption{Case numbers and overall amount of tagged data for each stickout cases.}
	\label{tab:chatter_case_number}
\end{table}
%----------------

%!TEX root = ../machining_ML_Similarity.tex
%-------------------------------
%************************************
\section{Wavelet Packet Transform with Recursive Feature Elimination (RFE)}
\label{sec:wpt}
%************************************

\begin{figure}[H]
\centering
\includegraphics[width=0.50\textwidth,height=.50\textheight,keepaspectratio]{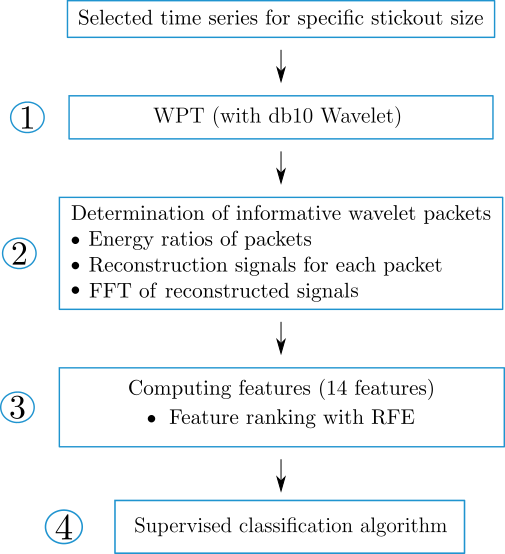}
\caption{Overview over the Wavelet Packet Transform (WPT) method with Recursive Feature Elimination (WPT).}
\label{fig:WPTSteps}
\end{figure}

In this section we describe the Wavelet Packet Transform (WPT) with Recursive Feature Elimination (RFE) for chatter detection in metal cutting. The method can be divided into four steps, which are summarized in Fig.~\ref{fig:WPTSteps}. The first step is the decomposition of the time series into wavelet packets. This is a technique from signal processing that is especially useful for a high resolution time-frequency analysis. The motivation for an additional decomposition of the signal is the increase of the signal-to-noise ratio and an increasing sensitivity for chatter features \cite{Chen2017}. The output of the WPT are different wavelet packets and the second step is the selection of the informative packets, based on the properties of the wavelet packets and the characteristics of chatter in the considered process. The third step is the feature extraction and its automatic ranking with the RFE method, which is used to distinguish between chatter and chatter-free motion. On the basis of the extracted features the fourth step is the classification into chatter/chatter-free cases via a Support Vector Machine (SVM), Logistic Regression (LR), Random Forest (RF) classification and Gradient Boosting (GB). 

%---------------
%---------------

%************************************* 
\subsection{Wavelet Packet Transform}
%*************************************

We follow Ref. \cite{Chen2017} and apply the WPT to the time series before feature extraction and classification. The WPT is an extension of the discrete wavelet transform. One level of the discrete wavelet transform decomposes the signal into a low and a high frequency component by passing it simultaneously through a low and a high pass filter. The properties of the two filters are related to each other and are determined by the chosen wavelet basis. According to \cite{Chen2017}, we use the Daubechies orthogonal wavelet db10 as the wavelet basis function. The outputs of the low and the high pass filter give the approximation coefficients and detailed coefficients denoted by $A_i$ and $D_i$, respectively, where the subscript $i$ specifies the level of the decomposition. The resulting signal after the decomposition is called wavelet packet and can be reconstructed from the approximation or detailed coefficients by using the filter properties \cite{Chen2017}. In the discrete wavelet transform only the output $A_i$ is passed again through both filters to generate two additional outputs $AA_{i+1}$ and $AD_{i+1}$ in the next level. 

\begin{figure}[h]
\centering
\includegraphics[width=\textwidth,height=.95\textheight,keepaspectratio]{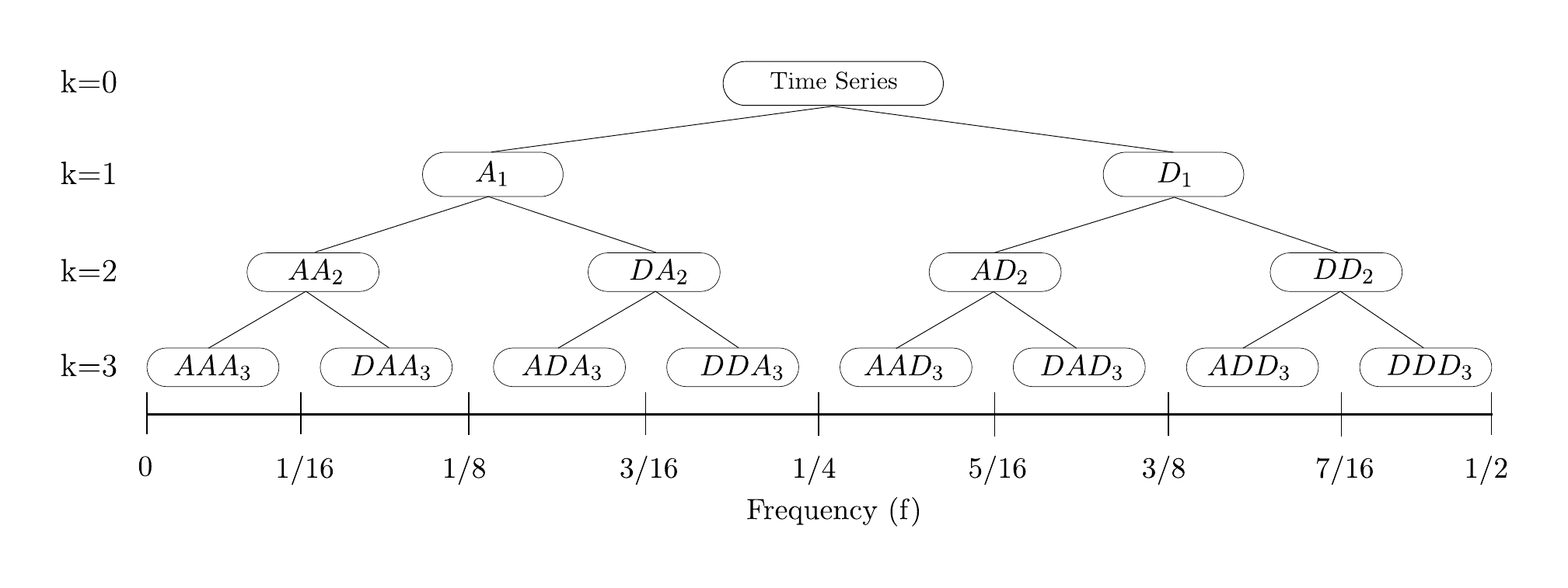}
\caption{3 Level Wavelet Packet Transform.}
\label{fig:wpt3}
\end{figure}

In contrast, in the WPT approach the output $A_i$ of the low pass filter as well as as the output $D_i$ of the high pass filter are both again low- and high-pass filtered to generate the wavelet packets $AA_{i+1}$, $AD_{i+1}$, $DA_{i+1}$ and $DD_{i+1}$ in the next level. This means that the WPT generates $2^{k}$ wavelet packets at the $k$th level, see Fig.~\ref{fig:wpt3} for a schematic of level $3$ WPT. In Fig.~\ref{fig:wpt3}, for example, $DAA_3$ denotes the packet in the third level, where in the first and the second level the low pass filter and in the third level the high pass filter have been applied. Before passing through the filters in the next level the signal is downsampled by a factor of two, which increases the frequency resolution. Moreover, since after each decomposition the two resulting wavelet packets contain only one-half of the frequencies of the input data this downsampling is possible without losing information. As a consequence the resulting wavelet packets in one level contain only a frequency band, which is mainly distinct from the bands of the other packets. Even if the frequency bands become narrower in each level, the packets contain rich information of the original signal due to the increase of the frequency resolution. The location of the frequency band is determined by the chronological order of the applied filters, which are used to generate the wavelet packet (cf. Fig.~\ref{fig:wpt3}). In the following, the wavelet packets are labeled according to the order of their frequency band beginning with 1 for the packet with the lowest frequencies ($A \ldots A_k$) resulting only from low pass filtering to $2^k$ for the packet containing the highest frequencies ($D \ldots D_k$) resulting from a successive application of the high pass filter.

%----------------
%----------------

\subsection{Selection of informative wavelet packets}
\label{sec:packetselection}

The next step is the selection of the informative wavelet packets, which are best suited to distinguish between stable cutting and chattering motion. The criteria for the selection of the informative wavelet packets are a high signal energy in comparison to other packets for a good signal-to-noise ratio, and a significant overlap of the frequency band of the packet with possible chatter frequencies.

\begin{figure}[h]
\centering
\includegraphics[width=0.95\textwidth,height=.85\textheight,keepaspectratio]{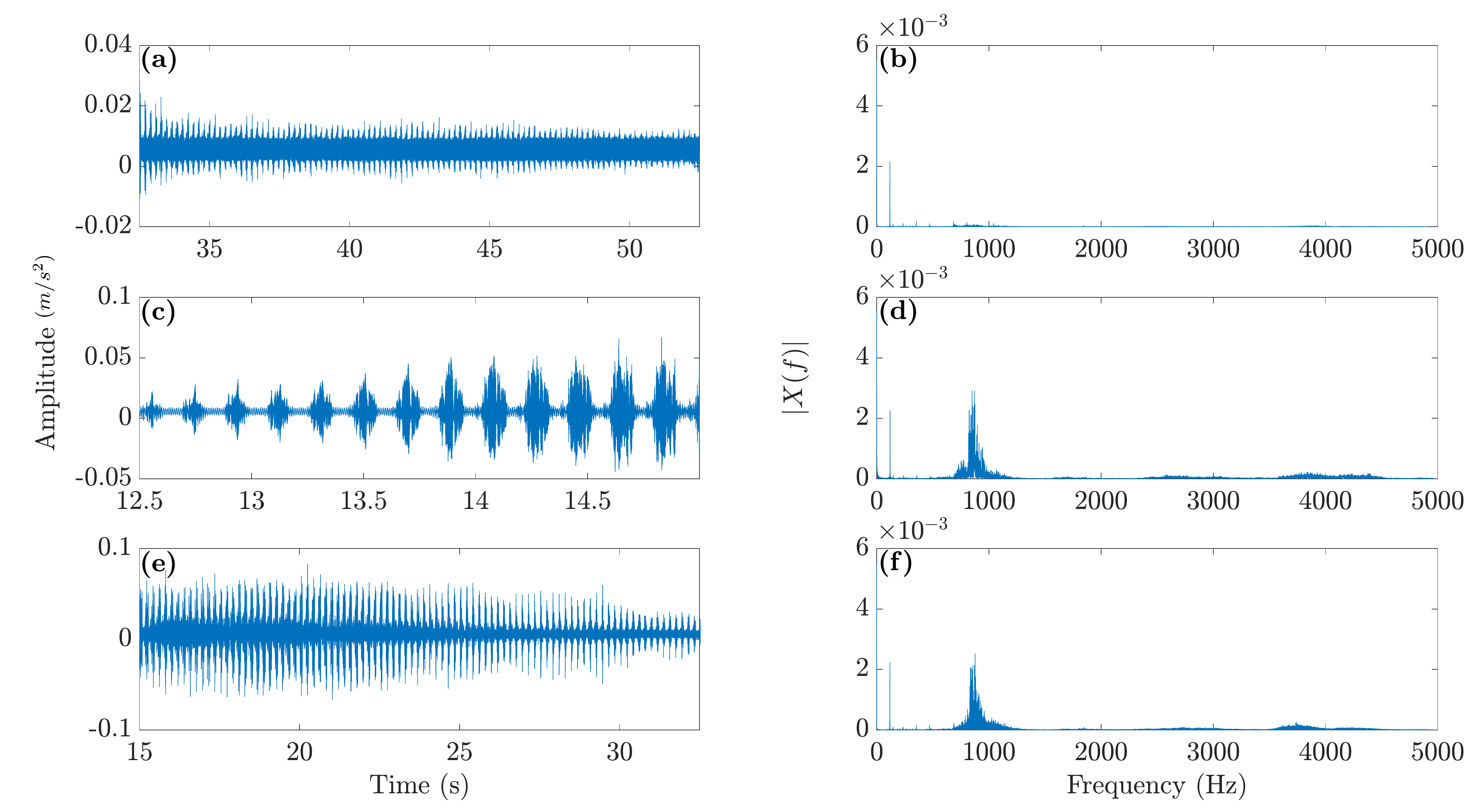}
\caption{Time domain and frequency domain of stable (a,b), intermediate (c,d) and chatter (e,f) regions for 5.08 cm (2 inch) stickout, 320 rpm, 0.002 inch depth of cut case.}
\label{fig:fft2320005}
\end{figure}

The identification of the band of chatter frequencies is done by examining the FFT of the signals tagged as stable, intermediate chatter, and chatter (see Section \ref{sec:data_labeling} for a description of the data labeling). Figure~\ref{fig:fft2320005} shows example time series and the corresponding Fourier spectra for three tagged signals for the case whose stickout length, rotational rpm, and depth of cut are $5.08$ cm ($2$ inch), $320$ rpm, and $0.127$ mm ($0.005$ inch), respectively. For stable cutting the dominant frequencies are low and correspond to the spindle rotation frequency. In addition, there is a significant peak at $120$ Hz, which can be found in all measurements and probably comes from an external source. For intermediate chatter and chatter a significant part of the energy in the signal is contained at high frequencies near $1000$ Hz, which is close to the eigenfrequency of the lateral tool vibration. As a consequence, these chatter frequencies become larger for increasing stickout length and for each of the four different stickout lengths a different range of chatter frequencies has been identified.

%---------------
%---------------

\begin{figure}[H]
\centering
\includegraphics[width=0.95\textwidth,height=.95\textheight,keepaspectratio]{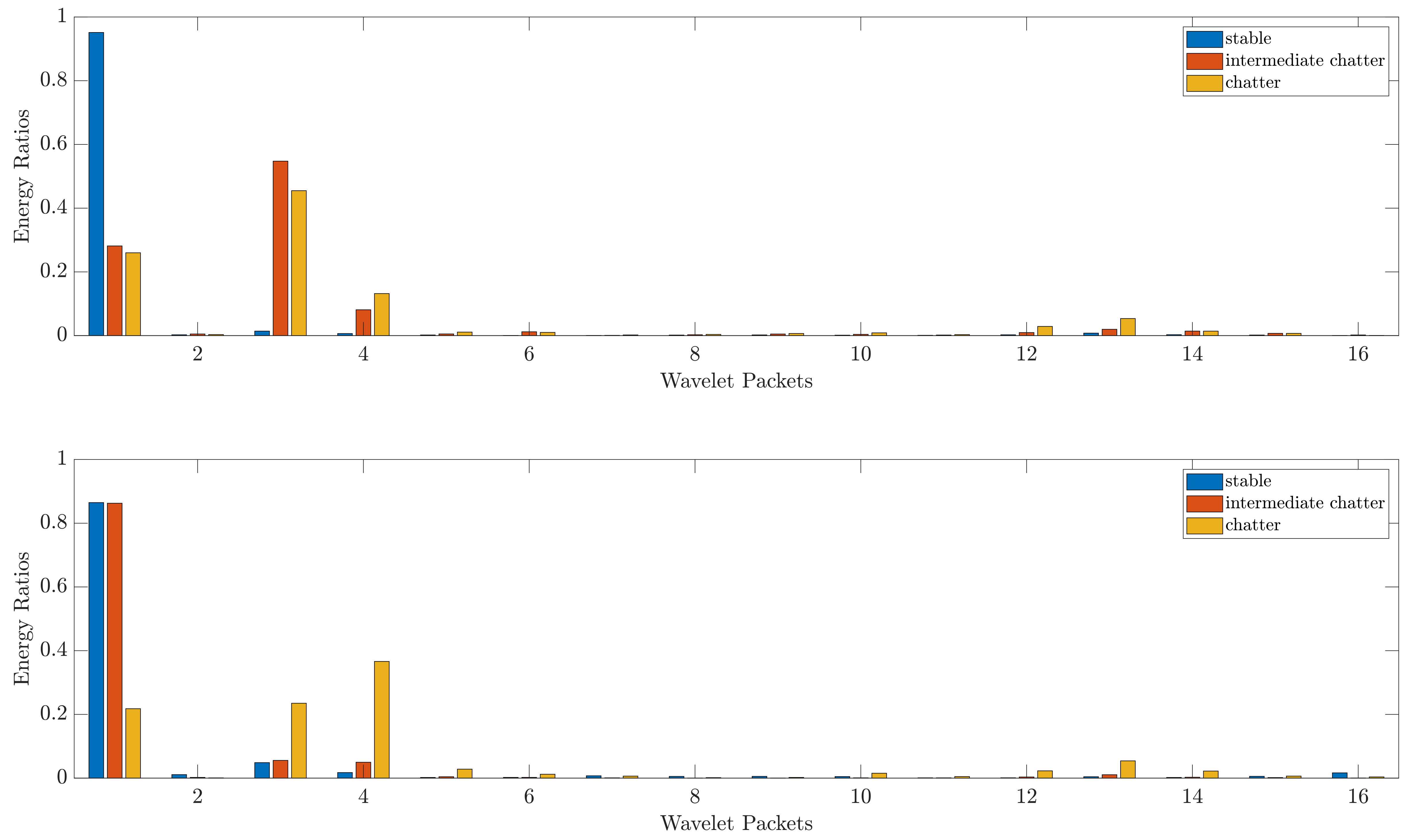}
\caption{Energy ratios of wavelet packets for two cases: (top) $5.08$ cm ($2$ inch) stickout, $320$ rpm and 0.0127 cm ($0.005$ inch) DOC, and (bottom) $5.08$ cm ($2$ inch) stickout, $570$ rpm and 0.00508 cm ($0.002$ inch) DOC. Note the differences in the scale of the vertical axis.}
\label{fig:energyratio}
\end{figure}

In order to analyze the properties of the wavelet packets, levels $1$, $2$, $3$, and $4$ WPT are obtained from the experimental data. Figure~\ref{fig:energyratio} shows the resulting level $4$ energy ratios of the wavelet packets for two example cases. The energy ratios represent the fraction of energy in each packet relative to the total energy in all the packets. It is obvious from the figure that for stable cutting most of the energy in this case is concentrated in the first wavelet. In contrast, for the intermediate chatter and the chatter regions the energy is concentrated mainly in the first, third and fourth wavelet packets. This is consistent with the behavior of the frequency spectrum of the original data in Figure~\ref{fig:fft2320005} since increasing the number of the wavelet packets corresponds to a higher frequency band. 

%---------------
%---------------

\begin{figure}[h]
\centering
\includegraphics[width=\textwidth,height=.95\textheight,keepaspectratio]{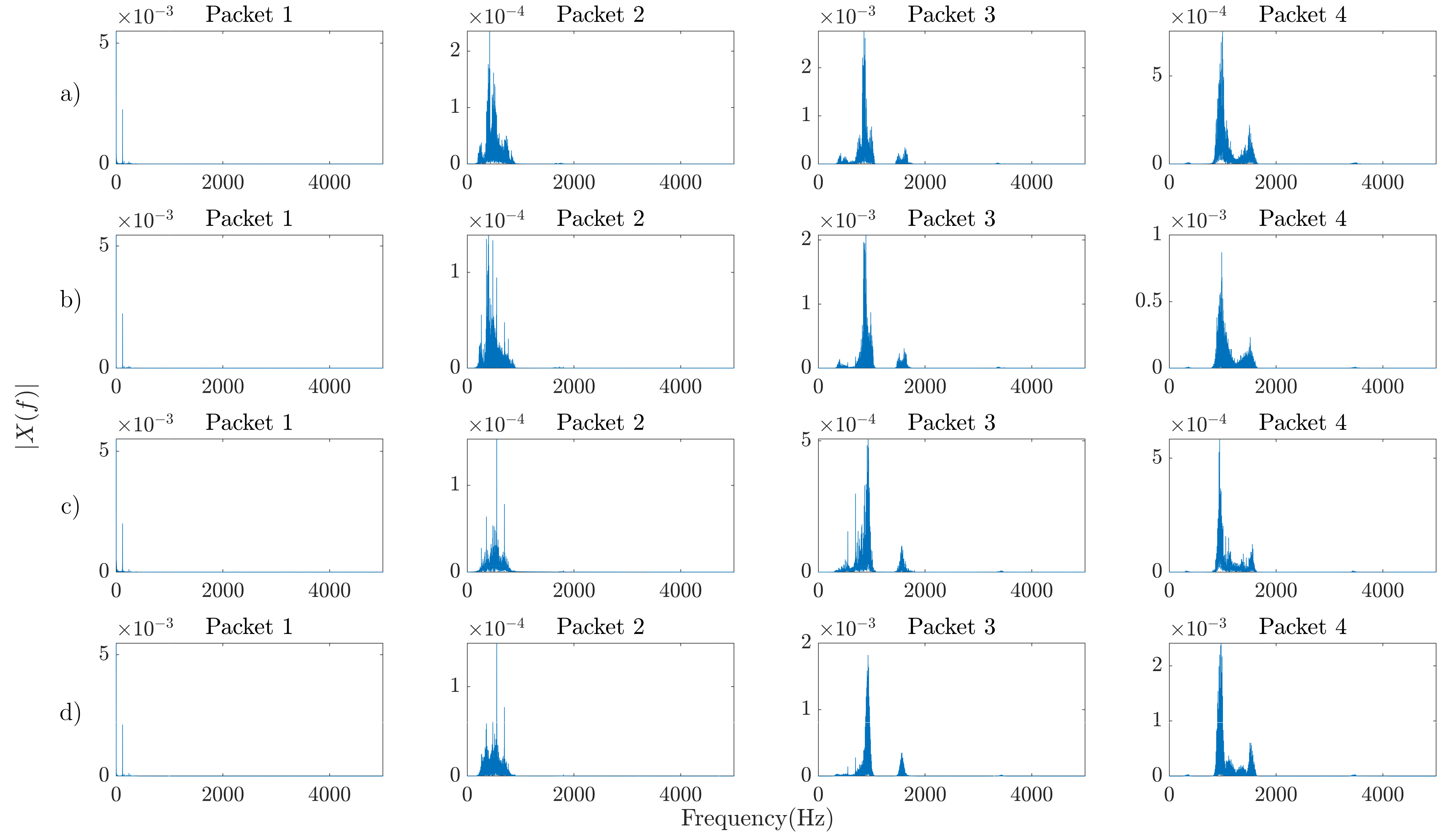}
\caption{Spectrum of first four wavelet packets of level 4 WPT for intermediate chatter (a),(c) and chatter (b),(d) in the case with 5.08 cm (2 inch) stickout size. The spindle speed and depth of cut is 320 rpm and 0.0127 cm (0.005 inch) in (a), (b) and 570 rpm and 0.00508 cm (0.002 inch) in (c), (d), respectively.}
\label{fig:intreconFFT}
\end{figure}

Upon identifying the wavelet packets whose energy ratios are relatively high with respect to the other packets, the third step is to identify the packets whose spectrum has significant peaks that overlap with the chatter frequencies given in Table~\ref{tab:chatter_freq_range} \cite{Chen2017}. Specifically, we reconstruct a time domain signal for each wavelet packet and obtain the corresponding FFT for each of the reconstructed signals. For the two examples with stickout length 5.08 cm ($2$ inch) the frequency spectrum of the reconstructed signals obtained from the first four wavelet packets for the intermediate chatter and chatter regions are provided in Fig.~\ref{fig:intreconFFT}. It can be seen that the peaks in the spectrum of the $3$rd and $4$th wavelet packet overlap with the band of the previously identified chatter frequency ($900$--$1000$ Hz, see Table~\ref{tab:chatter_freq_range} and Fig.~\ref{fig:fft2320005}). Since on average for the stickout length 5.08 cm ($2$ inch) the energy ratios and the amplitudes in corresponding FFT (see Fig.~\ref{fig:intreconFFT}) are slightly higher in the $3$rd wavelet packet than in the $4$th wavelet packet, we choose the $3$rd packet as the informative wavelet for chatter detection at level $4$ WPT. An overview of the selected informative wavelet packet for each level of the WPT can be found in Table~\ref{tab:chatter_freq_range}. For higher stickout length the dominant chatter frequencies increase, and therefore, in general, a wavelet packet with a higher frequency band is selected as the informative wavelet packet. 

%----------------
\begin{table}
\centering
	\begin{tabular}{cccc}
		\makecell{Stickout length\\ (cm (inch))} & \makecell{Chatter frequency range\\ (Hz)} & \makecell{Informative wavelet\\ packets} & \makecell{Informative\\ IMF} \\
		\toprule 
		5.08 (2)	  & $900$--$1000$ 	& Level 1 :1, Level 2: 1, Level 3: 2, Level 4: 3  & 2\\
		6.35 (2.5)	& $1200$--$1300$ 	& Level 1 :1, Level 2: 1, Level 3: 3, Level 4: 4  & 2\\
		8.89 (3.5) & $1600$--$1700$ 	& Level 1 :1, Level 2: 2, Level 3: 3, Level 4: 6  & 1\\
		11.43 (4.5) & $2900$--$3000$ 	& Level 1 :2, Level 2: 3, Level 3: 5, Level 4: 10 & 1\\
		\bottomrule
	\end{tabular}
	\caption{The chatter frequency ranges, the informative wavelet packets, and the informative IMFs corresponding to each stickout length of the cutting tool.}
	\label{tab:chatter_freq_range}
\end{table}
%----------------
%---------------
%---------------
%---------------
%\begin{figure}[H]
%\centering
%\includegraphics[width=\textwidth,height=.95\textheight,keepaspectratio]{figures/2inch_320_005_Chatter_WPT_Reconstruction_FFT}
%\caption{Spectrum of each wavelet packet for chatter state of 2 inch stickout, 320 rpm and 0.005 inch depth of cut}
%\label{fig:chatterreconFFT}
%\end{figure}
%---------------

We note that the informative wavelet packet is not necessarily the one with the highest energy because it is important that the range of possible chatter frequencies are in the frequency band of the informative wavelet packet. In fact, often the first packet has the highest energy ratio but its frequency band does not overlap with the chatter frequencies, which are mainly contained in packets with a higher index (cf. Table~\ref{tab:chatter_freq_range}). 

Since the frequency band of the wavelet packets can be predicted from the WPT tree in Fig.\ref{fig:wpt3}, it is also possible to predict the informative wavelet packet that contains information about chatter frequencies. For example, from the sampling rate $10$ kHz it follows that the first wavelet packet in level 3 corresponds to the frequency band $0$--$625$ Hz. The upper frequency limits for other packets in level 3 are equal to the corresponding wavelet packet number times the upper frequency level of the first wavelet packet (cf. Fig.\ref{fig:wpt3}). Table \ref{tab:Informative_packet_comparison} provides the predicted and the selected informative wavelet packets for the level $4$ WPT. For all cases, the selected informative wavelet packets are consistent with the predicted ones.
 
%----------------
\begin{table}[H]
\centering
	\begin{tabular}{cccc}
		\makecell{Stickout length\\ (cm (inch))} & \makecell{Chatter frequency range\\ (Hz)} & \makecell{Informative wavelet\\ packets (Predicted)} & \makecell{Informative wavelet\\ packets (Selected)} \\
		\toprule 
		5.08 (2)	  & $900$--$1000$ 	& Level 4: 3-4   & Level 4: 3\\
		6.35 (2.5)	& $1200$--$1300$ 	& Level 4: 4-5   & Level 4: 4\\
		8.89 (3.5)  & $1600$--$1700$ 	& Level 4: 6     & Level 4: 6\\
		11.43 (4.5) & $2900$--$3000$ 	& Level 4: 10    & Level 4: 10\\
		\bottomrule
	\end{tabular}
	\caption{Comparison between predicted and selected informative wavelet packet number for all stickout cases. 
	Predicted stickout cases are decided with by overlapping the chatter frequency with the wavelet packet frequency range obtained from the WPT tree (Fig. \ref{fig:wpt3}) for level 4.}
	\label{tab:Informative_packet_comparison}
\end{table}
%----------------

%In addition, since the wavelet packets in lower level WPT correspond to a broader frequency range, it is assumed that an informative packet selected in first level includes frequency information for upper levels. To validate this assumption, we implemented transfer learning whereby a classifier is trained with first level WPT feature matrices for 2 inch case and then, it is tested on other cutting configurations in level 1 and level 4. More detailed discussions on the results of this implementation is presented in section \ref{sec:results}.
 
%\textcolor{blue}{[AO: Another interesting point is the verification of the hypothesis that a lower level WPT leads to better results if the chatter frequencies have a broader range. I think this could be one result of the transfer learning. For example, if the method trained for 2inch stickout length is applied on 3.5ich stickout length data, I expect that the 2nd-level WPT is better than the 3rd- or 4th-level WPT because in a lower level the informative packet has a broader frequency band and may contain still a significant part of the chatter frequencies.]}

\subsection{Recursive Feature Elimination}
\label{sec:rfe}

The reconstructed signal from the informative wavelet packet allows the extraction of both frequency domain as well as time domain features for chatter identification. A collection of frequency domain and time domain features, which are taken from Ref.~\cite{Chen2017}, are tested in this paper and are provided in Table~\ref{tab:features}. 

We used Python to train an SVM classifier combined with Recursive Feature Elimination (RFE) where in this case we have a maximum of $14$ features at the level $4$ WPT. Recursive feature elimination is an iterative process that eliminates one of the features in each iteration until all the features are removed for classification \cite{Chen2017}, which means that the number of iterations for RFE equals the number of the considered features. Elimination of features is based on their influence on the classification: the feature with the smallest effect, is eliminated in each iteration \cite{Yan2015}. At the end, RFE returns a feature ranking list corresponding to one specific training set.

The ranked features are used to generate feature vectors where the first vector contains only the first ranked feature, while each consecutive feature vector adds the subsequent feature in the ranking until all the features are included in the $14$th vector at the fourth level of WPT, see Tables~\ref{tab:resultsforwholecases1}--\ref{tab:resultsforwholecases4} for examples. The classification accuracy is calculated for all $14$ feature vectors. In other words, in the first step only the top ranked feature is used, and in each further step, the next highest ranked feature was added to the feature matrix and the classification accuracy is computed again. 

%---------------
%\begin{figure}[h]
%\centering
%\includegraphics[width=\textwidth,height=.95\textheight,keepaspectratio]{figures/2inch_320_005_570_002_Chatter_Packet_4_ReconFFT}
%\caption{Spectrum of reconstructed signal for 5.08 cm (2 inch) stickout, 320 rpm and 0.0127 cm (0.005 inch) DOC (up) and 5.08 cm (2 inch) stickout, 570 rpm and 0.00508 cm (0.002 %inch) DOC (down).}
%\label{fig:SpectNewSignal}
%\end{figure}
%---------------

%---------------
\begin{table}[H]
\centering
\renewcommand{\arraystretch}{1.8} % Default value: 1
\caption{Time domain features ($a_{1},\ldots,a_{10}$) and frequency domain features ($a_{11,\ldots,14}$).}
\label{tab:features}
\begin{tabular}{|l  l|}
\hline
\multicolumn{2}{|c|}{Features} \\
\hline
$a_{1}= \frac{1}{N}\sum\limits_{m=1}^{N} x_{m}$ (Mean)& $a_{8}= \frac{a_{4}}{(\frac{1}{N}\sum\limits_{m=1}^{N} \sqrt{|x_{m}|})^{2}}$ (Clearance Factor)\\
$a_{2}= \sigma(x_{m})$ (Standard Deviation) &  $a_{9}= \frac{a_{3}}{\frac{1}{N}\sum\limits_{m=1}^{N} |x_{m}|}$ (Shape Factor) \\
$a_{3}= \sqrt{\frac{1}{N}\sum\limits_{m=1}^{N} x_{m}^{2}}$ (RMS) & $a_{10}= \frac{a_{4}}{\frac{1}{N}\sum\limits_{m=1}^{N} |x_{m}|}$ (Impulse Factor)\\
$a_{4}= max{(|x_{m}|)}$ (Peak) & $a_{11}= \frac{\sum\limits_{k=1}^{M} f_{k}^{2}|X(f_{i})|}{\sum\limits_{k=1}^{M} |X(f_{k})|}$ (Mean Square Frequency) \\
$a_{5}= \frac{\sum\limits_{m=1}^{N} (x_{m}-a_{1})^{3}}{(N-1)a_{3}^3}$ (Skewness) & $a_{12}= \frac{\sum\limits_{k=1}^{M} cos(2\pi f_{k}\Delta{t})|X(f_{k})|}{\sum\limits_{k=1}^{M} |X(f_{k})|}$ (One Step Auto Correlation Function)\\
$a_{6}= \frac{\sum\limits_{m=1}^{N} (x_{m}-a_{1})^{4}}{(N-1)a_{3}^4}$ (Kurtosis) & $a_{13}= \frac{\sum\limits_{k=1}^{M} f_{k}|X(f_{i})|}{\sum\limits_{k=1}^{M} |X(f_{k})|}$ (Frequency Center) \\
$a_{7}= \frac{a_{4}}{a_{3}}$ (Crest Factor)  & $a_{14}= \frac{\sum\limits_{k=1}^{M} (f_{k}-a_{13})^{2}|X(f_{i})|}{\sum\limits_{k=1}^{M} |X(f_{k})|}$ (Standard Frequency) \\

\hline
\end{tabular}
\end{table}
\section{Ensemble Empirical Mode Decomposition (EEMD) with Recursive Feature Elimination(RFE)}
\label{sec:eemd}
%*************************************
In this section we describe the Ensemble Empirical Mode Decomposition (EEMD) for chatter detection in metal cutting. The structure of the method is similar to the WPT based approach, which is described in Section~\ref{sec:wpt}. However, in contrast to the WPT method here the EEMD is used for the decomposition of the original time series and the output of the EEMD are intrinsic mode functions (IMF) instead of wavelet packets. After the decomposition, the informative IMF is selected and various features for chatter detection are extracted. The features are automatically ranked via the RFE method and SVM is used to classify into chatter/chatter-free cases.

\subsection{Ensemble Empirical Mode Decomposition}

EEMD is based on the Empirical Mode Decomposition (EMD), which is an elementary step in the Hilbert-Huang transform \cite{1998huang}. Similar to WPT, EMD is useful for non-stationary signals since the resulting IMFs contain time and frequency information of the signal. The main difference in contrast to WPT and other linear decomposition methods is that the expansion bases of EMD are not fixed but are rather adaptive and they are determined by the data. On the one hand, this means that EMD is a nonlinear decomposition and, on the other hand, it is suitable for analyzing nonlinear and non-stationary data \cite{1998huang}. 

The algorithm for the decomposition of a given time series $s(t)$ can be described as follows. The first residue $r_0(t)$ is equivalent to the original data, i.e. $r_0(t)=s(t)$. Then the IMFs $c_{i}(t)$ with $i\geq 1$ are generated from the residues $r_{i-1}(t)$ by repeated application of the so-called sifting process described below. After extracting the $i$th IMF $c_{i}(t)$, the next residue is calculated by
\begin{equation}
\label{eq:EMDiteration}
r_{i}(t)=r_{i-1}(t)-c_i(t).       
\end{equation}
This procedure is repeated until the result of Eq.~\eqref{eq:EMDiteration}, that is the $i$th residue $r_{i}(t)$, becomes a monotonic function and no more IMFs can be extracted. As a result the decomposition of the original data can be given by
%----------------
\begin{equation}
s(t) = \sum_{i=1}^{N} c_{i}(t)+r_{N},
\label{eq:EMD}
\end{equation}
%----------------

The sifting process for the generation of the $i$th IMF $c_i$ from the residue $r_{i-1}$ is done via the following iterative scheme. A lower and upper envelopes of the data are generated by using cubic splines for an interpolation between the local minima and maxima of the residue, respectively. The mean $m_1(t)$ of the lower and upper envelope is calculated. The first guess for the IMF is obtained by the difference between the residue $r_{i-1}$ and $m_1(t)$. Then the first guess for the IMF is treated as the new data and the sifting process is repeated until a given stoppage criterion is fulfilled. As a consequence of the iteration, the lower and the upper envelopes of the final IMF $c_i(t)$ are nearly symmetric and the mean of the latter is approximately zero. Moreover, the number of extrema and the number of zero crossings is equal or differs at most by one. IMFs with lower indices correspond to high frequency bands while the ones with higher indices correspond to lower frequency bands. These properties of the decomposition make it useful for further data analysis. 

However, one major problem with the original EMD is the occurrence of mode mixing, which means that one IMF contains two signals, whose frequency bands are totally different, or a signal of similar scale is observed inside different IMFs whose frequency bands are different \cite{ZHAOHUA2009}. EEMD was developed to solve the mode mixing problem in EMD \cite{Wang2012}. Accordingly, Wu and Huang \cite{ZHAOHUA2009} proposed the following steps for EEMD:
\begin{enumerate}[nosep]
  \item Create an ensemble from the original data by adding white noise.
  \item Decompose each member of the ensemble into IMFs.
  \item Compute the ensemble means of the corresponding IMFs.
\end{enumerate}
The added white noise amplitude must not exceed $20\%$ of the standard deviation of the original signal while the ensemble size for the EEMD can be selected as $200$ \cite{Chen2018}. For our analysis we used the Python package \href{https://github.com/wmayner/pyemd}{PyEMD} with the default stoppage criterion \cite{pele2008,pele2009}. We set the ensemble number and the noise width parameter to $200$ and $0.2$ ($20\%$), respectively.

% %************************************
\subsection{Selection of informative intrinsic mode function}
\label{sec:informative_imf} 
%************************************

\begin{figure}[H]
\centering
\includegraphics[width=0.90\textwidth,height=.95\textheight,keepaspectratio]{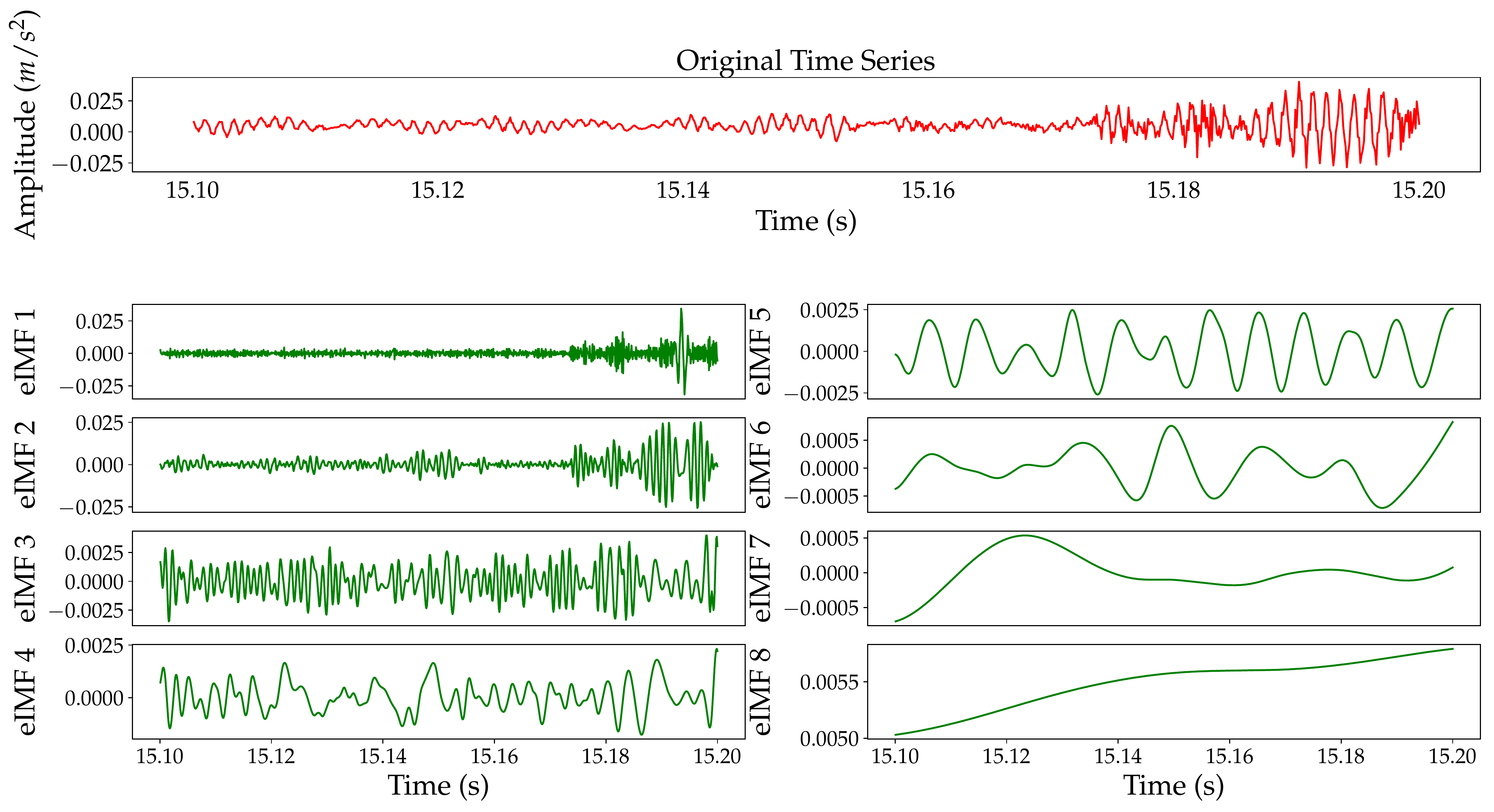}
\caption{The original time series and the corresponding intrinsic mode functions (IMFs) for the case of 5.08 cm ($2$ inch) stickout, $320$ rpm, and 0.0127 cm ($0.005$ inch) DOC.}
\label{fig:IMFs}
\end{figure}

\begin{figure}[H]
\centering
\includegraphics[width=0.80\textwidth,height=.95\textheight,keepaspectratio]{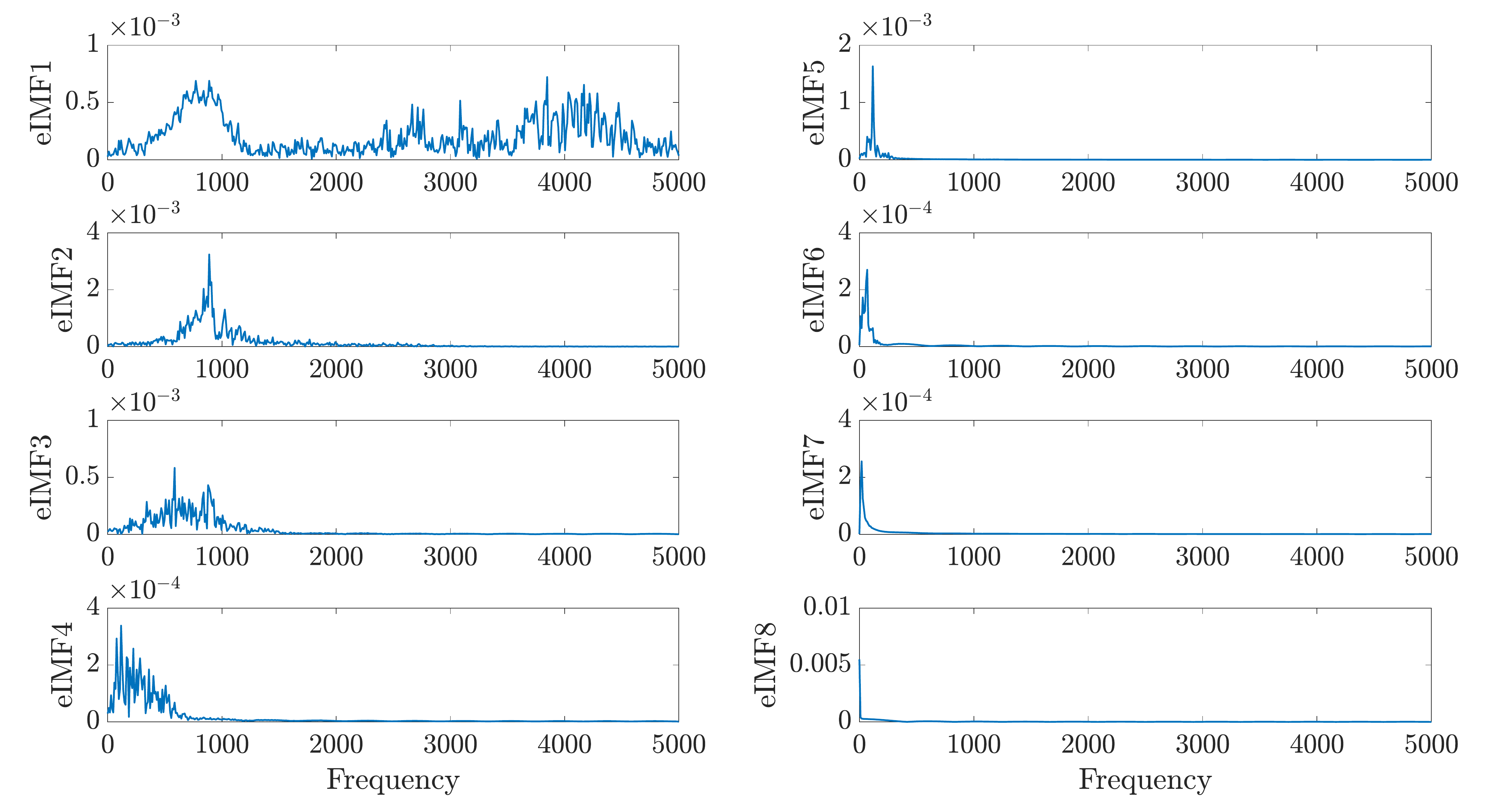}
\caption{The spectrum of each intrinsic mode function (IMF) for the case of 5.08 cm ($2$ inch) stickout, $320$ rpm and 0.0127 cm ($0.005$ inch) DOC.}
\label{fig:fftimfs}
\end{figure}

In order to obtain features for machine learning from vibration signals using EEMD, we first decompose the vibration signals into IMFs, see Fig.~\ref{fig:IMFs} for an example. For long time series, we reduced the computation time for this step by dividing the signal into shorter segments whose length is approximately $1000$ points. The informative IMF selection process is very similar to their WPT counterparts, see Sec.~\ref{sec:packetselection}. Specifically, the power spectrum in Fig.~\ref{fig:fftimfs} shows that the first IMF includes the high frequency vibrations while higher order IMFs include the low frequency ones. 
For example, for the 5.08 cm ($2$ inch) stickout case, the FFT of the second IMF matches the chatter frequency region ($900$--$1000$ Hz). Therefore, in this case, the second IMF is selected as the informative IMF. The informative IMFs for the other stickout cases are summarized in Table~\ref{tab:chatter_freq_range}.
%-----------------------
%-----------------------
%-----------------------
%-----------------------

% %************************************
\subsection{Feature extraction using EEMD} 
\label{sec:EEMDandFeatureRanking}
%************************************

Similar to Chen et al.~\cite{Chen2018}, we extract seven time domain features from the informative IMF. These features are listed in Table~\ref{tab:features7} and they include the energy ratio, peak to peak value, standard deviation, root mean square, crest factor, as well as skewness and kurtosis of the signals. The features are computed and then ranked using the Recursive Feature Elimination (RFE) method which was introduced in \cite{Guyon2002} and is described in Sec.~\ref{sec:rfe}. The feature matrix for classification is formed starting with the top-ranked feature by itself and then by concatenating, in descending order, the rest of the features one at a time. This results in seven combinations of features, which are then used for classification into chatter and chatter-free cases via four different classifiers similar to the WPT approach (cf. Sec.~\ref{sec:svm}).

%--------------
\begin{table}[H]
\centering
\renewcommand{\arraystretch}{2} % Default value: 1
\caption{Time domain features for the intrinsic mode functions $c_{i}(t_{k})$. The parameters $t_{k}$ and $\bar{c_{i}}$ represent, respectively, the $k$th discrete time and the mean of $i$th IMF.}
\label{tab:features7}
\begin{tabular}{|l  l|}
\hline
Feature &  Equation \\
\hline
Energy ratio & $f_{1} = \dfrac{\sum\limits_{k=1}^{n} c_{i}^{2}(t_{k})}{\sum\limits_{i=1}^{I} \sum\limits_{k=1}^{n} c_{i}^{2}(t_{k})}$\\
Peak to Peak & $f_{2} = max(c_{i}(t_{k}))-min(c_{i}(t_{k}))$\\
Standard Deviation & $f_{3} = \sigma(c_{i}(t_{k}))$\\
Root Means Square (RMS) & $f_{4} = \sqrt{\frac{1}{n}\sum\limits_{k=1}^{n} c_{i}^{2}(t_{k})}$\\
Crest Factor & $f_{5} = \frac{max(c_{i}(t_{k}))}{f_{4}}$\\
Skewness & $f_{6} = \frac{\sum\limits_{k=1}^{n} (c_{i}(t_{k})-\bar{c_{i}}(t_{k}))^{3}}{(n-1)f_{4}^3}$ \\
Kurtosis & $f_{7} = \frac{\sum\limits_{k=1}^{n} (c_{i}(t_{k})-\bar{c_{i}}(t_{k}))^{4}}{(n-1)f_{4}^4}$ \\
\hline
\end{tabular}
\end{table}
%--------------

%!TEX root = ../machining_ML_Wavelet_CIRP_Journal_Version.tex
%-------------------------------
\section{Classification Algorithms}
\label{sec:classifiers}
This section gives background information on the different classifiers used to test the performance of considered feature extraction methods, namely SVM, logistic regression, random forest classification, and gradient boosting. 
\subsection{Support Vector Machine}
\label{sec:svm}
\begin{figure}[H]
\centering
\includegraphics[width=0.60\textwidth,height=.40\textheight,keepaspectratio]{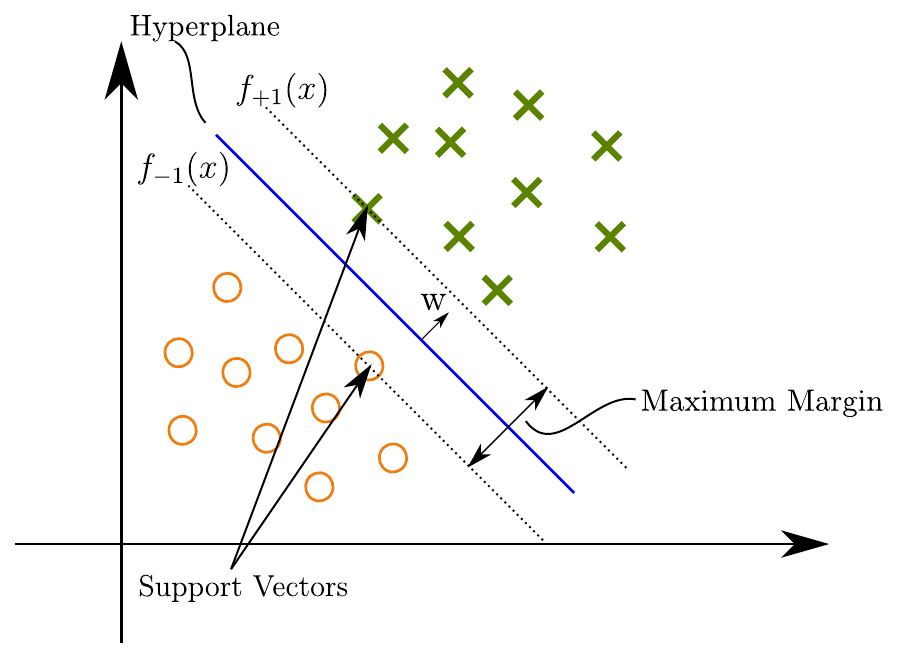}
\caption{Selected optimal hyperplane for linearly separable two class data set.}
\label{fig:SVM_hyperplane}
\end{figure}

A Support Vector Machine (SVM) is used to classify the time series by using the feature vectors. The Support Vector Machine algorithm is a supervised machine learning technique for finding the optimal hyperplane that separates two classes of a training data set. This hyperplane can then be used to classify the test data. The two dimensional case of a linear SVM is illustrated in Fig.~\ref{fig:SVM_hyperplane}. The feature vectors corresponding to two different classes, e.g. chatter (crosses) and no-chatter (circles), form two linearly separable data sets. The optimal hyperplane is selected such that the perpendicular distances from the feature vectors, which are closest to the hyperplane and also called support vectors, are equal. This means that the optimal hyperplane has the largest margin \cite{burges1998tutorial}. In general, it can be described by the set of points $\mathbf{x}$ satisfying 
\begin{equation}
\label{eq:optimal_hyperplane}
\mathbf{w} \cdot \mathbf{x} + c = 0,
\end{equation}
and the dashed lines where the support vectors lie on are defined according to
\begin{equation}
\label{eq:two_hyperplane}
f_{\pm 1}(\mathbf{x}) =  \mathbf{w} \cdot \mathbf{x} + c = \pm 1.
\end{equation}
Then, the margin of the optimal hyperplane can be denoted as $2/\Vert \mathbf{w} \Vert$. The two hyperplanes with Eq.~\ref{eq:two_hyperplane}, and therefore the optimal hyperplane from Eq.~\eqref{eq:optimal_hyperplane}, can be found by maximizing the distance $2/\Vert \mathbf{w} \Vert$ or by minimizing ${\Vert \mathbf{w} \Vert}^{2}$ with the constraints  
\begin{equation}
\begin{split}
\mathbf{w} \cdot \mathbf{x} + c &\geq +1, \quad \text{and } \\
\mathbf{w} \cdot \mathbf{x} + c &\leq -1.
\end{split}
\label{eq:constraints}
\end{equation}
The classification for a feature vector $\mathbf{x}_\text{test}$ of the test set can be made by checking the sign of the expression $\mathbf{w} \cdot \mathbf{x}_\text{test} + c$, which defines the label for the two classes. For the theory behind multi-class classification with SVM, one can refer to \cite{weston1998multi}. For some cases the training data are not separable by a linear hyperplane. In this case, the SVM can be extended to nonlinear classification with the help of kernel functions \cite{Maji2008}. 

%---------Logistic Regression--------------------

\subsection{Logistic Regression}
\begin{figure}[H]
\centering
\includegraphics[width=0.75\textwidth,height=.40\textheight,keepaspectratio]{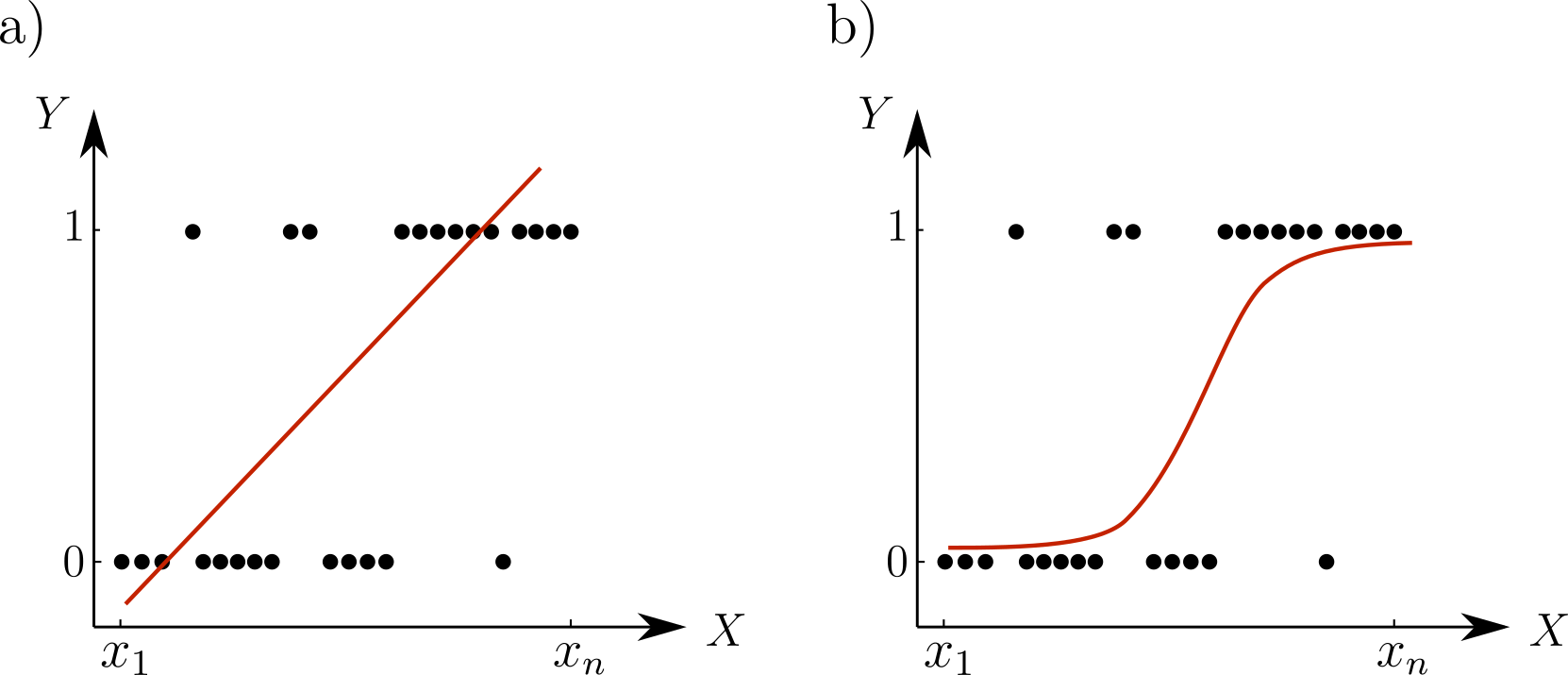}
\caption{Linear (a) and logistic (b) regression onto data set whose output is binary.}
\label{fig:linear_vs_logistic}
\end{figure}

Logistic regression is a supervised learning classification algorithm that computes the probability of two class labels for a given dependent variables \cite{Dreiseitl2002}. 
It is quite similar to linear regression but if differs in that is output is divided into two categories \cite{hosmer2013}. 
Figure \ref{fig:linear_vs_logistic} illustrates linear and logistic regression on a binary dataset. 
In this figure, $X=\{x_{1},x_{2},\ldots,x_{n}\}$ is the set of elements in the feature vector while $Y \in \{0, 1\}$ is the dichotomous outcome variable. 
For dichomotous output, linear regression can be applied but the model will not fit well as shown in Figure \ref{fig:linear_vs_logistic}a. 
There are are two main reasons for why the linear equation does not explain the relation between the variables $X$ and $Y$ \cite{Peng2002}: (1) the relationship between the variables does not have a linear trend and (2) the errors are not constant or they are not normally distributed. 
However, this problem can be solved by introducing the logit transformation. 

Let $\pi(x)=E(Y|x)$ be the expected value of $Y$ given the value of $x$. 
The regression model $g(x)$ and the logit transformation $\pi(x)$, respectively, are defined according to \cite{hosmer2013}
\begin{subequations}
\begin{equation}
\label{eq:logit_transform}
g(x) = \ln\Big(\frac{\pi(x)}{1-\pi(x)}\Big)=\beta_{0}+\beta_{1}x ,
\end{equation}

\begin{equation}
\label{eq:prob_sigmoid}
\pi(x)=\frac{e^{\beta_{0}+\beta_{1}x}}{1+e^{\beta_{0}+\beta_{1}x}},
\end{equation}
\end{subequations}

where $\pi(x)$ is defined as the sigmoid function for logistic regression in Eq.~\ref{eq:prob_sigmoid}.
The regression model is expressed as linear function, however it is converted into nonlinear probability function with logit transformation. 
Although Eq.~\ref{eq:logit_transform} is defined for only one independent input variable $x$, the model can be further extended to a multivariate version. 
To assign labels for a given input $x$, the decision boundary must first be formed. 
In Fig.~\ref{fig:linear_vs_logistic}b this boundary is the sigmoid function that splits tags $0$ and $1$. 
The $x$ values which satisfy $\beta_{0}+\beta_{1}x=0$ form the decision boundary \cite{Dreiseitl2002}, and the probability at the boundary, per Eq.~\eqref{eq:prob_sigmoid}, is $0.5$. 
The parameters $\beta_{0}$ and $\beta_{1}$ in the regression model can be identified using maximum likelihood estimators \cite{Dreiseitl2002}. 

% For two categorical output which are $1$ and $0$, probabilities can be expressed as $\pi(x)$ and $1-\pi(x)$ respectively. Then, the likelihood function is given as \cite{hosmer2013}:

% \begin{equation}
% \label{eq:likelihood}
% l(\beta_{0},\beta_{1}) = \prod_{i=1}^{n} \pi(x_{i})^{y_{i}}[1-\pi(x_{i})]^{1-y_{i}}
% \end{equation}
% where $y_{i}$ is either 0 or 1. $\beta_{0}$,$\beta_{1}$ are estimated by finding values which maximize Eq.\ref{eq:likelihood}.

\subsection{Random Forest Classification}
\begin{figure}[H]
\centering
\includegraphics[width=0.9\textwidth,height=.40\textheight,keepaspectratio]{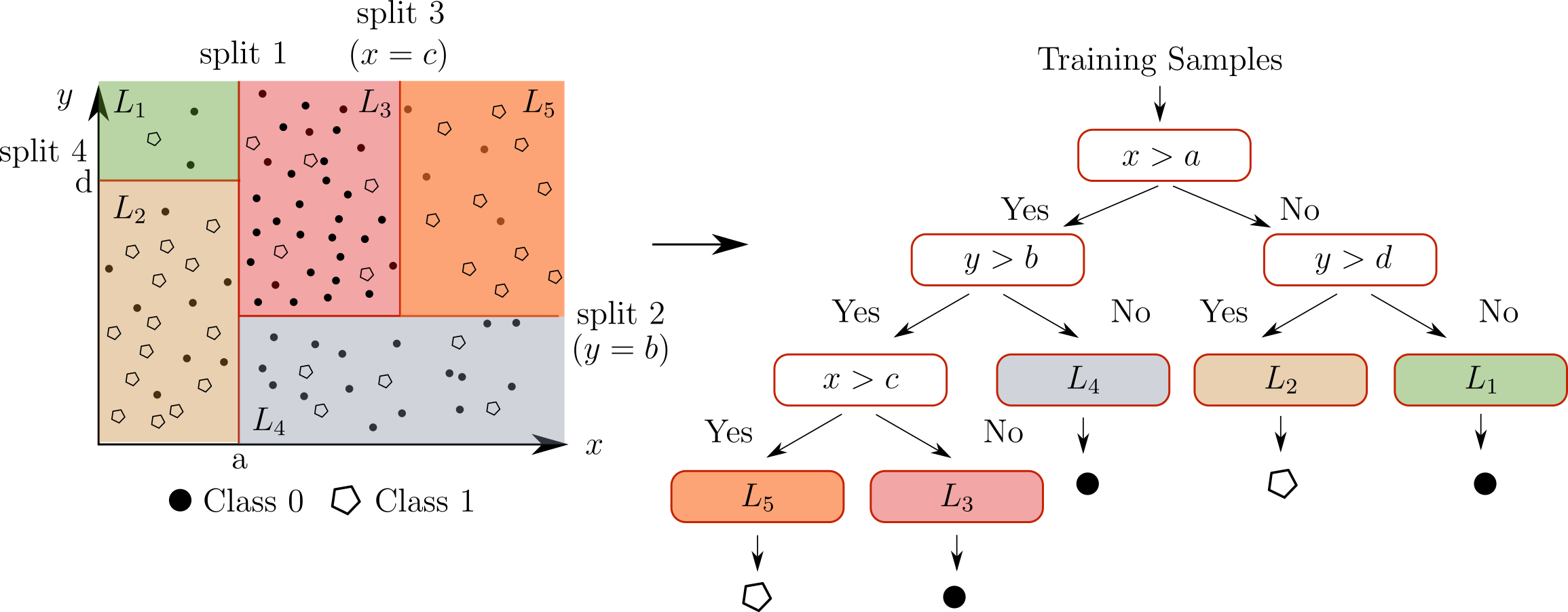}
\caption{Generation of decision tree}
\label{fig:Tree_generation}
\end{figure}

Ensemble learning is using multiple methods to get higher prediction rates for a given problem. Random forest is an ensemble learning method composed of decision trees where the number of these trees is part of the user input to the algorithm \cite{breiman2001}. 
Each of the decision trees is composed of branch nodes with two branches emanating from each root node; hence, they are called binary trees. 
The nodes that have no descendants are termed leaf nodes or leafs.  
Assuming numeric inputs, each branch node corresponds to one variable and its split point, while the leaf nodes correspond to output variables. 
Fig.~\ref{fig:Tree_generation} illustrates decision tree classification using two classes ($0$ and $1$) and two input variables ($x$ and $y$). 
The first step is to partition the input space of the training set into rectangles (or hyper-rectangles in higher dimensions), in this case $L_1$ through $L_5$. 
Selecting the partitions is based on making each subset of the training set purer, i.e., with fewer mixed labels, than the training set itself \cite{breiman2017}. 
The goodness of each partition is defined by an impurity function, see \cite{breiman2017} for a discussion on optimum splits. 
After defining the partitions for the training data set (left graph in Fig.~\ref{fig:Tree_generation}), a tree is formed (right graph in Fig.~\ref{fig:Tree_generation}). 

\begin{figure}[H]
\centering
\includegraphics[width=0.7\textwidth,height=.40\textheight,keepaspectratio]{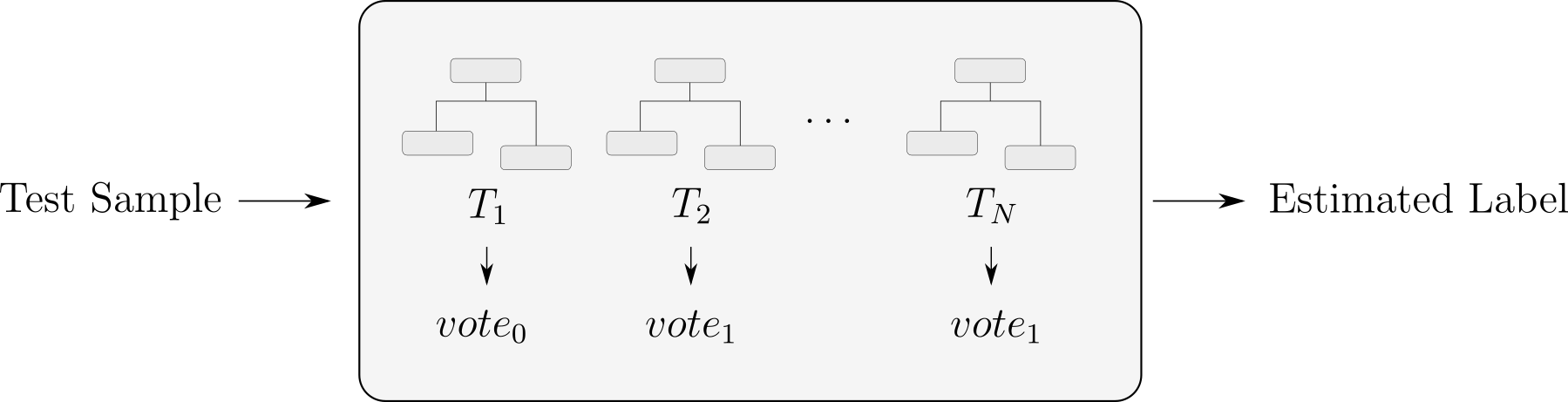}
\caption{Random Forest Classification ($vote_{0}$:Final decision of the corresponding tree is class 0. $vote_{1}$:Final decision of the corresponding tree is class 1.)}
\label{fig:RF_Classifier}
\end{figure}

The branch nodes of the tree correspond to conditions, either on $x$ or on $y$, such that the samples $L_1$ through $L_5$ can be placed in one of the leaf nodes. 
Each leaf node is then labeled by following the plurality rule \cite{breiman2017}: the most frequent labels in any node are assigned as the label for that node. 
For example in Fig.~\ref{fig:RF_Classifier}, leaf nodes $L_{1}$, $L_3$, and $L_4$ are labeled as class $0$, while leafs $L_2$ and $L_5$ are labeled as class $1$. 
Given a new input, a tag is generated by traversing the tree starting at the root node of the tree. 
The new input's label is then matched to the the leaf node it ends up in within the tree. 

In Random forest classification, there are $N$ decisions trees, and each tree votes for a label for a given test sample. 
The algorithm chooses a number of samples to generate the decision trees, and this is iterated until the desired number of trees is obtained. 
The estimation for the label of the sample is made with respect to the most frequent votes \cite{Belgiu2016} as shown in Fig.~\ref{fig:RF_Classifier}.

\subsection{Gradient Boosting } 
Gradient boosting algorithm was introduced by Schapire to answer the question of whether the performance of a single strong learner is equal to the set of weak learner performance \cite{Schapire1990}. 
Gradient boosting was proposed as an algorithm which provides more accurate predictions for regression and classification problems by generating new base models which can be linear models, smooth models and decision trees \cite{Natekin2013}. 
Gradient Boosting aims to correct the previous models by adding new base models to minimize the loss function. 
When the decision trees are used as new base models, a new decision tree is added after computing the loss function. 
The new decision tree is generated by parametrizing it, so that it can decrease the loss of the existing model.
Specifically, the gradient descent is used to minimize the loss function value and it is applied in functional space since each of tree (base learner) can be represented as functions.
Gradient boosting algorithm fits the new base models to the negative gradient of the loss function, where the choice of the loss function is user-dependent, to increase accuracy of the overall model \cite{Friedman2002}.

%!TEX root = ../machining_ML_Similarity.tex
%-------------------------------
%*******************************
\section{Results}
\label{sec:results}
%*******************************
This section shows the classification accuracy for the different featurization methods discussed in this paper. 
Specifically, Sections \ref{sec:WaveletResults} and \ref{sec:eemd_results} show the WPT-based and the EEMD-based results, respectively. 
The results are obtained by randomly splitting the data from each stickout case into $67\%$ training and $33\%$ testing sets. 
As described in Sections \ref{sec:wpt} and \ref{sec:eemd}, we extract the features from the informative wavelet packet or informative IMF and use four different classification algorithm on the training set to obtain classifiers. 
Then, we test the accuracy of the classifier using the corresponding test set. 
We repeat this split-train-test process $10$ times, and we tabulate the averages and standard deviations of the resulting classification accuracy. 
In addition, transfer learning results for several cutting configurations are provided for both WPT and EEMD methods in Sec.~\ref{sec:transferlearning}. 

%*******************************
\subsection{Wavelet Packet Transform with RFE}
 \label{sec:WaveletResults}
%*******************************

\begin{figure}[h]
\centering
\includegraphics[width=\textwidth,height=.95\textheight,keepaspectratio]{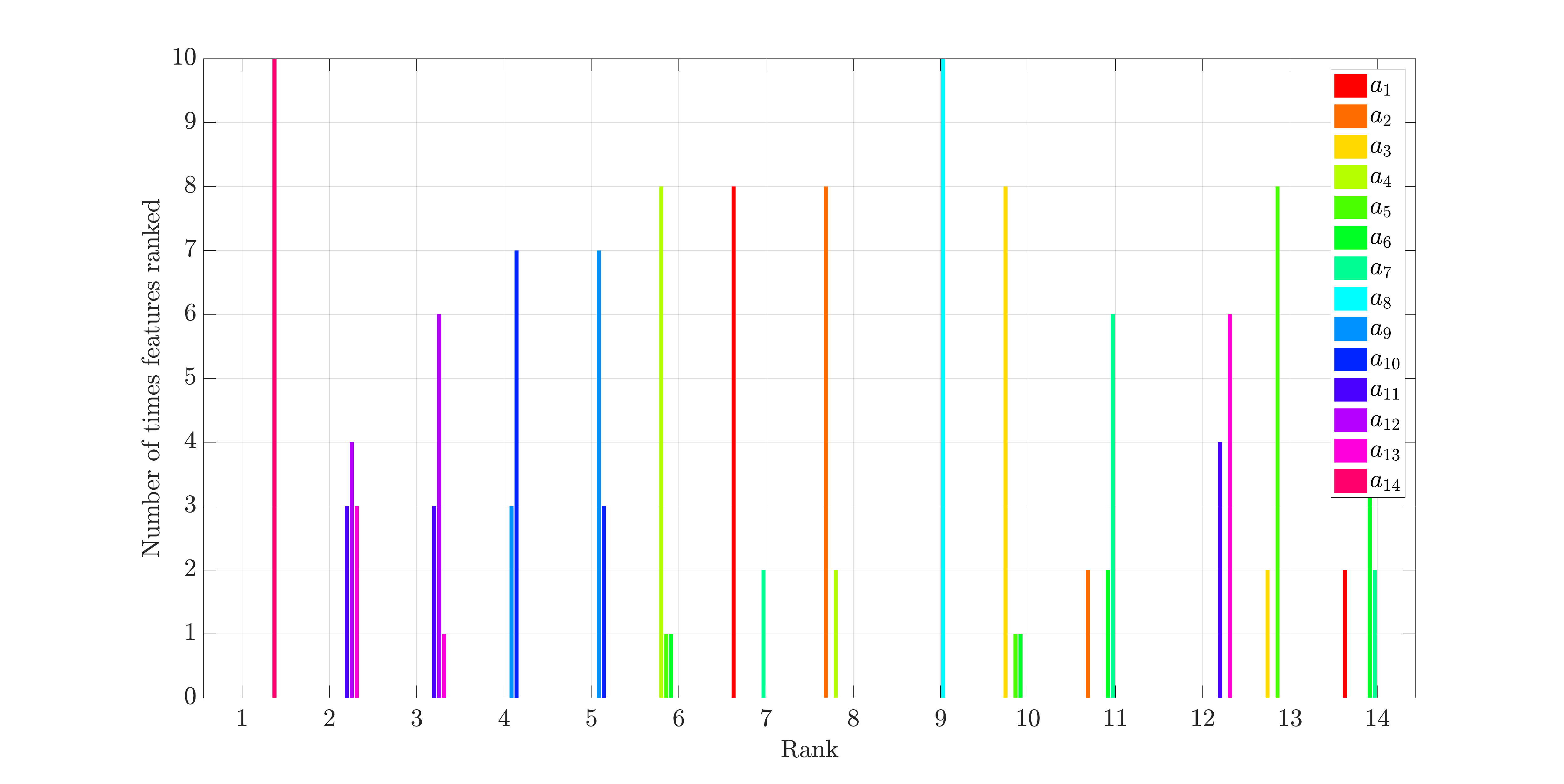}
\caption{Bar plot for feature ranking for 5.08 cm (2 inch) stickout case at level 4 WPT}
\label{fig:feature_ranking}
\end{figure}

In each realization of training data and test data, we repeat the feature ranking vial RFE as described in Sec.~\ref{sec:rfe}. 
Since the training and test sets are different in each realization, ten different rankings of the features are obtained. 
Figure \ref{fig:feature_ranking} shows the ranking for the $10$ iterations where each bar corresponds to a feature whose equation is provided in Table~\ref{tab:features}. 
The height of the bar in the figure shows the number of times each feature is ranked for the corresponding rank number. 
For instance, feature $a_{14}$ (standard frequency) is the feature with the most influence on the classification in all realizations. 
On the other hand, features $a_{11}$, $a_{12}$ and $a_{13}$ are ranked second, respectively, in three, four and three out of ten split-train-test realizations. 
In general, the features based on the frequency domain are higher ranked than the time domain features. 
Feature ranking plots for other stickout cases are provided in Figs.~\ref{fig:feature_ranking_2p5}--\ref{fig:feature_ranking_4p5} of the Appendix. 

%------------------------------
%------------------------------

\begin{figure}[h]
\centering
\includegraphics[width=\textwidth,height=.95\textheight,keepaspectratio]{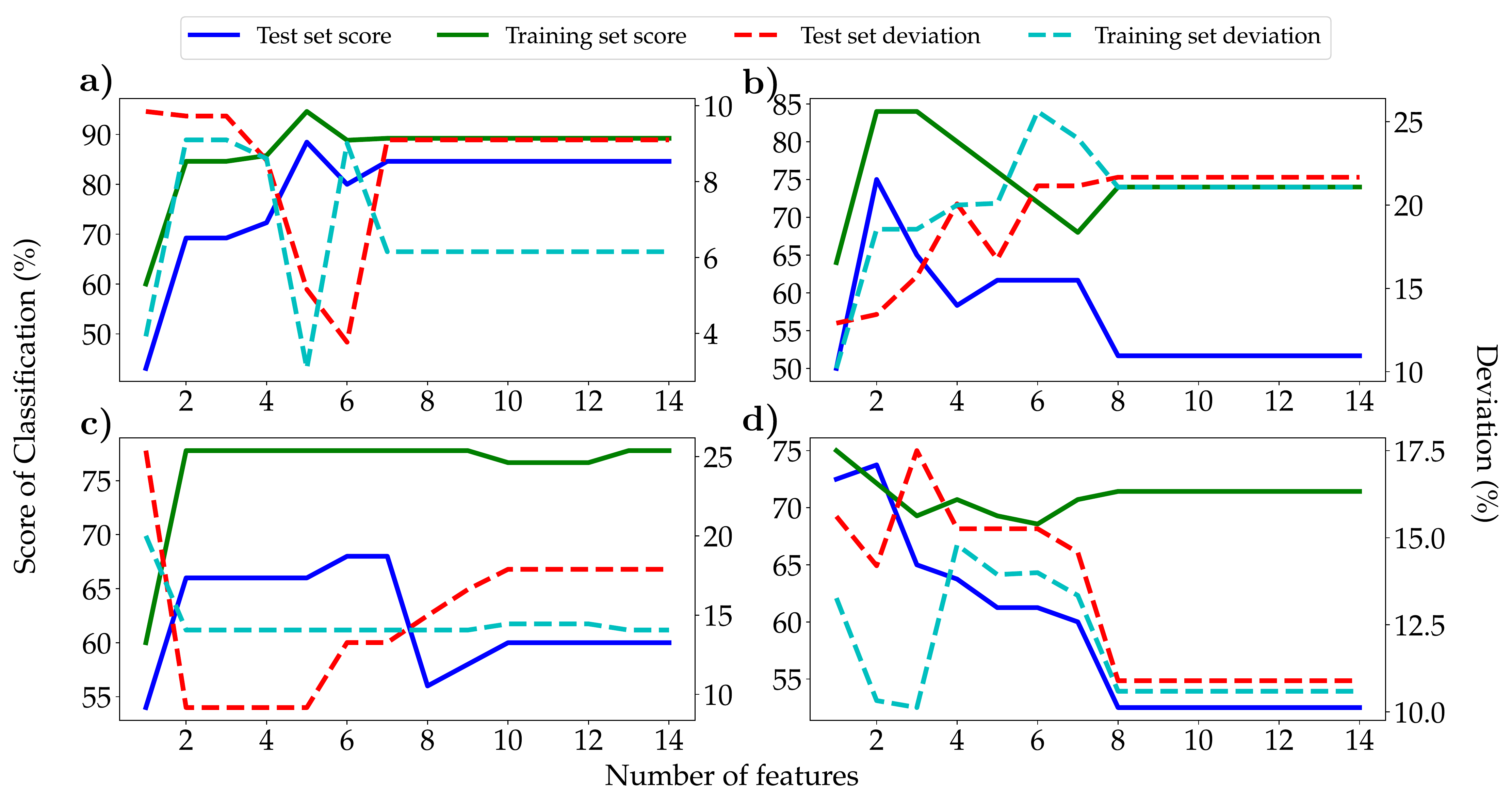}
\caption{Level 4 Wavelet Packet Transform(WPT) feature extraction method results for all stickout cases. (a) 5.08 cm ($2$ inch), (b) 6.35 cm ($2.5$ inch), (c) 8.89 cm ($3.5$ inch), and (d) 11.43 cm ($4.5$ inch).}
\label{fig:results_WPT}
\end{figure}

The mean and the standard deviation of the accuracy of the classification for the $10$ realizations of training and test sets based on the level $4$ WPT method are presented in Fig.~\ref{fig:results_WPT} for all stickout cases. In this figure, it is seen that when the number of the features is $8$ or $10$, adding lower ranked features into the feature vector does not affect the result. This shows that RFE ranked the features properly and that lower ranked features do not have influence on the results.

%------------------------------
%------------------------------

One difference between the WPT-based approach that we describe in this paper and the one described in \cite{Chen2017} is that we investigate the accuracy of the classifier using informative wavelet functions computed at each level of the WPT. 
On average the level $1$ and level $2$ WPT leads to better classification results in the test sets than the level $3$ and level $4$ WPT. 
This might be attributed to the fact that the lower level WPT contain information in a broader frequency range than the higher level WPT, and for chatter detection only the detection of chatter frequencies in the spectrum is relevant but not their frequency value or the exact shape of the peaks. 
We tabulate the full classification results for each level of the WPT up to level $4$ in Tables~\ref{tab:resultsforwholecases1}--\ref{tab:resultsforwholecases4} of the Appendix. 
Since the feature ranking is different for each realization of the splitting into training and test data, the $i$th ranked feature is only denoted by $r_{i}$. 
Below in Table~\ref{tab:comparison_of_results_times} we report the WPT results with the highest average accuracy out of all the different combinations of WPT levels and feature vectors and compare them to the results of the EEMD method.
 We also test the performance of the both method WPT and EEMD with the classifiers explained in Sec.\ref{sec:classifiers}. 
Tables~\ref{tab:4_Classifiers_Results_EEMD_WPT_Level1}--~\ref{tab:4_Classifiers_Results_EEMD_WPT_Level2} provide the accuracies obtained from Level 1 and Level 2 WPT and EEMD feature extraction methods with four different classifiers and compare the methods to each other.

%------------------------------
%\begin{figure}[H]
%\begin{subfigure}{0.5\textwidth}
  %\centering
  %\includegraphics[width=1\linewidth]{figures/Number_of_features_vs_deviation_accuracyWPT_2inch}
  %\caption{}
  %\label{fig:2_inch}
%\end{subfigure}%
%\begin{subfigure}{.5\textwidth}
  %\centering
  %\includegraphics[width=1\linewidth]{figures/Number_of_features_vs_deviation_accuracyWPT_2_5inch}
  %\caption{}
  %\label{fig:2.5_inch}
%\end{subfigure}
%\vskip\baselineskip
%\begin{subfigure}{.5\textwidth}
  %\centering
  %\includegraphics[width=1\linewidth]{figures/Number_of_features_vs_deviation_accuracyWPT_3_5inch}
  %\caption{}
  %\label{fig:3.5_inch}
%\end{subfigure}%
%\begin{subfigure}{.5\textwidth}
  %\centering
  %\includegraphics[width=1\linewidth]{figures/Number_of_features_vs_deviation_accuracyWPT_4_5inch}
  %\caption{}
  %\label{fig:4.5_inch}
%\end{subfigure}
%\caption{Classification results using level $4$ WPT for the different stickout cases: (a) 5.08 cm ($2$ inch), (b) 6.35 cm ($2.5$ inch), (c) 8.89 cm ($3.5$ inch), and (d) 11.43 cm ($4.5$ inch).}
%\label{fig:AccandDev}
%\end{figure}
%------------------------------

%**********************************************************
\subsection{Ensemble Empirical Mode Decomposition with RFE}
\label{sec:eemd_results}
%**********************************************************

Similar to Section \ref{sec:WaveletResults}, we combine EEMD with RFE and utilize four different classifiers in each realization of the splitting into test and train data set. The classification accuracy is on average better than the results from the level $3$ and level $4$ WPT and comparable to the accuracy of the lower level WPT. Table \ref{tab:EEMDresults} of the Appendix lists the resulting mean accuracies and standard deviations for all stickout cases and feature vectors. The combination with the best accuracies in each cutting case is reported when comparing the different methods in Table~\ref{tab:comparison_of_results_times}. In this table, the results highlighted with dark blue represent the highest accuracy across a given row while those highlighted in light blue have an average accuracy which is encapsulated by the error bars of the method with the highest average accuracy.

%------------------------------Comparison of the Level 1-Level2 WPT and EEMD Results
\begin{table}[H]
\centering
\caption{Comparison of the classification results obtained with SVM classifier and run times for each chatter detection method. Given run times include feature computation and classification for EEMD and level 4 WPT.}
\label{tab:comparison_of_results_times}
\begin{tabular}{|l|c|c|c|c|c|}
\hline
\multicolumn{1}{|c|}{} & \multicolumn{3}{c|}{Classification Results} &  \multicolumn{2}{c|}{Time Comparison (seconds)}\\
\hline
Stickout Length& WPT Level & WPT & EEMD  &WPT &EEMD\\
\hline
5.08 cm (2 inch)	&1	&\cellcolor[rgb]{0.13,0.67,0.8}$\SI{93.9 \pm 5.8}{\percent}$ &$\SI{84.2 \pm 0.8}{\percent}$  &\cellcolor[rgb]{0.13,0.67,0.8}115.99 &14540.06 \\
\hline
6.35 cm (2.5 inch)&2	&\cellcolor[rgb]{0.13,0.67,0.8}$\SI{100.0 \pm 0.0}{\percent}$ &$\SI{78.6 \pm 1.2}{\percent}$ &\cellcolor[rgb]{0.13,0.67,0.8}36.65   &3371.58  \\
\hline
8.89 cm (3.5 inch)&1	&\cellcolor[rgb]{0.63,0.79,0.95}$\SI{84.0 \pm 15.0}{\percent}$&\cellcolor[rgb]{0.13,0.67,0.8}$\SI{90.7 \pm 1.4}{\percent}$ &\cellcolor[rgb]{0.13,0.67,0.8}4.51  &1583.38   \\
\hline
11.43 cm (4.5 inch)&2	&\cellcolor[rgb]{0.13,0.67,0.8}$\SI{87.5 \pm 11.2}{\percent}$&\cellcolor[rgb]{0.63,0.79,0.95}$\SI{79.1 \pm 1.2}{\percent}$&\cellcolor[rgb]{0.13,0.67,0.8}6.53  &3096.07 \\
\hline
\end{tabular}
\end{table}
%------------------------------

%------------------------------Four Different Classifiers with Level 1 WPT and EEMD
\begin{table}[H]
\centering
\caption{Results obtained by using Level 1 WPT and EEMD feature extraction methods with four different classifiers.}
\label{tab:4_Classifiers_Results_EEMD_WPT_Level1}
\resizebox{\textwidth}{!} {  % make the table a bit smaller so it doesn't spill out
\begin{tabular}{|c|c|c|c|c|c|c|c|c|c|}
\hline
\multicolumn{1}{|c|}{} & \multicolumn{4}{c|}{WPT} &  \multicolumn{4}{c|}{EEMD}\\
\hline
\makecell{Stickout\\Length} & SVM                            & \makecell{Logistic\\Regression} & \makecell{Random\\Forest}       &\makecell{Gradient\\Boosting}   & SVM                           & \makecell{Logistic\\Regression} & \makecell{Random\\Forest}    &\makecell{Gradient\\Boosting} \\
\hline
\makecell{5.08 cm \\ (2 inch)}    & $\SI{93.9 }{\percent}$ & $\SI{84.6}{\percent}$  &$\SI{93.1}{\percent}$  &$\SI{90.0}{\percent}$  &$\SI{84.2} {\percent}$  &$\SI{93.5 }{\percent}$   &\cellcolor[rgb]{0.63,0.79,0.95}$\SI{94.8 }{\percent}$ &\cellcolor[rgb]{0.13,0.67,0.8}$\SI{94.9 }{\percent}$\\
\hline
\makecell{6.35 cm \\ (2.5 inch)}  & \cellcolor[rgb]{0.63,0.79,0.95}$\SI{85.0}{\percent}$ & $\SI{71.7}{\percent}$  &\cellcolor[rgb]{0.13,0.67,0.8}$\SI{91.7 }{\percent}$  &\cellcolor[rgb]{0.13,0.67,0.8}$\SI{91.7 }{\percent}$  &$\SI{78.6 }{\percent}$  &$\SI{79.4 }{\percent}$   &$\SI{80.1 }{\percent}$ &$\SI{82.2 }{\percent}$\\
\hline
\makecell{8.89 cm \\ (3.5 inch)}  & $\SI{84.0 }{\percent}$ & $\SI{94.0}{\percent}$  &\cellcolor[rgb]{0.13,0.67,0.8}$\SI{100.0 }{\percent}$  &$\SI{96.0 }{\percent}$  &$\SI{90.7 }{\percent}$  &$\SI{89.0 }{\percent}$   &$\SI{93.5 }{\percent}$ &$\SI{94.5 }{\percent}$\\
\hline
\makecell{11.43 cm \\ (4.5 inch)} & \cellcolor[rgb]{0.63,0.79,0.95}$\SI{78.8 }{\percent}$ & \cellcolor[rgb]{0.63,0.79,0.95}$\SI{81.3}{\percent}$  &\cellcolor[rgb]{0.63,0.79,0.95}$\SI{86.3 }{\percent}$  &\cellcolor[rgb]{0.13,0.67,0.8}$\SI{87.5 }{\percent}$  &\cellcolor[rgb]{0.63,0.79,0.95}$\SI{79.1 }{\percent}$  &\cellcolor[rgb]{0.63,0.79,0.95}$\SI{78.7 }{\percent}$   &\cellcolor[rgb]{0.63,0.79,0.95}$\SI{81.6 }{\percent}$ &\cellcolor[rgb]{0.63,0.79,0.95}$\SI{81.4 }{\percent}$\\
\hline
\end{tabular}
}
\end{table}
%------------------------------

%------------------------------Four Different Classifiers with Level 2 WPT and EEMD
\begin{table}[H]
\centering
\caption{Results obtained by using Level 2 WPT and EEMD feature extraction methods with four different classifiers.}
\label{tab:4_Classifiers_Results_EEMD_WPT_Level2}
\resizebox{\textwidth}{!} {  % make the table a bit smaller so it doesn't spill out
\begin{tabular}{|c|c|c|c|c|c|c|c|c|c|}
\hline
\multicolumn{1}{|c|}{} & \multicolumn{4}{c|}{WPT} &  \multicolumn{4}{c|}{EEMD}\\
\hline
\makecell{Stickout\\Length} & SVM                            & \makecell{Logistic\\Regression} & \makecell{Random\\Forest}       &\makecell{Gradient\\Boosting}   & SVM                           & \makecell{Logistic\\Regression} & \makecell{Random\\Forest}    &\makecell{Gradient\\Boosting} \\
\hline
\makecell{5.08 cm \\ (2 inch)}    & $\SI{91.5}{\percent}$& $\SI{87.7}{\percent}$  &$\SI{93.8}{\percent}$  &$\SI{90.0}{\percent}$  &$\SI{84.2} {\percent}$  &$\SI{93.5 }{\percent}$   &\cellcolor[rgb]{0.63,0.79,0.95}$\SI{94.8 }{\percent}$ &\cellcolor[rgb]{0.13,0.67,0.8}$\SI{94.9 }{\percent}$\\
\hline
\makecell{6.35 cm \\ (2.5 inch)}  & \cellcolor[rgb]{0.13,0.67,0.8}$\SI{100.0}{\percent}$ & $\SI{80.0}{\percent}$  &$\SI{95.0 }{\percent}$  &$\SI{96.7 }{\percent}$  &$\SI{78.6 }{\percent}$  &$\SI{79.4 }{\percent}$   &$\SI{80.1 }{\percent}$ &$\SI{82.2 }{\percent}$\\
\hline
\makecell{8.89 cm \\ (3.5 inch)}  & $\SI{78.0}{\percent}$ & $\SI{58.0}{\percent}$  &\cellcolor[rgb]{0.63,0.79,0.95}$\SI{94.0 }{\percent}$  &$\SI{78.0 }{\percent}$  &$\SI{90.7 }{\percent}$  &$\SI{89.0 }{\percent}$   &\cellcolor[rgb]{0.63,0.79,0.95}$\SI{93.5 }{\percent}$ &\cellcolor[rgb]{0.13,0.67,0.8}$\SI{94.5 }{\percent}$\\
\hline
\makecell{11.43 cm \\ (4.5 inch)} & \cellcolor[rgb]{0.63,0.79,0.95}$\SI{87.5 }{\percent}$ & \cellcolor[rgb]{0.63,0.79,0.95}$\SI{78.8}{\percent}$  &\cellcolor[rgb]{0.13,0.67,0.8}$\SI{88.8 }{\percent}$  &\cellcolor[rgb]{0.63,0.79,0.95}$\SI{80.0 }{\percent}$  &\cellcolor[rgb]{0.63,0.79,0.95}$\SI{79.1 }{\percent}$  &\cellcolor[rgb]{0.63,0.79,0.95}$\SI{78.7 }{\percent}$   &\cellcolor[rgb]{0.63,0.79,0.95}$\SI{81.6 }{\percent}$ &\cellcolor[rgb]{0.63,0.79,0.95}$\SI{81.4 }{\percent}$\\
\hline
\end{tabular}
}
\end{table}
\begin{figure}[htbp]
\centering
\includegraphics[width=1\textwidth,height=.85\textheight,keepaspectratio]{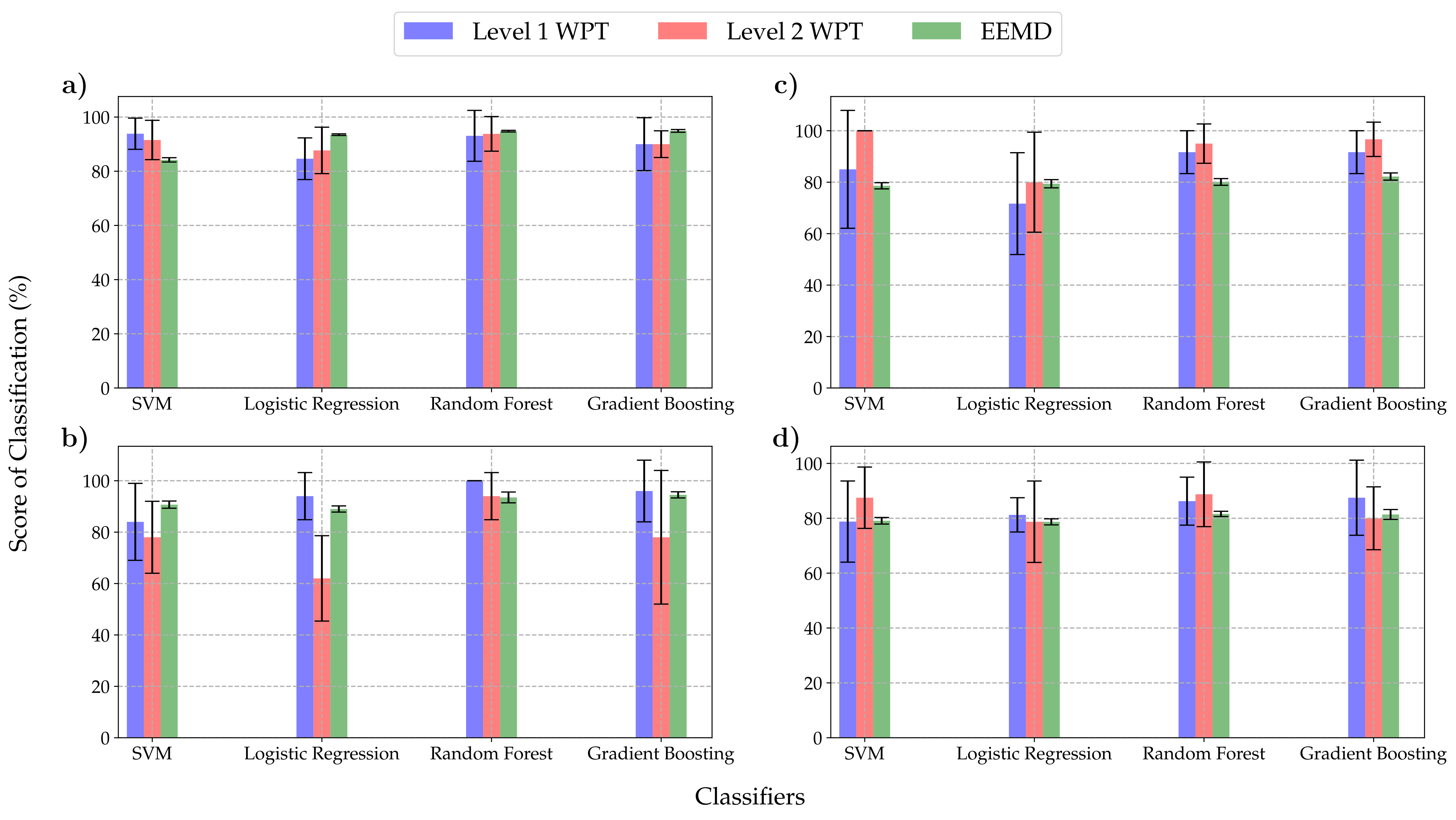}
\caption{Bar plot including the error bars of the classification results for Level 1 WPT, Level 2 WPT and EEMD with four different classifier. a) 5.08 cm (2 inch) stickout size, b)6.35 cm (2.5 inch) stickout size, c)8.89 cm (3.5 inch) stickout size, d) 11.43 cm (4.5 inch) stickout size.}
\label{fig:4_Classifier_WPT_EEMD_Bar_Plot}
\end{figure}
%--------------

Table~\ref{tab:comparison_of_results_times} shows that features based on the WPT algorithm give the highest accuracy for three stickout cases out of four cutting configurations. 
Specifically, feature extraction with WPT and RFE is the most accurate for the 5.08 and 6.35 cm ($2$ $2.5$ and $4.5$) stickout cases scoring $93.9\%$, $100.0\%$ and $87.5\%$ respectively. 
While the results from EEMD give the highest accuracies for 8.89 cm (3.5 inch) stickout cases, WPT result for this case still lies within the error bars of EEMD results. 
We also see the results when we used different classifiers other than SVM in Tab.\ref{tab:4_Classifiers_Results_EEMD_WPT_Level1} and \ref{tab:4_Classifiers_Results_EEMD_WPT_Level2}.
In Table \ref{tab:4_Classifiers_Results_EEMD_WPT_Level1}, performance of Level 1 WPT is better than EEMD since WPT has the highest accuracies in three cutting configuration cases and EEMD results are in error bars of WPT results. 
On the other hand, Table \ref{tab:4_Classifiers_Results_EEMD_WPT_Level2} indicates that both methods has the highest accuracy for two cutting configurations. 
These two table also provide the evidence that lower level (Level 1) WPT outperforms EEMD. 
Further, 100\% accuracy is observed in Tab.\ref{tab:4_Classifiers_Results_EEMD_WPT_Level1} and \ref{tab:4_Classifiers_Results_EEMD_WPT_Level2} for two different cutting configurations. 
These cutting configurations has the lowest number of time series as experimental data.
Since time series are not split into smaller pieces for WPT method, so the size of the test set is quite small and it is possible to get such high results.

The standard deviation of the WPT results is quite high as it is seen from Tab.\ref{tab:comparison_of_results_times} and Fig.\ref{fig:4_Classifier_WPT_EEMD_Bar_Plot} since the computation time for this method does not require splitting a long time series into smaller pieces. 
Therefore, the total number of samples for identical stickout cases is smaller in comparison to the EEMD method where long time series were split into shorter ones of approximately $1000$ points, thus increasing the number of samples and resulting in tighter error bars. 
Therefore, the amount of deviation can be reduced, especially for the WPT-based approach, by increasing the size and the number of the training sets.
In addition, Table~\ref{tab:comparison_of_results_times} compares the run time in seconds for each of the different featurization methods for chatter detection. 
These comparisons were performed using a Dell Optilex 7050 desktop with Intel Core i7-7700 CPU and 16.0 GB RAM. 
It can be seen that feature extraction with WPT and RFE is the fastest across all of the stickout cases. 
We point out that the built-in WPT package that we used is highly optimized, whereas in comparison the EEMD does not enjoy the same level of code optimization. 
Moreover, for EEMD the EMD is performed for an ensemble of time series with ensemble size $200$, which needs much higher computation effort and can be reduced by varying the ensemble parameters of the EEMD.

%------TRANSFER LEARNING-------------------
\subsection{Transfer learning capabilities}
\label{sec:transferlearning}

Transfer learning was applied to the WPT and the EEMD: A classifier is trained on the data of the 5.08 cm (2 inch) or the 11.43 cm (4.5 inch) cases and tested on the data of another stickout length. 
The reason why these two cutting configurations were chosen for the training data set is that these are the ones with the largest number of cases (see Table \ref{tab:chatter_case_number}). 
The classifier is trained with $70\%$ of the training data set and tested on 70 \% of the test set. 
Classification is repeated for $10$ realizations of the split-train-test process. 
The classification has been done for the level $1$ WPT, the level $4$ WPT and the EEMD methods. 
The  mean and the standard deviation of the classification accuracies for WPT and EEMD methods can be found in Tables \ref{tab:transfer_learning_WPT_Level1_2_4p5_inch}--\ref{tab:transfer_learning_EEMD_2_4p5_inch} of the Appendix. 
The main result can be seen already in the best results for transfer learning between the 5.08 cm (2 inch) and the 11.43 cm (4.5 inch) stickout cases presented in Table \ref{tab:transfer_learning_WPT_EEMD_FFT_best_4classifier}. 

%---------------
%\begin{table}[H]
%\centering
%\caption{Highest accuracies obtained with SVM from transfer learning results for 5.08 cm (2 inch) and 11.43 cm (4.5 inch) cases.}
%\label{tab:transfer_learning_WPT_EEMD_best}
%\begin{tabular}{|l|c|c|c|c|}
%\hline
%\multicolumn{1}{|c|}{} & \multicolumn{2}{c|}{\shortstack{Training Set: 5.08 cm (2 inch) \\ Test Set: 11.43 cm (4.5 inch)}} & \multicolumn{2}{c|}{\shortstack{Training Set: 11.43 cm (4.5 inch) \\ Test Set: 5.08 cm (2 inch)}} \\
%\hline
%\multicolumn{1}{|c|}{Method} & Test Set & Training Set & Test Set& Training Set \\
%\hline
%WPT Level 1 &$\SI{59.4 \pm 14.6}{\percent}$&$\SI{87.7 \pm 4.5}{\percent}$&$\SI{62.9 \pm 21.5}{\percent}$&$\SI{81.4 \pm 12.0}{\%}$\\
%WPT Level 4	&$\SI{59.4 \pm 22.2}{\percent}$&$\SI{73.8 \pm 5.7}{\percent}$&$\SI{53.6 \pm 7.3}{\percent}$&$\SI{73.6 \pm 8.5}{\%}$\\
%EEMD        &\cellcolor[rgb]{0.13,0.67,0.8}$\SI{64.0 \pm 0.7}{\percent}$&$\SI{93.8 \pm 0.3}{\percent}$&\cellcolor[rgb]{0.13,0.67,0.8}$\SI{83.6 \pm 0.4}{\percent}$&$\SI{77.0 \pm 0.6}{\%}$\\
%\hline
%\end{tabular}
%\end{table}
%---------------

%---------------
\begin{table}[H]
\centering
\caption{Comparison highest accuracies obtained with four different classifier by applying of transfer learning results for 5.08 cm (2 inch) and 11.43 cm (4.5 inch) cases.}
\label{tab:transfer_learning_WPT_EEMD_FFT_best_4classifier}
\resizebox{\textwidth}{!} {  % make the table a bit smaller so it doesn't spill out
\begin{tabular}{|l|c|c|c|c|c|c|c|c|}
\hline
\multicolumn{1}{|c|}{} & \multicolumn{4}{c|}{\shortstack{Training Set: 5.08 cm (2 inch) \\ Test Set: 11.43 cm (4.5 inch)}} & \multicolumn{4}{c|}{\shortstack{Training Set: 11.43 cm (4.5 inch) \\ Test Set: 5.08 cm (2 inch)}} \\
\hline
\makecell{Method} & SVM                    & \makecell{Logistic\\Regression} & \makecell{Random\\Forest}       &\makecell{Gradient\\Boosting}   & SVM                           & \makecell{Logistic\\Regression} & \makecell{Random\\Forest}    &\makecell{Gradient\\Boosting} \\
\hline
WPT Level 1				&$\SI{59.4}{\percent}$   & $\SI{56.3}{\percent}$   & $\SI{67.5}{\percent}$   & $\SI{66.3}{\percent}$   & $\SI{62.9}{\percent}$   & $\SI{51.1}{\percent}$   & $\SI{62.1}{\percent}$   & $\SI{62.9}{\percent}$  \\
WPT Level 4				&$\SI{59.4}{\percent}$   & $\SI{61.9}{\percent}$   & \cellcolor[rgb]{0.13,0.67,0.8}$\SI{85.6}{\percent}$   & \cellcolor[rgb]{0.63,0.79,0.95}$\SI{73.8}{\percent}$   & $\SI{53.6}{\percent}$   & $\SI{47.5}{\percent}$   & $\SI{59.6}{\percent}$   & $\SI{62.5}{\percent}$  \\
EEMD       				&$\SI{64.0}{\percent}$   & \cellcolor[rgb]{0.63,0.79,0.95}$\SI{78.1}{\percent}$   & \cellcolor[rgb]{0.63,0.79,0.95}$\SI{81.1}{\percent}$   & \cellcolor[rgb]{0.63,0.79,0.95}$\SI{79.4}{\percent}$   & $\SI{83.6}{\percent}$   & $\SI{92.6}{\percent}$   & \cellcolor[rgb]{0.13,0.67,0.8}$\SI{94.9}{\percent}$   & $\SI{94.6}{\percent}$  \\
\hline
\end{tabular}
}
\end{table}
%---------------

%---------------
\begin{figure}[htbp]
\centering
\includegraphics[width=1\textwidth,height=0.8\textheight,keepaspectratio]{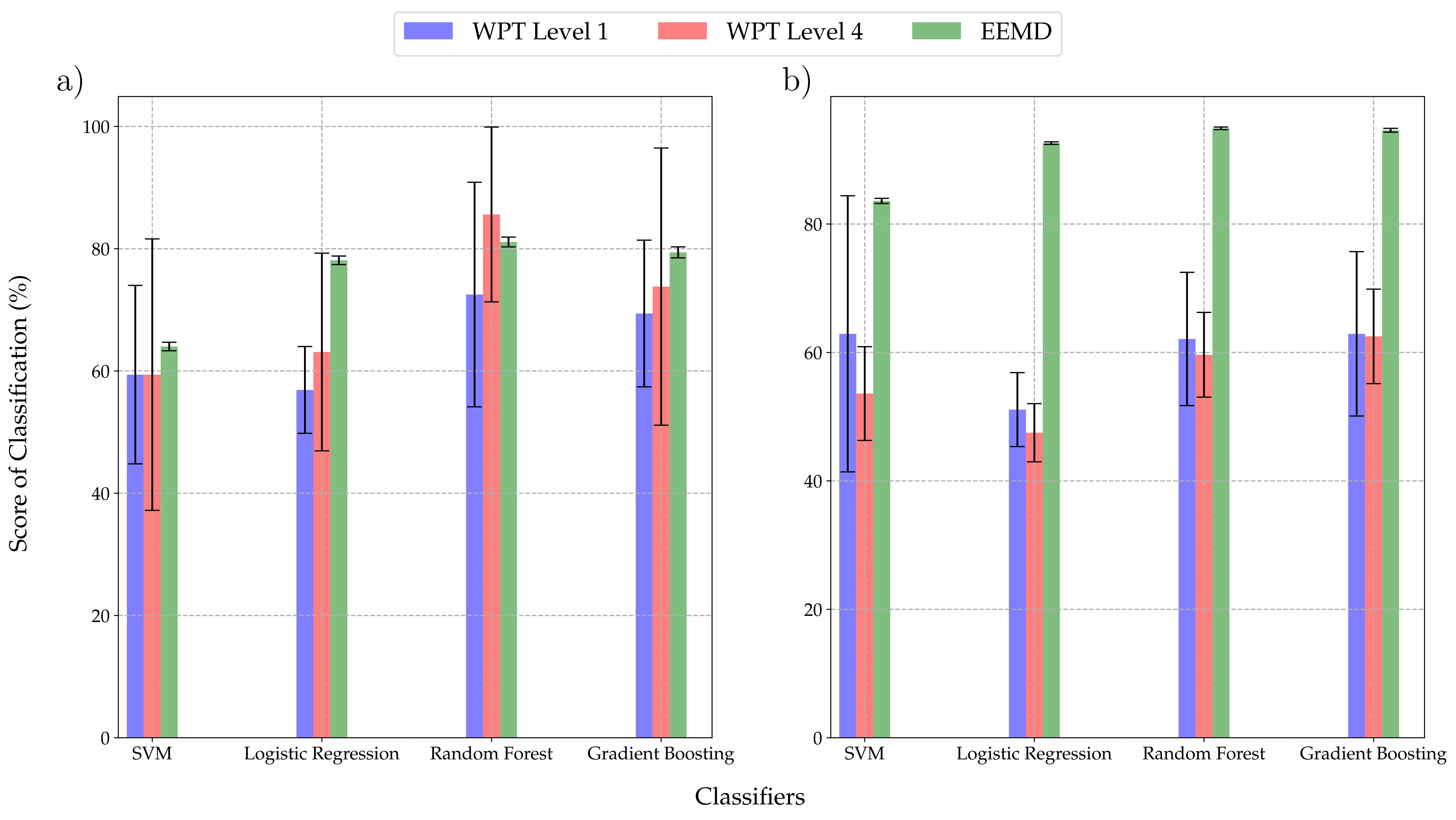}
\caption{Transfer learning results based on four different classifiers for WPT/EEMD approaches. a) Training set: 5.08 cm (2 inch) stickout size, Test set: 11.43 cm (4.5 inch)  stickout size.  b)Training set: 11.43 (4.5 inch) cm stickout size, Test set: 5.08 cm (2 inch) stickout size.}
\label{fig:comparison_high_acc_4_classifier_transfer_learning}
\end{figure}
%---------------

Table \ref{tab:transfer_learning_WPT_EEMD_FFT_best_4classifier} shows that even the highest accuracies for level 4 transform are either equal to or smaller than the ones for level 1 WPT , which means that lower level WPT achieve better transfer learning classification rates with SVM.
This statement is true for the results obtained with other classifiers except in the first application of transfer learning where we train classifier on 5.08 cm (2 inch) case and test it on 11.43 cm (4.5 inch).   
In general, lower level wavelet packets have broader frequency range compared to higher ones, so wavelet packets in the lower level transform are more likely to include the chatter frequency information for the cases whose informative wavelet packet number is different. 
When the informative wavelet packet number of the training and test sets are the same, Tables \ref{tab:transfer_learning_WPT_Level1_2_4p5_inch}--\ref{tab:transfer_learning_WPT_Level4_2_4p5_inch} show that higher transfer learning classification scores are achieved. 
For instance, a classifier trained with 5.08 cm (2 inch) stickout case has the first wavelet packet as the informative one in level 1 WPT.
When this classifier is tested on other stickout cases, it is seen that it gives high accuracies for 6.35 and 8.89 cm (2.5 and 3.5 inch) cases as expected since these cases have also the first wavelet packet as the informative one. 
However, when the classifier is tested on the 11.43 cm (4.5 inch) case, where the second packet is the informative one (see Table \ref{tab:chatter_freq_range}), the classification accuracy dramatically decreases. 
For level 4 WPT, all cutting configurations have different informative packet numbers. 
Therefore, classification results are not as high as in level 1 WPT for all stickout cases. 

Further, EEMD has better performance than WPT in transfer learning, although the best accuracy obtained for the classifier trained with 5.08 cm (2 inch) belongs to WPT. 
Fig.\ref{fig:comparison_high_acc_4_classifier_transfer_learning} shows that the WPT method provides the highest accuracy for that application of transfer learning with high amount of deviation.
In addition, we see the same trend when we train a classifier with two different cutting size data set features and tested it on other remaining two cutting size data set features in Tab. \ref{tab:transfer_learning_WPT_EEMD_best_4classifier_2_Train_2_Test}.
The plot of the best accuracies with error bands are provided in Fig.\ref{fig:transfer_learning_trained_with2case_tested_on2case}.
In Table \ref{tab:transfer_learning_WPT_EEMD_best_4classifier_2_Train_2_Test}, first application of transfer learning provides the highest accuracy with EEMD methods whose deviations is 0.6\% while WPT outperforms the EEMD with accuracy difference of 3.7\% and WPT has the 12.1 \% for the highest accuracy when training set and test set cases are interchanged. 
Moreover, it is seen that there is drops and sudden increases in accuracy when we use different classifiers in right hand side of Tab.\ref{tab:transfer_learning_WPT_EEMD_best_4classifier_2_Train_2_Test}. 
The reason for this is that the classifiers which provides low accuracies has over fitting problem even though the RFE is utilized during classification. 
It is worth to note that we have used the default parameters for all classifiers except Random Forest Classification. 
Number of decision trees and maximum depth of trees are selected as 100 and 2, respectively. The reason why EEMD has low accuracy for Random Forest and Gradient Boosting can be explained with the classifier parameters which are not tuned.
If one increase the maximum depth of decision trees, which will lead to generate purer leaves, or tune the parameters for Gradient Boosting, the accuracy can increase and over fitting can be solved.
However, we keep parameters for all classifiers fixed in all classifications. 

Low order IMFs include the chatter frequency in all the studied cases, and the difference between informative IMF numbers is not large as in the WPT-based approach.
Recall that the informative IMF is the second IMF for the 5.08 cm (2 inch) and the 6.35 cm (2.5 inch) case and the first IMF for the 8.89 cm (3.5 inch) and the 11.43 cm (4.5 inch) case. 
It is expected that information on machine tool chatter is contained in both IMFs for all stickout cases, which may explain the high accuracies for some transfer learning cases, where the IMF is not the same. 
Specifically, this can be seen for the EEMD based classifier, which was trained on the 5.08 cm (2 inch) case and tested on the 8.89 cm (3.5 inch) case, as well as the one, which was trained on the 11.43 cm (4.5 inch) case and tested on the 5.08 cm (2 inch) case (cf. Table~\ref{tab:transfer_learning_EEMD_2_4p5_inch} of the Appendix). 

%---------------
\begin{table}[H]
\centering
\caption{Highest accuracies obtained with transfer learning application based on training a classifier on two different stickout size and testing on remaining other two stickout size}
\label{tab:transfer_learning_WPT_EEMD_best_4classifier_2_Train_2_Test}
\resizebox{\textwidth}{!} {  % make the table a bit smaller so it doesn't spill out
\begin{tabular}{|l|c|c|c|c|c|c|c|c|}
\hline
\multicolumn{1}{|c|}{} & \multicolumn{4}{c|}{\shortstack{Training Set: 5.08 cm (2 inch) and  6.35 cm (2.5 inch) \\ Test Set: 8.89 cm (3.5 inch) and 11.43 cm (4.5 inch)}} & \multicolumn{4}{c|}{\shortstack{Training Set: 8.89 cm (3.5 inch) and 11.43 cm (4.5 inch) \\ Test Set: 5.08 cm (2 inch) and  6.35 cm (2.5 inch)}} \\
\hline
\makecell{Method} & SVM                    & \makecell{Logistic\\Regression} & \makecell{Random\\Forest}       &\makecell{Gradient\\Boosting}   & SVM                           & \makecell{Logistic\\Regression} & \makecell{Random\\Forest}    &\makecell{Gradient\\Boosting} \\
\hline
WPT Level 4				&$\SI{60.4}{\percent}$   & $\SI{58.4}{\percent}$   & $\SI{69.6}{\percent}$   & $\SI{73.1}{\percent}$   & $\SI{51.5}{\percent}$   & $\SI{51.5}{\percent}$   &  \cellcolor[rgb]{0.13,0.67,0.8}$\SI{84.6}{\percent}$   & \cellcolor[rgb]{0.63,0.79,0.95}$\SI{73.8}{\percent}$  \\
EEMD       				&\cellcolor[rgb]{0.63,0.79,0.95}$\SI{81.1}{\percent}$   &  \cellcolor[rgb]{0.13,0.67,0.8}$\SI{81.7}{\percent}$   & \cellcolor[rgb]{0.63,0.79,0.95}$\SI{81.2}{\percent}$   & $\SI{78.8}{\percent}$   &  \cellcolor[rgb]{0.63,0.79,0.95}$\SI{80.9}{\percent}$   &  \cellcolor[rgb]{0.63,0.79,0.95}$\SI{79.9}{\percent}$   & $\SI{51.6}{\percent}$   & $\SI{49.9}{\percent}$  \\
\hline
\end{tabular}
}
\end{table}
%---------------

%!TEX root = ../machining_ML_Similarity.tex
%*****************************
\section{Conclusion} 
\label{sec:conclusion}
%*****************************
Two advanced chatter detection methods, i.e. the Wavelet Packet Transform (WPT) and the Ensemble Empirical Mode Decomposition (EEMD) with Recursive Feature Elimination (RFE) have been used for the classification of recorded acceleration signals from a turning process into chatter-free cutting or chattering motion. 
We use the two algorithms not only to classify measured test data with the same cutting conditions as used in the training phase but also for transfer learning, which means that the test data originates from a cutting process with different cutting conditions. 
In particular, the chatter frequencies between the training data and the test data differ significantly.

Our results in Table~\ref{tab:comparison_of_results_times} show that WPT have the highest accuracies for three cutting configurations when the classifier is trained and tested on the same data set while EEMD provides the best results for one cutting configuration. 
In addition, training different classifiers other than SVM leads to increase in mean accuracies for both method as shown in Tab. \ref{tab:4_Classifiers_Results_EEMD_WPT_Level1} and \ref{tab:4_Classifiers_Results_EEMD_WPT_Level2}.
These two tables is the evidence that WPT performance decreases as we increase the level of transform. 
On the other hand, Table \ref{tab:transfer_learning_WPT_EEMD_FFT_best_4classifier} and \ref{tab:transfer_learning_WPT_EEMD_best_4classifier_2_Train_2_Test} show that EEMD performs better when the transfer learning approach is used, although WPT provides the best results for a few transfer learning applications in these tables.
This is due to the fact that the selected informative intrinsic mode functions (IMFs) of the EEMD contain information of a broader frequency band than the informative wavelet packets of the WPT. 
For transfer learning applications where the WPT methods has the highest accuracy (see Tab. \ref{tab:transfer_learning_WPT_EEMD_FFT_best_4classifier} and \ref{tab:transfer_learning_WPT_EEMD_best_4classifier_2_Train_2_Test}), it seen that the WPT provides the high accuracy with larger deviations and the EEMD results are in the error range of these best results.
For similar reasons, in general, low level WPT (level 1 or 2) performs better than higher level WPT (see Table \ref{tab:comparison_of_results_times}) because the informative wavelet packets of a lower level contains a broader frequency band and it is more likely that the signatures of chattering motion or chatter-free cutting can be accurately detected. 
Nevertheless, for a specific process without many changes in the dynamics during the process also a high level WPT with a narrow frequency band may be useful. 
In addition to accuracy comparisons, we also recorded the overall runtime or each method. 
Table~\ref{tab:comparison_of_results_times} shows that WPT has the fastest runtime, while EEMD method clocks the longest runtime. 
This slowdown is mostly related to the computation of the ensemble of IMFs and can be reduced by changing the ensemble parameters and optimizing the code.

There are two main drawbacks of the methods. 
1) the WPT featurization process is cumbersome since it requires taking the WPT of the signal, investigating the packets that contain the chatter frequencies, and then choosing the packet that has considerably high energy ratio and that includes chatter frequency. 
Once these packets are found they are fixed for the investigated process and are used for chatter classification. 
However, inherent to this process is the a priori identification of chatter frequency and the assumption that the chosen packets (referred to as the informative packets) will always contain it. 
This is a limitation since (a) it requires highly skilled users for analyzing the signal and extracting the informative packets, and (b) the chatter frequency band can move during the cutting process which will yield the informative packets ineffective for chatter classification. 
Further, Section \ref{sec:wpt} points out that the informative packets are not necessarily the ones with the higher energy ratio. This makes automating the feature selection process more difficult in the WPT approach. 
The EEMD also suffers some of these drawbacks since the process for choosing the informative IMFs and the informative packets in WPT is quite similar. 
The second drawback is that 2) it is not always possible to differentiate between intermediate and full chatter. 
Specifically, although the intermediate chatter time series (Fig.~\ref{fig:fft2320005}c) and the chatter time series (Fig.~\ref{fig:fft2320005}e) are visually very different in the time domain, their energy content shown in the top graph of Fig.~\ref{fig:energyratio} can be too close to distinguish between the two cases.

%While in this paper we performed the classification by considering the intermediate chatter and the full chatter as one case labeled ``chatter'', it can be desirable to distinguish between three classes: no chatter, intermediate chatter, and full chatter.
%The reason for possibly being interested in identifying intermediate chatter is that controlled chatter can still yield parts with acceptable quality while reducing the machining time \cite{Honeycutt2017}.
%Further, the intentional texturing provided by intermediate chatter can be a desirable surface finish.

\section*{Acknowledgement}
This material is based upon work supported by the National Science Foundation under Grant Nos.~CMMI-1759823 and DMS-1759824 with PI FAK.
%
% \clearpage
\bibliographystyle{ieeetr}
\bibliography{ML-WPT_EEMD}

\newpage
%!TEX root = ../machining_ML_Wavelet.tex
%-------------------------------
\appendix

% \appendixpage
\section{Supplemental classification results}
\label{sec:appendixA}

\begin{table}[H]
\centering
\caption{Results obtained with Level 1 Wavelet Packet Transform feature extraction for 2, 2.5, 3.5 and 4.5 inch stickout case.}
\label{tab:resultsforwholecases1}
\begin{tabular}{|l|c|c|c|c|}
\hline
\multicolumn{1}{|c|}{Classifier: SVM} & \multicolumn{2}{c|}{5.08 cm (2 inch)} & \multicolumn{2}{c|}{6.35 cm (2.5 inch)} \\
\hline
\multicolumn{1}{|c|}{Features} & Test Set & Training Set & Test Set& Training Set \\
\hline
$r_{1}$                   & $\SI{93.1 \pm 6.4 }{\percent}$  & $\SI{85.0 \pm 3.6 }{\percent}$  & $\SI{73.3 \pm 21.3}{\percent}$ & $\SI{83.0  \pm 13.5}{\percent}$ \\
$r_{1}$,$r_{2}$           & \cellcolor[rgb]{0.13,0.67,0.8}$\SI{93.9 \pm 5.8 }{\percent}$  & $\SI{85.8 \pm 3.0 }{\percent}$  & \cellcolor[rgb]{0.13,0.67,0.8}$\SI{85.0 \pm 22.9}{\percent}$ & $\SI{100.0 \pm 0.0 }{\percent}$  \\
$r_{1}$,$r_{2}$,$r_{3}$   & $\SI{92.3 \pm 11.4}{\percent}$  & $\SI{89.6 \pm 1.8 }{\percent}$  & \cellcolor[rgb]{0.13,0.67,0.8} $\SI{85.0 \pm 22.9}{\percent}$ & $\SI{100.0 \pm 0.0 }{\percent}$  \\
$r_{1}$,$\ldots$,$r_{4}$  & $\SI{90.8 \pm 13.7}{\percent}$  & $\SI{90.0 \pm 3.1 }{\percent}$  & \cellcolor[rgb]{0.13,0.67,0.8} $\SI{85.0 \pm 22.9}{\percent}$ & $\SI{100.0 \pm 0.0 }{\percent}$  \\
$r_{1}$,$\ldots$,$r_{5}$  & $\SI{89.2 \pm 13.9}{\percent}$  & $\SI{89.2 \pm 3.4 }{\percent}$  & \cellcolor[rgb]{0.13,0.67,0.8}$\SI{85.0 \pm 22.9}{\percent}$ & $\SI{100.0 \pm 0.0 }{\percent}$  \\
$r_{1}$,$\ldots$,$r_{6}$  & $\SI{81.5 \pm 15.1}{\percent}$  & $\SI{85.0 \pm 4.7 }{\percent}$  & \cellcolor[rgb]{0.13,0.67,0.8}$\SI{85.0 \pm 22.9}{\percent}$ & $\SI{100.0 \pm 0.0 }{\percent}$   \\
$r_{1}$,$\ldots$,$r_{7}$  & $\SI{76.9 \pm 14.6}{\percent}$  & $\SI{85.0 \pm 4.7 }{\percent}$  & \cellcolor[rgb]{0.13,0.67,0.8}$\SI{85.0 \pm 22.9}{\percent}$ & $\SI{100.0 \pm 0.0 }{\percent}$  \\
$r_{1}$,$\ldots$,$r_{8}$  & $\SI{78.5 \pm 13.7}{\percent}$  & $\SI{85.8 \pm 6.0 }{\percent}$  & \cellcolor[rgb]{0.13,0.67,0.8}$\SI{85.0 \pm 22.9}{\percent}$ & $\SI{100.0 \pm 0.0 }{\percent}$  \\
$r_{1}$,$\ldots$,$r_{9}$  & $\SI{78.5 \pm 13.7}{\percent}$  & $\SI{86.2 \pm 5.5 }{\percent}$  & \cellcolor[rgb]{0.13,0.67,0.8}$\SI{85.0 \pm 22.9}{\percent}$ & $\SI{100.0 \pm 0.0 }{\percent}$  \\
$r_{1}$,$\ldots$,$r_{10}$ & $\SI{78.5 \pm 13.7}{\percent}$  & $\SI{86.2 \pm 5.5 }{\percent}$  & \cellcolor[rgb]{0.13,0.67,0.8}$\SI{85.0 \pm 22.9}{\percent}$ & $\SI{100.0 \pm 0.0 }{\percent}$  \\
$r_{1}$,$\ldots$,$r_{11}$ & $\SI{78.5 \pm 13.7}{\percent}$  & $\SI{86.2 \pm 5.5 }{\percent}$  & \cellcolor[rgb]{0.13,0.67,0.8}$\SI{85.0 \pm 22.9}{\percent}$ & $\SI{100.0 \pm 0.0 }{\percent}$  \\
$r_{1}$,$\ldots$,$r_{12}$ & $\SI{78.5 \pm 13.7}{\percent}$  & $\SI{86.2 \pm 5.5 }{\percent}$  & \cellcolor[rgb]{0.13,0.67,0.8}$\SI{85.0 \pm 22.9}{\percent}$ & $\SI{100.0 \pm 0.0 }{\percent}$  \\
$r_{1}$,$\ldots$,$r_{13}$ & $\SI{78.5 \pm 13.7}{\percent}$  & $\SI{86.2 \pm 5.5 }{\percent}$  & \cellcolor[rgb]{0.13,0.67,0.8}$\SI{85.0 \pm 22.9}{\percent}$ & $\SI{100.0 \pm 0.0 }{\percent}$  \\
$r_{1}$,$\ldots$,$r_{14}$ & $\SI{78.5 \pm 13.7}{\percent}$  & $\SI{86.2 \pm 5.5 }{\percent}$  & \cellcolor[rgb]{0.13,0.67,0.8}$\SI{85.0 \pm 22.9}{\percent}$ & $\SI{100.0 \pm 0.0 }{\percent}$  \\
\hline
\multicolumn{1}{|c|}{Classifier: SVM} & \multicolumn{2}{c|}{8.89 cm (3.5 inch)} & \multicolumn{2}{c|}{11.43 cm (4.5 inch)}\\
\hline 
$r_{1}$                   & $\SI{56.0 \pm 17.4}{\percent}$ & $\SI{83.3  \pm 11.4}{\percent}$ & \cellcolor[rgb]{0.13,0.67,0.8}$\SI{78.8 \pm 14.8 }{\percent}$& $\SI{83.6 \pm 8.5 }{\percent}$ \\
$r_{1}$,$r_{2}$           & \cellcolor[rgb]{0.13,0.67,0.8}$\SI{84.0 \pm 15.0}{\percent}$ & $\SI{100.0 \pm 0.0 }{\percent}$ & $\SI{68.8 \pm 17.0 }{\percent}$& $\SI{82.9 \pm 9.7 }{\percent}$\\
$r_{1}$,$r_{2}$,$r_{3}$   & \cellcolor[rgb]{0.13,0.67,0.8}$\SI{84.0 \pm 15.0}{\percent}$ & $\SI{100.0 \pm 0.0 }{\percent}$ & $\SI{67.5 \pm 13.9 }{\percent}$& $\SI{80.7 \pm 12.4}{\percent}$ \\
$r_{1}$,$\ldots$,$r_{4}$  & \cellcolor[rgb]{0.13,0.67,0.8}$\SI{84.0 \pm 15.0}{\percent}$ & $\SI{100.0 \pm 0.0 }{\percent}$ & $\SI{68.8 \pm 15.1 }{\percent}$& $\SI{82.9 \pm 9.7 }{\percent}$\\
$r_{1}$,$\ldots$,$r_{5}$  & \cellcolor[rgb]{0.13,0.67,0.8}$\SI{84.0 \pm 15.0}{\percent}$ & $\SI{100.0 \pm 0.0 }{\percent}$ & $\SI{68.8 \pm 15.1 }{\percent}$& $\SI{82.9 \pm 9.7 }{\percent}$\\
$r_{1}$,$\ldots$,$r_{6}$  & \cellcolor[rgb]{0.13,0.67,0.8}$\SI{84.0 \pm 15.0}{\percent}$ & $\SI{100.0 \pm 0.0 }{\percent}$ & $\SI{73.8 \pm 10.4 }{\percent}$& $\SI{83.6 \pm 10.1}{\percent}$ \\
$r_{1}$,$\ldots$,$r_{7}$  & \cellcolor[rgb]{0.13,0.67,0.8}$\SI{84.0 \pm 15.0}{\percent}$ & $\SI{100.0 \pm 0.0 }{\percent}$ & $\SI{76.3 \pm 10.4 }{\percent}$& $\SI{84.3 \pm 7.7 }{\percent}$ \\
$r_{1}$,$\ldots$,$r_{8}$  & \cellcolor[rgb]{0.13,0.67,0.8}$\SI{84.0 \pm 15.0}{\percent}$ & $\SI{100.0 \pm 0.0 }{\percent}$ & $\SI{76.3 \pm 10.4 }{\percent}$& $\SI{84.3 \pm 7.7 }{\percent}$ \\
$r_{1}$,$\ldots$,$r_{9}$  & \cellcolor[rgb]{0.13,0.67,0.8}$\SI{84.0 \pm 15.0}{\percent}$ & $\SI{100.0 \pm 0.0 }{\percent}$ & $\SI{76.3 \pm 10.4 }{\percent}$& $\SI{84.3 \pm 7.7 }{\percent}$ \\
$r_{1}$,$\ldots$,$r_{10}$ & \cellcolor[rgb]{0.13,0.67,0.8}$\SI{84.0 \pm 15.0}{\percent}$ & $\SI{100.0 \pm 0.0 }{\percent}$ & $\SI{76.3 \pm 10.4 }{\percent}$& $\SI{84.3 \pm 7.7 }{\percent}$ \\
$r_{1}$,$\ldots$,$r_{11}$ & \cellcolor[rgb]{0.13,0.67,0.8}$\SI{84.0 \pm 15.0}{\percent}$ & $\SI{100.0 \pm 0.0 }{\percent}$ & $\SI{76.3 \pm 10.4 }{\percent}$& $\SI{84.3 \pm 7.7 }{\percent}$ \\
$r_{1}$,$\ldots$,$r_{12}$ & \cellcolor[rgb]{0.13,0.67,0.8}$\SI{84.0 \pm 15.0}{\percent}$ & $\SI{100.0 \pm 0.0 }{\percent}$ & $\SI{76.3 \pm 10.4 }{\percent}$& $\SI{84.3 \pm 7.7 }{\percent}$ \\
$r_{1}$,$\ldots$,$r_{13}$ & \cellcolor[rgb]{0.13,0.67,0.8}$\SI{84.0 \pm 15.0}{\percent}$ & $\SI{100.0 \pm 0.0 }{\percent}$ & $\SI{76.3 \pm 10.4 }{\percent}$& $\SI{84.3 \pm 7.7 }{\percent}$ \\
$r_{1}$,$\ldots$,$r_{14}$ & \cellcolor[rgb]{0.13,0.67,0.8}$\SI{84.0 \pm 15.0}{\percent}$ & $\SI{100.0 \pm 0.0 }{\percent}$ & $\SI{76.3 \pm 10.4 }{\percent}$& $\SI{84.3 \pm 7.7 }{\percent}$ \\
\hline
\end{tabular}
\end{table}
%---------------

%---------------
\begin{table}[H]
\centering
\caption{Results obtained with Level 2 Wavelet Packet Transform feature extraction for 2, 2.5, 3.5 and 4.5 inch stickout case.}
\label{tab:resultsforwholecases2}
\begin{tabular}{|l|c|c|c|c|}
\hline
\multicolumn{1}{|c|}{Classifier: SVM} & \multicolumn{2}{c|}{5.08 cm (2 inch)} & \multicolumn{2}{c|}{6.35 cm (2.5 inch)} \\
\hline
\multicolumn{1}{|c|}{Features} & Test Set & Training Set & Test Set& Training Set \\
\hline
$r_{1}$                   & \cellcolor[rgb]{0.13,0.67,0.8}$\SI{92.3 \pm  6.0  }{\percent}$ & $\SI{93.8 \pm  1.9 }{\percent}$  & \cellcolor[rgb]{0.13,0.67,0.8}$\SI{100.0\pm 0.0 }{\percent}$  & $\SI{100.0 \pm 0.0 }{\percent}$\\
$r_{1}$,$r_{2}$           & $\SI{90.8 \pm  8.3  }{\percent}$ & $\SI{93.5 \pm  1.8 }{\percent}$  & $\SI{95.0 \pm 7.6 }{\percent}$  & $\SI{100.0 \pm 0.0 }{\percent}$\\
$r_{1}$,$r_{2}$,$r_{3}$   & $\SI{90.0 \pm  6.9  }{\percent}$ & $\SI{91.5 \pm  4.1 }{\percent}$  & $\SI{95.0 \pm  7.6 }{\percent}$  & $\SI{100.0 \pm 0.0 }{\percent}$\\
$r_{1}$,$\ldots$,$r_{4}$  & $\SI{90.0 \pm  6.9  }{\percent}$ & $\SI{91.5 \pm  4.1 }{\percent}$  & $\SI{95.0 \pm  7.6 }{\percent}$  & $\SI{100.0  \pm0.0 }{\percent}$\\
$r_{1}$,$\ldots$,$r_{5}$  & $\SI{91.5 \pm  7.3  }{\percent}$ & $\SI{92.7 \pm  3.6 }{\percent}$  & $\SI{95.0 \pm  7.6 }{\percent}$  & $\SI{100.0 \pm0.0 }{\percent}$\\
$r_{1}$,$\ldots$,$r_{6}$  & $\SI{76.2 \pm  11.1 }{\percent}$ & $\SI{78.5 \pm  8.8 }{\percent}$  & $\SI{95.0 \pm  7.6 }{\percent}$  & $\SI{100.0 \pm 0.0 }{\percent}$\\
$r_{1}$,$\ldots$,$r_{7}$  & $\SI{83.8 \pm  7.3  }{\percent}$ & $\SI{78.5 \pm  7.5 }{\percent}$  & $\SI{95.0 \pm  7.6 }{\percent}$  & $\SI{100.0\pm  0.0 }{\percent}$\\
$r_{1}$,$\ldots$,$r_{8}$  & $\SI{84.6 \pm  8.4  }{\percent}$ & $\SI{77.3 \pm  6.8 }{\percent}$  & $\SI{95.0 \pm  7.6 }{\percent}$  & $\SI{100.0 \pm 0.0 }{\percent}$\\
$r_{1}$,$\ldots$,$r_{9}$  & $\SI{81.5 \pm  7.8  }{\percent}$ & $\SI{76.5 \pm  6.1 }{\percent}$  & $\SI{95.0 \pm  7.6 }{\percent}$  & $\SI{100.0 \pm 0.0 }{\percent}$\\
$r_{1}$,$\ldots$,$r_{10}$ & $\SI{81.5 \pm  7.8  }{\percent}$ & $\SI{76.5 \pm  6.1 }{\percent}$  & $\SI{95.0 \pm   7.6 }{\percent}$  & $\SI{100.0\pm  0.0 }{\percent}$\\
$r_{1}$,$\ldots$,$r_{11}$ & $\SI{81.5 \pm  7.8  }{\percent}$ & $\SI{76.5 \pm  6.1 }{\percent}$  & $\SI{95.0 \pm  7.6 }{\percent}$  & $\SI{100.0 \pm 0.0 }{\percent}$\\
$r_{1}$,$\ldots$,$r_{12}$ & $\SI{81.5 \pm  7.8  }{\percent}$ & $\SI{76.5 \pm  6.1 }{\percent}$  & $\SI{95.0 \pm  7.6 }{\percent}$  & $\SI{100.0 \pm 0.0 }{\percent}$ \\
$r_{1}$,$\ldots$,$r_{13}$ & $\SI{81.5 \pm  7.8  }{\percent}$ & $\SI{76.5 \pm  6.1 }{\percent}$  & $\SI{95.0 \pm  7.6 }{\percent}$  & $\SI{100.0 \pm 0.0 }{\percent}$ \\
$r_{1}$,$\ldots$,$r_{14}$ & $\SI{81.5 \pm  7.8  }{\percent}$ & $\SI{76.5 \pm  6.1 }{\percent}$  & $\SI{95.0 \pm  7.6 }{\percent}$  & $\SI{100.0 \pm 0.0 }{\percent}$\\
\hline
\multicolumn{1}{|c|}{Classifier: SVM} & \multicolumn{2}{c|}{8.89 cm (3.5 inch)} & \multicolumn{2}{c|}{11.43 cm (4.5 inch)}\\
\hline 
$r_{1}$                   & $\SI{70.0 \pm  22.4}{\percent}$ & $\SI{78.9  \pm 11.6 }{\percent}$ & $\SI{66.3  \pm 16.8 }{\percent}$ &$\SI{ 69.3 \pm  15.7}{\percent}$ \\
$r_{1}$,$r_{2}$           & $\SI{72.0 \pm  16.0}{\percent}$ & $\SI{92.2  \pm 10.0 }{\percent}$ & \cellcolor[rgb]{0.13,0.67,0.8}$\SI{87.5  \pm 11.2 }{\percent}$ &$\SI{ 82.1 \pm  8.6 }{\percent}$\\
$r_{1}$,$r_{2}$,$r_{3}$   & $\SI{72.0 \pm  16.0}{\percent}$ & $\SI{93.3  \pm 8.9  }{\percent}$ & \cellcolor[rgb]{0.13,0.67,0.8}$\SI{87.5  \pm 11.2 }{\percent}$ &$\SI{ 82.1 \pm  8.6 }{\percent}$\\
$r_{1}$,$\ldots$,$r_{4}$  & $\SI{68.0 \pm  25.6}{\percent}$ & $\SI{88.9  \pm 14.1 }{\percent}$ & \cellcolor[rgb]{0.13,0.67,0.8}$\SI{87.5  \pm 11.2 }{\percent}$ &$\SI{ 82.1 \pm 8.6 }{\percent}$\\
$r_{1}$,$\ldots$,$r_{5}$  & $\SI{74.0 \pm  12.8}{\percent}$ & $\SI{87.8  \pm 15.3 }{\percent}$ & \cellcolor[rgb]{0.13,0.67,0.8}$\SI{87.5  \pm 11.2 }{\percent}$ &$\SI{ 82.1  \pm 8.6 }{\percent}$\\
$r_{1}$,$\ldots$,$r_{6}$  & \cellcolor[rgb]{0.13,0.67,0.8}$\SI{78.0 \pm  14.0}{\percent}$ & $\SI{90.0  \pm 10.5 }{\percent}$ & \cellcolor[rgb]{0.13,0.67,0.8}$\SI{87.5  \pm 11.2 }{\percent}$ &$\SI{ 82.1  \pm  8.6 }{\percent}$\\
$r_{1}$,$\ldots$,$r_{7}$  & $\SI{76.0 \pm  12.0}{\percent}$ & $\SI{86.7  \pm 15.6 }{\percent}$ & \cellcolor[rgb]{0.13,0.67,0.8}$\SI{87.5  \pm 11.2 }{\percent}$ &$\SI{ 82.1  \pm 8.6 }{\percent}$\\
$r_{1}$,$\ldots$,$r_{8}$  & $\SI{76.0 \pm  12.0}{\percent}$ & $\SI{86.7  \pm 15.6}{\percent}$  & \cellcolor[rgb]{0.13,0.67,0.8}$\SI{87.5  \pm 11.2 }{\percent}$ &$\SI{ 82.1  \pm 8.6 }{\percent}$\\
$r_{1}$,$\ldots$,$r_{9}$  & $\SI{76.0 \pm  12.0}{\percent}$ & $\SI{86.7  \pm 15.6}{\percent}$  & \cellcolor[rgb]{0.13,0.67,0.8}$\SI{87.5  \pm 11.2 }{\percent}$ &$\SI{ 82.1  \pm 8.6 }{\percent}$\\
$r_{1}$,$\ldots$,$r_{10}$ & $\SI{76.0 \pm  12.0}{\percent}$ & $\SI{86.7  \pm 15.6}{\percent}$  & \cellcolor[rgb]{0.13,0.67,0.8}$\SI{87.5  \pm 11.2 }{\percent}$ &$\SI{ 82.1  \pm 8.6 }{\percent}$\\
$r_{1}$,$\ldots$,$r_{11}$ & $\SI{76.0 \pm  12.0}{\percent}$ & $\SI{86.7  \pm 15.6}{\percent}$  & \cellcolor[rgb]{0.13,0.67,0.8}$\SI{87.5  \pm 11.2 }{\percent}$ &$\SI{ 82.1  \pm 8.6 }{\percent}$\\
$r_{1}$,$\ldots$,$r_{12}$ & $\SI{76.0 \pm  12.0}{\percent}$ & $\SI{86.7  \pm 15.6}{\percent}$  & \cellcolor[rgb]{0.13,0.67,0.8}$\SI{87.5  \pm 11.2 }{\percent}$ &$\SI{ 82.1  \pm 8.6 }{\percent}$\\
$r_{1}$,$\ldots$,$r_{13}$ & $\SI{76.0 \pm  12.0}{\percent}$ & $\SI{86.7  \pm 15.6}{\percent}$  & \cellcolor[rgb]{0.13,0.67,0.8}$\SI{87.5  \pm 11.2 }{\percent}$ &$\SI{ 82.1  \pm 8.6 }{\percent}$\\
$r_{1}$,$\ldots$,$r_{14}$ & $\SI{76.0 \pm  12.0}{\percent}$ & $\SI{86.7  \pm 15.6}{\percent}$  & \cellcolor[rgb]{0.13,0.67,0.8}$\SI{87.5  \pm 11.2 }{\percent}$ &$\SI{ 82.1  \pm 8.6 }{\percent}$\\
\hline
\end{tabular}
\end{table}
%---------------

%---------------
\begin{table}[H]
\centering
\caption{Results obtained with Level 3 Wavelet Packet Transform feature extraction for 2, 2.5, 3.5 and 4.5 inch stickout case.}
\label{tab:resultsforwholecases3}
\begin{tabular}{|l|c|c|c|c|}
\hline
\multicolumn{1}{|c|}{Classifier: SVM} & \multicolumn{2}{c|}{5.08 cm (2 inch)} & \multicolumn{2}{c|}{6.35 cm (2.5 inch)} \\
\hline
\multicolumn{1}{|c|}{Features} & Test Set & Training Set & Test Set& Training Set \\
\hline
$r_{1}$                   & $\SI{70.8  \pm 13.2 }{\percent}$  & $\SI{71.5 \pm  4.3 }{\percent}$   & $\SI{63.3 \pm  20.8 }{\percent}$  & $\SI{73.0  \pm 19.0}{\percent}$  \\
$r_{1}$,$r_{2}$           & $\SI{77.7  \pm 8.7  }{\percent}$  & $\SI{85.4 \pm  3.4 }{\percent}$   & $\SI{78.3 \pm  7.6  }{\percent}$  & $\SI{94.0  \pm 12.0}{\percent}$  \\
$r_{1}$,$r_{2}$,$r_{3}$   & \cellcolor[rgb]{0.13,0.67,0.8}$\SI{78.5  \pm 6.7  }{\percent}$  & $\SI{85.8 \pm  2.5 }{\percent}$   & $\SI{76.7 \pm  8.2  }{\percent}$  & $\SI{95.0  \pm 10.2}{\percent}$  \\
$r_{1}$,$\ldots$,$r_{4}$  & $\SI{77.7  \pm 7.3  }{\percent}$  & $\SI{84.6 \pm  3.8 }{\percent}$   & $\SI{78.3 \pm  10.7 }{\percent}$  & $\SI{97.0  \pm 6.4 }{\percent}$ \\
$r_{1}$,$\ldots$,$r_{5}$  & $\SI{76.9  \pm 4.9  }{\percent}$  & $\SI{84.6 \pm  6.9 }{\percent}$   & $\SI{78.3 \pm  10.7 }{\percent}$  & $\SI{97.0  \pm 6.4 }{\percent}$ \\
$r_{1}$,$\ldots$,$r_{6}$  & $\SI{73.8  \pm 13.0 }{\percent}$  & $\SI{78.1 \pm  7.5 }{\percent}$   & $\SI{78.3 \pm  10.7 }{\percent}$  & $\SI{97.0  \pm 6.4 }{\percent}$ \\
$r_{1}$,$\ldots$,$r_{7}$  & $\SI{62.3  \pm 11.6 }{\percent}$  & $\SI{74.2 \pm  8.9 }{\percent}$   & $\SI{78.3 \pm  10.7 }{\percent}$  & $\SI{97.0  \pm 6.4 }{\percent}$ \\
$r_{1}$,$\ldots$,$r_{8}$  & $\SI{60.8  \pm 11.1 }{\percent}$  & $\SI{72.3 \pm  7.7 }{\percent}$   & \cellcolor[rgb]{0.13,0.67,0.8}$\SI{81.7 \pm  11.7 }{\percent}$  & $\SI{98.0 \pm  4.0 }{\percent}$ \\
$r_{1}$,$\ldots$,$r_{9}$  & $\SI{60.8  \pm 11.1 }{\percent}$  & $\SI{72.3 \pm  7.7 }{\percent}$   & \cellcolor[rgb]{0.13,0.67,0.8}$\SI{81.7 \pm  11.7 }{\percent}$  & $\SI{98.0 \pm  4.0 }{\percent}$ \\
$r_{1}$,$\ldots$,$r_{10}$ & $\SI{60.8  \pm 11.1 }{\percent}$  & $\SI{72.3 \pm  7.7 }{\percent}$   & \cellcolor[rgb]{0.13,0.67,0.8}$\SI{81.7 \pm  11.7 }{\percent}$  & $\SI{98.0 \pm  4.0 }{\percent}$ \\
$r_{1}$,$\ldots$,$r_{11}$ & $\SI{60.8  \pm 11.1 }{\percent}$  & $\SI{72.3 \pm  7.7 }{\percent}$   & \cellcolor[rgb]{0.13,0.67,0.8}$\SI{81.7 \pm  11.7 }{\percent}$  & $\SI{98.0  \pm 4.0 }{\percent}$ \\
$r_{1}$,$\ldots$,$r_{12}$ & $\SI{60.8  \pm 11.1 }{\percent}$  & $\SI{72.3 \pm  7.7 }{\percent}$   & \cellcolor[rgb]{0.13,0.67,0.8}$\SI{81.7 \pm  11.7 }{\percent}$  & $\SI{98.0 \pm  4.0 }{\percent}$ \\
$r_{1}$,$\ldots$,$r_{13}$ & $\SI{60.8  \pm 11.1 }{\percent}$  & $\SI{72.3 \pm  7.7 }{\percent}$   & \cellcolor[rgb]{0.13,0.67,0.8}$\SI{81.7 \pm  11.7 }{\percent}$  & $\SI{98.0 \pm  4.0 }{\percent}$ \\
$r_{1}$,$\ldots$,$r_{14}$ & $\SI{60.8  \pm 11.1 }{\percent}$  & $\SI{72.3 \pm  7.7 }{\percent}$   & \cellcolor[rgb]{0.13,0.67,0.8}$\SI{81.7 \pm  11.7 }{\percent}$  & $\SI{98.0 \pm  4.0 }{\percent}$ \\
\hline
\multicolumn{1}{|c|}{Classifier: SVM} & \multicolumn{2}{c|}{8.89 cm (3.5 inch)} & \multicolumn{2}{c|}{11.43 cm (4.5 inch)}\\
\hline
$r_{1}$                   & $\SI{62.0 \pm  14.0}{\percent}$  & $\SI{73.3  \pm 16.6}{\percent}$  & \cellcolor[rgb]{0.13,0.67,0.8}$\SI{85.0 \pm 9.4 }{\percent}$  & $\SI{83.6  \pm 10.6 }{\percent}$\\
$r_{1}$,$r_{2}$           & $\SI{60.0 \pm  15.5}{\percent}$  & $\SI{80.0  \pm 18.5}{\percent}$  & $\SI{72.5 \pm  16.6}{\percent}$  & $\SI{81.4 \pm  14.7}{\percent}$ \\
$r_{1}$,$r_{2}$,$r_{3}$   & $\SI{66.0 \pm  9.2 }{\percent}$  & $\SI{83.3  \pm 17.4}{\percent}$  & $\SI{72.5 \pm  16.6}{\percent}$  & $\SI{81.4 \pm  14.7 }{\percent}$\\
$r_{1}$,$\ldots$,$r_{4}$  & $\SI{66.0 \pm  9.2 }{\percent}$  & $\SI{83.3  \pm 17.4}{\percent}$  & $\SI{75.0 \pm  18.5}{\percent}$  & $\SI{82.1 \pm  14.7 }{\percent}$\\
$r_{1}$,$\ldots$,$r_{5}$  & $\SI{64.0 \pm  12.0}{\percent}$  & $\SI{82.2  \pm 18.7}{\percent}$  & $\SI{75.0 \pm  18.5}{\percent}$  & $\SI{82.1 \pm  14.7 }{\percent}$\\
$r_{1}$,$\ldots$,$r_{6}$  & $\SI{58.0 \pm  6.0 }{\percent}$  & $\SI{81.1  \pm 20.5}{\percent}$  & $\SI{75.0 \pm  18.5}{\percent}$  & $\SI{82.1 \pm  14.7 }{\percent}$\\
$r_{1}$,$\ldots$,$r_{7}$  & \cellcolor[rgb]{0.13,0.67,0.8}$\SI{68.0 \pm  13.3}{\percent}$  & $\SI{83.3  \pm 15.9}{\percent}$  & $\SI{77.5 \pm  14.6 }{\percent}$ & $\SI{84.3 \pm  13.5 }{\percent}$\\
$r_{1}$,$\ldots$,$r_{8}$  & $\SI{64.0 \pm  15.0}{\percent}$  & $\SI{78.9  \pm 19.5}{\percent}$  & $\SI{76.3 \pm  15.3 }{\percent}$ & $\SI{87.1 \pm  7.7 }{\percent}$\\
$r_{1}$,$\ldots$,$r_{9}$  & $\SI{64.0 \pm  15.0}{\percent}$  & $\SI{78.9  \pm 19.5}{\percent}$  & $\SI{76.3 \pm  15.3}{\percent}$  & $\SI{87.1 \pm  7.7 }{\percent}$\\
$r_{1}$,$\ldots$,$r_{10}$ & $\SI{64.0 \pm  15.0}{\percent}$  & $\SI{78.9  \pm 19.5}{\percent}$  & $\SI{76.3 \pm  15.3}{\percent}$  & $\SI{87.1 \pm  7.7 }{\percent}$\\
$r_{1}$,$\ldots$,$r_{11}$ & $\SI{64.0 \pm  15.0}{\percent}$  & $\SI{78.9  \pm 19.5}{\percent}$  & $\SI{76.3 \pm  15.3 }{\percent}$ & $\SI{87.1 \pm  7.7 }{\percent}$\\
$r_{1}$,$\ldots$,$r_{12}$ & $\SI{64.0 \pm  15.0}{\percent}$  & $\SI{78.9  \pm 19.5}{\percent}$  & $\SI{76.3 \pm  15.3}{\percent}$ & $\SI{87.1  \pm 7.7 }{\percent}$\\
$r_{1}$,$\ldots$,$r_{13}$ & $\SI{64.0 \pm  15.0}{\percent}$  & $\SI{78.9  \pm 19.5}{\percent}$  & $\SI{76.3 \pm  15.3}{\percent}$  & $\SI{87.1 \pm  7.7}{\percent}$ \\
$r_{1}$,$\ldots$,$r_{14}$ & $\SI{64.0 \pm  15.0}{\percent}$  & $\SI{78.9  \pm 19.5}{\percent}$  & $\SI{76.3 \pm  15.3 }{\percent}$ & $\SI{87.1 \pm  7.7 }{\percent}$\\
\hline
\end{tabular}
\end{table}
%---------------

%---------------

\begin{table}[H]
\centering
\caption{Results obtained with Level 4 Wavelet Packet Transform feature extraction for 2, 2.5, 3.5 and 4.5 inch stickout case.}
\label{tab:resultsforwholecases4}
\begin{tabular}{|l|c|c|c|c|}
\hline
\multicolumn{1}{|c|}{Classifier: SVM} & \multicolumn{2}{c|}{5.08 cm (2 inch)} & \multicolumn{2}{c|}{6.35 cm (2.5 inch)} \\
\hline
\multicolumn{1}{|c|}{Features} & Test Set & Training Set & Test Set& Training Set \\
\hline
$r_{1}$										&$\SI{43.1 \pm 9.9}{\percent}$&$\SI{60.0 \pm 3.9}{\percent}$& $\SI{50.0 \pm  12.9}{\percent}$  & $\SI{64.0 \pm  10.2}{\percent}$ \\
$r_{1}$,$r_{2}$						&$\SI{69.2 \pm 9.7}{\percent}$&$\SI{84.6 \pm 9.1}{\percent}$&\cellcolor[rgb]{0.13,0.67,0.8} $\SI{75.0 \pm  13.4}{\percent}$  & $\SI{84.0 \pm  18.5}{\percent}$ \\
$r_{1}$,$r_{2}$,$r_{3}$		&$\SI{69.2 \pm 9.7}{\percent}$&$\SI{84.6 \pm 9.1}{\percent}$& $\SI{65.0 \pm  15.7}{\percent}$  & $\SI{84.0 \pm  18.5}{\percent}$ \\
$r_{1}$,$\ldots$,$r_{4}$	&$\SI{72.3 \pm 8.6}{\percent}$&$\SI{85.7 \pm 8.6}{\percent}$& $\SI{58.3 \pm  20.1}{\percent}$  & $\SI{80.0 \pm  20.0}{\percent}$ \\
$r_{1}$,$\ldots$,$r_{5}$	&\cellcolor[rgb]{0.13,0.67,0.8}$\SI{88.5 \pm 5.1}{\percent}$& $\SI{94.6 \pm  3.1 }{\percent}$  & $\SI{61.7 \pm  16.8}{\percent}$ &$\SI{76.0 \pm  20.1}{\percent}$ \\
$r_{1}$,$\ldots$,$r_{6}$	&$\SI{80.0 \pm 3.7}{\percent}$&$\SI{88.8 \pm 9.0}{\percent}$& $\SI{61.7 \pm  21.1}{\percent}$  & $\SI{72.0 \pm  25.6}{\percent}$ \\
$r_{1}$,$\ldots$,$r_{7}$	&$\SI{84.6 \pm 9.1}{\percent}$&$\SI{89.2 \pm 6.2}{\percent}$& $\SI{61.7 \pm  21.1}{\percent}$  & $\SI{68.0 \pm  24.0}{\percent}$ \\
$r_{1}$,$\ldots$,$r_{8}$	&$\SI{84.6 \pm 9.1}{\percent}$&$\SI{89.2 \pm 6.2}{\percent}$& $\SI{51.7 \pm  21.7}{\percent}$  & $\SI{74.0 \pm  21.1}{\percent}$ \\
$r_{1}$,$\ldots$,$r_{9}$	&$\SI{84.6 \pm 9.1}{\percent}$&$\SI{89.2 \pm 6.2}{\percent}$& $\SI{51.7 \pm  21.7}{\percent}$  & $\SI{74.0 \pm  21.1}{\percent}$ \\
$r_{1}$,$\ldots$,$r_{10}$	&$\SI{84.6 \pm 9.1}{\percent}$&$\SI{89.2 \pm 6.2}{\percent}$& $\SI{51.7 \pm  21.7}{\percent}$  & $\SI{74.0 \pm  21.1}{\percent}$ \\ 
$r_{1}$,$\ldots$,$r_{11}$	&$\SI{84.6 \pm 9.1}{\percent}$&$\SI{89.2 \pm 6.2}{\percent}$& $\SI{51.7 \pm  21.7}{\percent}$  & $\SI{74.0 \pm  21.1}{\percent}$ \\
$r_{1}$,$\ldots$,$r_{12}$	&$\SI{84.6 \pm 9.1}{\percent}$&$\SI{89.2 \pm 6.2}{\percent}$& $\SI{51.7 \pm  21.7}{\percent}$  & $\SI{74.0 \pm  21.1}{\percent}$ \\
$r_{1}$,$\ldots$,$r_{13}$	&$\SI{84.6 \pm 9.1}{\percent}$&$\SI{89.2 \pm 6.2}{\percent}$& $\SI{51.7 \pm  21.7}{\percent}$  & $\SI{74.0 \pm  21.1}{\percent}$ \\ 
$r_{1}$,$\ldots$,$r_{14}$	&$\SI{84.6 \pm 9.1}{\percent}$&$\SI{89.2 \pm 6.2}{\percent}$& $\SI{51.7 \pm  21.7}{\percent}$  & $\SI{74.0 \pm  21.1}{\percent}$ \\
\hline
\multicolumn{1}{|c|}{Classifier: SVM} & \multicolumn{2}{c|}{8.89 cm (3.5 inch)} & \multicolumn{2}{c|}{11.43 cm (4.5 inch)}\\
\hline
$r_{1}$                   & $\SI{54.0 \pm  25.4 }{\percent}$ & $\SI{60.0  \pm 20.0 }{\percent}$ & \cellcolor[rgb]{0.13,0.67,0.8}$\SI{72.5  \pm 15.6 }{\percent}$ & $\SI{75.0 \pm  13.3}{\percent}$ \\
$r_{1}$,$r_{2}$           & $\SI{66.0 \pm  9.2  }{\percent}$ & $\SI{77.8  \pm 14.1 }{\percent}$ & \cellcolor[rgb]{0.13,0.67,0.8}$\SI{73.8  \pm 14.2 }{\percent}$ & $\SI{72.1 \pm  10.3}{\percent}$ \\
$r_{1}$,$r_{2}$,$r_{3}$   & $\SI{66.0 \pm  9.2  }{\percent}$ & $\SI{77.8  \pm 14.1 }{\percent}$ & $\SI{65.0  \pm 17.5 }{\percent}$ & $\SI{69.3 \pm  10.1}{\percent}$ \\
$r_{1}$,$\ldots$,$r_{4}$  & $\SI{66.0 \pm  9.2  }{\percent}$ & $\SI{77.8  \pm 14.1 }{\percent}$ & $\SI{63.8  \pm 15.3 }{\percent}$ & $\SI{70.7 \pm  14.8}{\percent}$ \\
$r_{1}$,$\ldots$,$r_{5}$  & $\SI{66.0 \pm  9.2  }{\percent}$ & $\SI{77.8  \pm 14.1 }{\percent}$ & $\SI{61.3  \pm 15.3 }{\percent}$ & $\SI{69.3 \pm  13.9}{\percent}$ \\
$r_{1}$,$\ldots$,$r_{6}$  & \cellcolor[rgb]{0.13,0.67,0.8}$\SI{68.0 \pm  13.3 }{\percent}$ & $\SI{77.8  \pm 14.1 }{\percent}$ & $\SI{61.3  \pm 15.3 }{\percent}$ & $\SI{68.6 \pm  14.0}{\percent}$ \\
$r_{1}$,$\ldots$,$r_{7}$  & \cellcolor[rgb]{0.13,0.67,0.8}$\SI{68.0 \pm  13.3 }{\percent}$ & $\SI{77.8  \pm 14.1 }{\percent}$ & $\SI{60.0  \pm 14.6 }{\percent}$ & $\SI{70.7 \pm  13.3}{\percent}$ \\
$r_{1}$,$\ldots$,$r_{8}$  & $\SI{56.0 \pm  15.0 }{\percent}$ & $\SI{77.8  \pm 14.1 }{\percent}$ & $\SI{52.5  \pm 10.9 }{\percent}$ & $\SI{71.4 \pm  10.6}{\percent}$ \\
$r_{1}$,$\ldots$,$r_{9}$  & $\SI{58.0 \pm  16.6 }{\percent}$ & $\SI{77.8  \pm 14.1 }{\percent}$ & $\SI{52.5  \pm 10.9 }{\percent}$ & $\SI{71.4 \pm  10.6}{\percent}$ \\
$r_{1}$,$\ldots$,$r_{10}$ & $\SI{60.0 \pm  17.9 }{\percent}$ & $\SI{76.7  \pm 14.4 }{\percent}$ & $\SI{52.5  \pm 10.9 }{\percent}$ & $\SI{71.4 \pm  10.6}{\percent}$ \\
$r_{1}$,$\ldots$,$r_{11}$ & $\SI{60.0 \pm  17.9 }{\percent}$ & $\SI{76.7  \pm 14.4 }{\percent}$ & $\SI{52.5  \pm 10.9 }{\percent}$ & $\SI{71.4 \pm  10.6}{\percent}$\\
$r_{1}$,$\ldots$,$r_{12}$ & $\SI{60.0 \pm  17.9 }{\percent}$ & $\SI{76.7  \pm 14.4 }{\percent}$ & $\SI{52.5  \pm 10.9 }{\percent}$ & $\SI{71.4 \pm  10.6}{\percent}$\\
$r_{1}$,$\ldots$,$r_{13}$ & $\SI{60.0 \pm  17.9 }{\percent}$ & $\SI{77.8  \pm 14.1 }{\percent}$ & $\SI{52.5  \pm 10.9 }{\percent}$ & $\SI{71.4 \pm  10.6}{\percent}$ \\
$r_{1}$,$\ldots$,$r_{14}$ & $\SI{60.0 \pm  17.9 }{\percent}$ & $\SI{77.8  \pm 14.1 }{\percent}$ & $\SI{52.5  \pm 10.9 }{\percent}$ & $\SI{71.4 \pm  10.6}{\percent}$\\
\hline
\end{tabular}
\end{table}
%---------------

%---------------

\begin{table}[H]
\centering
\caption{Results obtained with EEMD feature extraction method for 2, 2.5, 3.5 and 4.5 inch stickout case.}
\label{tab:EEMDresults}
\begin{tabular}{|l|c|c|c|c|}
\hline
\multicolumn{1}{|c|}{Classifier: SVM} & \multicolumn{2}{c|}{5.08 cm (2 inch)} & \multicolumn{2}{c|}{6.35 cm (2.5 inch)} \\
\hline
\multicolumn{1}{|c|}{Features} & Test Set & Training Set & Test Set& Training Set \\
\hline
$r_{1}$										&$\SI{74.4 \pm 0.9}{\percent}$&$\SI{74.4 \pm 0.5}{\percent}$&$\SI{78.3 \pm 1.2}{\percent}$&$\SI{78.6 \pm 0.5}{\percent}$\\
$r_{1}$,$r_{2}$						&\cellcolor[rgb]{0.13,0.67,0.8}$\SI{84.2 \pm 0.8}{\percent}$&$\SI{84.2 \pm 0.4}{\percent}$&$\SI{78.3 \pm 1.2}{\percent}$&$\SI{78.6 \pm 0.5}{\percent}$\\
$r_{1}$,$r_{2}$,$r_{3}$		&\cellcolor[rgb]{0.13,0.67,0.8}$\SI{84.2 \pm 0.8}{\percent}$&$\SI{84.2 \pm 0.4}{\percent}$&$\SI{78.3 \pm 1.4}{\percent}$&$\SI{78.6 \pm 0.5}{\percent}$\\
$r_{1}$,$\ldots$,$r_{4}$	&\cellcolor[rgb]{0.13,0.67,0.8}$\SI{84.2 \pm 0.8}{\percent}$&$\SI{84.2 \pm 0.4}{\percent}$&$\SI{78.5 \pm 1.4}{\percent}$&$\SI{78.7 \pm 0.4}{\percent}$\\
$r_{1}$,$\ldots$,$r_{5}$	&\cellcolor[rgb]{0.13,0.67,0.8}$\SI{84.2 \pm 0.8}{\percent}$&$\SI{84.2 \pm 0.4}{\percent}$&$\SI{78.5 \pm 1.4}{\percent}$&$\SI{78.7 \pm 0.4}{\percent}$\\
$r_{1}$,$\ldots$,$r_{6}$	&\cellcolor[rgb]{0.13,0.67,0.8}$\SI{84.2 \pm 0.8}{\percent}$&$\SI{84.2 \pm 0.4}{\percent}$&$\SI{78.5 \pm 1.2}{\percent}$&$\SI{78.7 \pm 0.5}{\percent}$\\
$r_{1}$,$\ldots$,$r_{7}$	&\cellcolor[rgb]{0.13,0.67,0.8}$\SI{84.2 \pm 0.8}{\percent}$&$\SI{84.2 \pm 0.4}{\percent}$&\cellcolor[rgb]{0.13,0.67,0.8}$\SI{78.6 \pm 1.2}{\percent}$&$\SI{78.8 \pm 0.5}{\percent}$\\

\hline
\multicolumn{1}{|c|}{Classifier: SVM} & \multicolumn{2}{c|}{8.89 cm (3.5 inch)} & \multicolumn{2}{c|}{11.43 cm (4.5 inch)}\\
\hline
$r_{1}$										&\cellcolor[rgb]{0.13,0.67,0.8}$\SI{90.7 \pm 1.4}{\percent}$&$\SI{91.1 \pm 0.6}{\percent}$&$\SI{77.1 \pm 1.2}{\percent}$&$\SI{77.1 \pm 0.5}{\percent}$\\
$r_{1}$,$r_{2}$						&\cellcolor[rgb]{0.13,0.67,0.8}$\SI{90.7 \pm 1.4}{\percent}$&$\SI{91.1 \pm 0.7}{\percent}$&$\SI{77.3 \pm 1.1}{\percent}$&$\SI{77.4 \pm 0.6}{\percent}$\\
$r_{1}$,$r_{2}$,$r_{3}$		&$\SI{90.6 \pm 1.4}{\percent}$&$\SI{91.0 \pm 0.6}{\percent}$&$\SI{77.4 \pm 1.0}{\percent}$&$\SI{77.5 \pm 0.6}{\percent}$\\
$r_{1}$,$\ldots$,$r_{4}$	&$\SI{90.6 \pm 1.4}{\percent}$&$\SI{91.0 \pm 0.6}{\percent}$&$\SI{78.9 \pm 1.1}{\percent}$&$\SI{78.9 \pm 0.7}{\percent}$\\
$r_{1}$,$\ldots$,$r_{5}$	&$\SI{90.5 \pm 1.4}{\percent}$&$\SI{90.9 \pm 0.6}{\percent}$&$\SI{78.9 \pm 1.2}{\percent}$&$\SI{79.0 \pm 0.7}{\percent}$\\
$r_{1}$,$\ldots$,$r_{6}$	&$\SI{90.5 \pm 1.4}{\percent}$&$\SI{91.0 \pm 0.6}{\percent}$&$\SI{78.9 \pm 1.2}{\percent}$&$\SI{79.0 \pm 0.7}{\percent}$\\
$r_{1}$,$\ldots$,$r_{7}$	&$\SI{90.5 \pm 1.4}{\percent}$&$\SI{90.8 \pm 0.6}{\percent}$&\cellcolor[rgb]{0.13,0.67,0.8}$\SI{79.1 \pm 1.2}{\percent}$&$\SI{79.0 \pm 0.6}{\percent}$\\
\hline
\end{tabular}
\end{table}
%---------------

%Transfer Learning Results

%---------------
\begin{table}[H]
\centering
\caption{Transfer learning results for SVM classifiers trained with 5.08 cm (2 inch) and 11.43 cm (4.5 inch) cases and tested on remaining cases for features extracted from level 1 WPT.}
\label{tab:transfer_learning_WPT_Level1_2_4p5_inch}
\begin{tabular}{|l|c|c|c|c|}
\hline
\multicolumn{1}{|c|}{} & \multicolumn{3}{c|}{Training Set: 5.08 cm (2 inch)}\\
\hline
\multicolumn{1}{|c|}{} & \multicolumn{1}{c|}{Test Set: 6.35 cm (2.5 inch)}  & \multicolumn{1}{c|}{Test Set: 8.89 cm (3.5 inch)} &  \multicolumn{1}{c|}{Training Set: 11.43 cm (4.5 inch)}\\
\hline
\multicolumn{1}{|c|}{Features} & \multicolumn{3}{c|}{Test set (Validation Set)} \\
\hline
$r_{1}$	                  &$\SI{86.7\pm	8.5}{\percent}$	&$\SI{82.0	\pm6.0 }{\percent}$ &	$\SI{40.6\pm	5.0}{\percent}$\\
$r_{1}$,$r_{2}$	          &$\SI{84.2\pm	8.7}{\percent}$	&$\SI{79.0\pm	8.3	}{\percent}$  &$\SI{40.6\pm	5.0}{\percent}$\\
$r_{1}$,$r_{2}$,$r_{3}$	  &$\SI{83.3	\pm7.5}{\percent}$&	$\SI{80.0\pm	8.9	}{\percent}$&  $\SI{40.6\pm	5.0}{\percent}$\\
$r_{1}$,$\ldots$,$r_{4}$	&$\SI{86.7\pm	5.5}{\percent}$	&$\SI{78.0\pm	8.7	}{\percent}$  &$\SI{43.1\pm	8.1}{\percent}$\\
$r_{1}$,$\ldots$,$r_{5}$	&$\SI{85.0\pm	5.0}{\percent}$	&$\SI{74.0\pm	12.8}{\percent}$	&$\SI{44.4\pm	10.6}{\percent}$\\
$r_{1}$,$\ldots$,$r_{6}$	&$\SI{82.5	\pm5.8}{\percent}$&	$\SI{72.0\pm	11.7}{\percent}$&	$\SI{55.6\pm	10.6}{\percent}$\\
$r_{1}$,$\ldots$,$r_{7}$	&$\SI{78.3	\pm8.5}{\percent}$&	$\SI{72.0\pm	11.7}{\percent}$&	$\SI{55.6\pm	14.4}{\percent}$\\
$r_{1}$,$\ldots$,$r_{8}$	&$\SI{80.0	\pm7.6}{\percent}$&	$\SI{72.0\pm	11.7}{\percent}$&	$\SI{56.9\pm	15.9}{\percent}$\\
$r_{1}$,$\ldots$,$r_{9}$	&$\SI{80.8	\pm7.5}{\percent}$&	$\SI{73.0	\pm11.9}{\percent}$	&$\SI{59.4	\pm14.6}{\percent}$\\
$r_{1}$,$\ldots$,$r_{10}$	&$\SI{80.8	\pm7.5}{\percent}$&	$\SI{74.0\pm	11.1}{\percent}$&	$\SI{56.9	\pm15.9}{\percent}$\\
$r_{1}$,$\ldots$,$r_{11}$	&$\SI{80.8	\pm7.5}{\percent}$&	$\SI{74.0	\pm11.1}{\percent}$	&$\SI{56.9	\pm15.9}{\percent}$\\
$r_{1}$,$\ldots$,$r_{12}$	&$\SI{80.8	\pm7.5}{\percent}$&	$\SI{74.0\pm	11.1}{\percent}$&	$\SI{56.9	\pm15.9}{\percent}$\\
$r_{1}$,$\ldots$,$r_{13}$	&$\SI{80.0	\pm8.5}{\percent}$&	$\SI{75.0	\pm11.2}{\percent}$	&$\SI{56.9	\pm15.9}{\percent}$\\
$r_{1}$,$\ldots$,$r_{14}$ &$\SI{80.0	\pm8.5}{\percent}$&	$\SI{75.0	\pm11.2}{\percent}$	&$\SI{56.9\pm	15.9}{\percent}$\\
\hline
\multicolumn{1}{|c|}{} & \multicolumn{3}{c|}{Training Set: 11.43 cm (4.5 inch)}\\
\hline
\multicolumn{1}{|c|}{} & \multicolumn{1}{c|}{Test Set:5.08 cm (2 inch)}  & \multicolumn{1}{c|}{Test Set: 6.35 cm (2.5 inch)} &  \multicolumn{1}{c|}{Test Set: 8.89 cm (3.5 inch)}\\
\hline
$r_{1}$	                 & $\SI{ 62.9\pm	21.5}{\percent}$&$\SI{	57.5\pm	15.1}{\percent}$ &$\SI{48.0	\pm21.4}{\percent}$\\
$r_{1}$,$r_{2}$          &	$\SI{51.4\pm	4.8	}{\percent}$ &$\SI{ 50.0\pm	9.1	 }{\percent}$&$\SI{ 35.0\pm	11.2}{\percent}$\\
$r_{1}$,$r_{2}$,$r_{3}$	 & $\SI{ 51.4\pm	5.1	}{\percent}$ & $\SI{50.0\pm	9.1	 }{\percent}$&$\SI{ 41.0\pm	16.4}{\percent}$\\
$r_{1}$,$\ldots$,$r_{4}$ & $\SI{ 61.4	\pm15.9}{\percent}$&$\SI{	55.0	\pm14.0 }{\percent}$&$\SI{ 45.0	\pm16.3}{\percent}$\\
$r_{1}$,$\ldots$,$r_{5}$ & $\SI{ 50.7	\pm5.0	}{\percent}$ &$\SI{ 55.8\pm	14.9 }{\percent}$&$\SI{ 47.0\pm	18.5}{\percent}$\\
$r_{1}$,$\ldots$,$r_{6}$ & $\SI{ 47.9	\pm4.6	}{\percent}$ &$\SI{ 57.5\pm	14.2 }{\percent}$&$\SI{ 57.0\pm	17.3}{\percent}$\\
$r_{1}$,$\ldots$,$r_{7}$ & $\SI{ 57.1	\pm19.0}{\percent}$	&$\SI{51.7	\pm9.0	 }{\percent}$&$\SI{ 66.0\pm	11.1}{\percent}$\\
$r_{1}$,$\ldots$,$r_{8}$ & $\SI{ 53.2	\pm15.7}{\percent}$&	$\SI{51.7	\pm9.0	 }{\percent}$&$\SI{ 63.0\pm	13.5}{\percent}$\\
$r_{1}$,$\ldots$,$r_{9}$ &  $\SI{59.6	\pm17.8}{\percent}$&$\SI{	53.3	\pm8.5}{\percent}$&$\SI{ 62.0\pm	13.3}{\percent}$\\
$r_{1}$,$\ldots$,$r_{10}$&$\SI{	51.1	\pm11.3}{\percent}$&	$\SI{51.7	\pm9.0	 }{\percent}$&$\SI{ 55.0\pm	18.0}{\percent}$\\
$r_{1}$,$\ldots$,$r_{11}$&$\SI{	56.4	\pm14.6}{\percent}$&$\SI{	51.7	\pm9.0}{\percent}$&$\SI{ 58.0\pm	17.8}{\percent}$\\
$r_{1}$,$\ldots$,$r_{12}$&$\SI{	55.0	\pm12.6}{\percent}$&$\SI{	50.8	\pm7.9}{\percent}$&$\SI{ 58.0	\pm17.8}{\percent}$\\
$r_{1}$,$\ldots$,$r_{13}$&$\SI{	55.0	\pm12.6}{\percent}$&	$\SI{48.3\pm	9.0	 }{\percent}$& $\SI{52.0\pm	19.4}{\percent}$\\
$r_{1}$,$\ldots$,$r_{14}$&$\SI{	51.1	\pm11.5}{\percent}$&	$\SI{53.3	\pm8.5	 }{\percent}$&$\SI{ 64.0\pm	15.0}{\percent}$\\
\hline
\end{tabular}
\end{table}
%---------------

\begin{table}[H]
\centering
\caption{Transfer learning results for SVM classifiers trained with 5.08 cm (2 inch) and 11.43 cm (4.5 inch) cases and tested on remaining cases for features extracted from level 4 WPT.}
\label{tab:transfer_learning_WPT_Level4_2_4p5_inch}
\begin{tabular}{|l|c|c|c|c|}
\hline
\multicolumn{1}{|c|}{} & \multicolumn{3}{c|}{Training Set: 5.08 cm (2 inch)}\\
\hline
\multicolumn{1}{|c|}{} & \multicolumn{1}{c|}{Test Set: 6.35 cm (2.5 inch)}  & \multicolumn{1}{c|}{Test Set: 8.89 cm (3.5 inch)} &  \multicolumn{1}{c|}{Training Set: 11.43 cm (4.5 inch)}\\
\hline
\multicolumn{1}{|c|}{Features} & \multicolumn{3}{c|}{Test set (Validation Set)} \\
\hline
$r_{1}$	                  &$\SI{55.8\pm	14.5}{\percent}$&$\SI{ 64.0\pm	10.2}{\percent}$&$\SI{	57.5\pm	16.5}{\percent}$ \\
$r_{1}$,$r_{2}$	          &$\SI{60.8\pm	16.3}{\percent}$&$\SI{ 64.0\pm	13.6}{\percent}$&$\SI{	59.4\pm	22.2}{\percent}$\\
$r_{1}$,$r_{2}$,$r_{3}$	  &$\SI{60.8\pm	17.1}{\percent}$&$\SI{ 60.0\pm	16.1}{\percent}$&$\SI{	55.0\pm	19.9}{\percent}$\\
$r_{1}$,$\ldots$,$r_{4}$	&$\SI{62.5\pm  17.2}{\percent}$&$\SI{ 56.0\pm	16.2}{\percent}$&$\SI{	50.0\pm	17.9}{\percent}$\\
$r_{1}$,$\ldots$,$r_{5}$	&$\SI{41.7\pm	9.1 }{\percent}$&$\SI{ 37.0	\pm9.0	}{\percent}$&$\SI{ 38.1	\pm7.1}{\percent}$\\
$r_{1}$,$\ldots$,$r_{6}$	&$\SI{41.7\pm	9.1	}{\percent}$&$\SI{ 37.0	\pm9.0	}{\percent}$&$\SI{ 38.1	\pm7.1}{\percent}$\\
$r_{1}$,$\ldots$,$r_{7}$	&$\SI{41.7\pm	9.1 }{\percent}$&$\SI{ 37.0	\pm9.0	}{\percent}$&$\SI{ 38.1	\pm7.1}{\percent}$\\
$r_{1}$,$\ldots$,$r_{8}$	&$\SI{41.7\pm	9.1 }{\percent}$&$\SI{ 37.0	\pm9.0	}{\percent}$&$\SI{ 38.1	\pm7.1}{\percent}$\\
$r_{1}$,$\ldots$,$r_{9}$	&$\SI{41.7\pm	9.1 }{\percent}$&$\SI{ 37.0	\pm9.0	}{\percent}$&$\SI{ 38.1\pm	7.1}{\percent}$\\
$r_{1}$,$\ldots$,$r_{10}$	&$\SI{41.7\pm	9.1 }{\percent}$&$\SI{ 37.0	\pm9.0}{\percent}$	&$\SI{ 38.1\pm	7.1}{\percent}$\\
$r_{1}$,$\ldots$,$r_{11}$	&$\SI{41.7\pm	9.1 }{\percent}$&$\SI{ 37.0	\pm9.0}{\percent}$ &$\SI{ 38.1	\pm7.1}{\percent}$\\
$r_{1}$,$\ldots$,$r_{12}$	&$\SI{41.7\pm	9.1}{\percent}$ &$\SI{ 37.0	\pm9.0}{\percent}$ &$\SI{	38.1	\pm7.1}{\percent}$\\
$r_{1}$,$\ldots$,$r_{13}$	&$\SI{41.7\pm	9.1}{\percent}$ &$\SI{ 37.0	\pm9.0}{\percent}$ &$\SI{	38.1	\pm7.1}{\percent}$\\
$r_{1}$,$\ldots$,$r_{14}$	&$\SI{41.7\pm	9.1}{\percent}$ &$\SI{ 37.0	\pm9.0}{\percent}$	&$\SI{ 38.1	\pm7.1}{\percent}$\\
\hline
\multicolumn{1}{|c|}{} & \multicolumn{3}{c|}{Training Set: 11.43 cm (4.5 inch)}\\
\hline
\multicolumn{1}{|c|}{} & \multicolumn{1}{c|}{Test Set:5.08 cm (2 inch)}  & \multicolumn{1}{c|}{Test Set: 6.35 cm (2.5 inch)} &  \multicolumn{1}{c|}{Test Set: 8.89 cm (3.5 inch)}\\
\hline
$r_{1}$	                 & $\SI{50.4\pm	8.4	}{\percent}$&$\SI{45.0\pm	9.3	 }{\percent}$&$\SI{61.0\pm	13.0}{\percent}$\\
$r_{1}$,$r_{2}$	         & $\SI{47.9\pm	6.6	}{\percent}$&$\SI{54.2\pm	10.7 }{\percent}$&$\SI{55.0	\pm16.3}{\percent}$\\
$r_{1}$,$r_{2}$,$r_{3}$	 & $\SI{50.7\pm	7.3	}{\percent}$&$\SI{52.5	\pm11.2 }{\percent}$&$\SI{57.0\pm	15.5}{\percent}$\\
$r_{1}$,$\ldots$,$r_{4}$	&$\SI{50.0\pm	6.6	}{\percent}$&$\SI{49.2	\pm9.5	 }{\percent}$&$\SI{57.0\pm	15.5}{\percent}$\\
$r_{1}$,$\ldots$,$r_{5}$	&$\SI{49.3\pm	5.7	}{\percent}$&$\SI{49.2\pm	9.5	 }{\percent}$&$\SI{57.0\pm	15.5}{\percent}$\\
$r_{1}$,$\ldots$,$r_{6}$	&$\SI{49.3\pm	5.7	}{\percent}$&$\SI{50.0\pm	9.9	 }{\percent}$&$\SI{57.0\pm	15.5}{\percent}$\\
$r_{1}$,$\ldots$,$r_{7}$	&$\SI{49.3\pm	5.7	}{\percent}$&$\SI{50.8\pm	10.2 }{\percent}$&$\SI{61.0\pm	13.0}{\percent}$\\
$r_{1}$,$\ldots$,$r_{8}$	&$\SI{50.0\pm	6.6	}{\percent}$&$\SI{55.0\pm	8.5	 }{\percent}$&$\SI{61.0	\pm13.0}{\percent}$\\
$r_{1}$,$\ldots$,$r_{9}$	&$\SI{50.7\pm	5.7	}{\percent}$&$\SI{55.0\pm	8.5	 }{\percent}$&$\SI{57.0	\pm15.5}{\percent}$\\
$r_{1}$,$\ldots$,$r_{10}$&$\SI{	50.0\pm	5.8}{\percent}$ &$\SI{55.0	\pm8.5	}{\percent}$& $\SI{59.0	\pm14.5}{\percent}$\\
$r_{1}$,$\ldots$,$r_{11}$&	$\SI{51.4\pm	8.0	}{\percent}$&$\SI{52.5\pm	8.4	}{\percent}$& $\SI{55.0	\pm16.3}{\percent}$\\
$r_{1}$,$\ldots$,$r_{12}$	&$\SI{51.4\pm	8.0	}{\percent}$&$\SI{51.7	\pm8.2	}{\percent}$& $\SI{55.0\pm	16.3}{\percent}$\\
$r_{1}$,$\ldots$,$r_{13}$	&$\SI{53.6	\pm7.3	}{\percent}$&$\SI{52.5	\pm8.4	}{\percent}$& $\SI{59.0	\pm14.5}{\percent}$\\
$r_{1}$,$\ldots$,$r_{14}$	&$\SI{46.8	\pm4.6	}{\percent}$&$\SI{51.7	\pm9.0	}{\percent}$& $\SI{59.0	\pm14.5}{\percent}$\\
\hline
\end{tabular}
\end{table}
%---------------

\begin{table}[H]
\centering
\caption{Transfer learning results for SVM classifiers trained with 5.08 cm (2 inch) and 11.43 cm (4.5 inch) cases and tested on remaining cases for features extracted from EEMD.}
\label{tab:transfer_learning_EEMD_2_4p5_inch}
\begin{tabular}{|l|c|c|c|c|}
\hline
\multicolumn{1}{|c|}{} & \multicolumn{3}{c|}{Training Set: 5.08 cm (2 inch)}\\
\hline
\multicolumn{1}{|c|}{} & \multicolumn{1}{c|}{Test Set: 6.35 cm (2.5 inch)}  & \multicolumn{1}{c|}{Test Set: 8.89 cm (3.5 inch)} &  \multicolumn{1}{c|}{Training Set: 11.43 cm (4.5 inch)}\\
\hline
\multicolumn{1}{|c|}{Features} & \multicolumn{3}{c|}{Test set (Validation Set)} \\
\hline
$r_{1}$	                  &$\SI{76.6\pm	0.6}{\percent}$	&$\SI{81.8\pm 0.8}{\percent}$&$\SI{	64.0\pm	0.7}{\percent}$\\
$r_{1}$,$r_{2}$	          &$\SI{75.9\pm	0.6}{\percent}$	&$\SI{81.8\pm	0.8}{\percent}$&$\SI{	64.0\pm	0.7}{\percent}$\\
$r_{1}$,$r_{2}$,$r_{3}$	  &$\SI{75.9\pm	0.6}{\percent}$	&$\SI{81.8\pm	0.8}{\percent}$&$\SI{	64.0\pm	0.7}{\percent}$\\
$r_{1}$,$\ldots$,$r_{4}$	&$\SI{75.9\pm	0.6}{\percent}$	&$\SI{81.8\pm	0.8}{\percent}$&$\SI{	64.0\pm	0.7}{\percent}$\\
$r_{1}$,$\ldots$,$r_{5}$	&$\SI{75.9\pm	0.6}{\percent}$	&$\SI{81.8\pm	0.8}{\percent}$&$\SI{	64.0\pm	0.7}{\percent}$\\
$r_{1}$,$\ldots$,$r_{6}$	&$\SI{75.9\pm	0.6}{\percent}$	&$\SI{81.8\pm	0.8}{\percent}$&$\SI{	64.0\pm	0.7}{\percent}$\\
$r_{1}$,$\ldots$,$r_{7}$	&$\SI{76.0\pm	0.6}{\percent}$	&$\SI{81.8\pm	0.8}{\percent}$&$\SI{	64.0\pm	0.7}{\percent}$\\
\hline
\multicolumn{1}{|c|}{} & \multicolumn{3}{c|}{Training Set: 11.43 cm (4.5 inch)}\\
\hline
\multicolumn{1}{|c|}{} & \multicolumn{1}{c|}{Test Set:5.08 cm (2 inch)}  & \multicolumn{1}{c|}{Test Set: 6.35 cm (2.5 inch)} &  \multicolumn{1}{c|}{Test Set: 8.89 cm (3.5 inch)}\\
\hline
$r_{1}$	                  &$\SI{83.6\pm	0.4}{\percent}$	&$\SI{63.5	\pm1.0}{\percent}$	&$\SI{90.9	\pm0.8}{\percent}$\\
$r_{1}$,$r_{2}$	          &$\SI{83.8\pm	0.5}{\percent}$	&$\SI{63.5	\pm1.0}{\percent}$	&$\SI{91.0	\pm0.7}{\percent}$\\
$r_{1}$,$r_{2}$,$r_{3}$	  &$\SI{83.8\pm	0.5}{\percent}$	&$\SI{63.5	\pm1.0}{\percent}$	&$\SI{91.0	\pm0.7}{\percent}$\\
$r_{1}$,$\ldots$,$r_{4}$	&$\SI{85.1\pm	0.6}{\percent}$	&$\SI{64.6	\pm0.9}{\percent}$	&$\SI{90.1	\pm0.9}{\percent}$\\
$r_{1}$,$\ldots$,$r_{5}$	&$\SI{85.1\pm  0.6}{\percent}$	&$\SI{64.6	\pm0.9}{\percent}$	&$\SI{90.0	\pm0.9}{\percent}$\\
$r_{1}$,$\ldots$,$r_{6}$	&$\SI{85.1\pm	0.6}{\percent}$	&$\SI{64.7	\pm0.9}{\percent}$	&$\SI{90.0	\pm0.9}{\percent}$\\
$r_{1}$,$\ldots$,$r_{7}$	&$\SI{85.1\pm  0.5}{\percent}$	&$\SI{64.7	\pm0.9}{\percent}$	&$\SI{90.0	\pm0.9}{\percent}$\\
\hline
\end{tabular}
\end{table}

%Feature ranking figures for 2.5, 3.5, 4.5 inch cases

%------------------------------
\begin{figure}[H]
\centering
\includegraphics[width=\textwidth,height=.95\textheight,keepaspectratio]{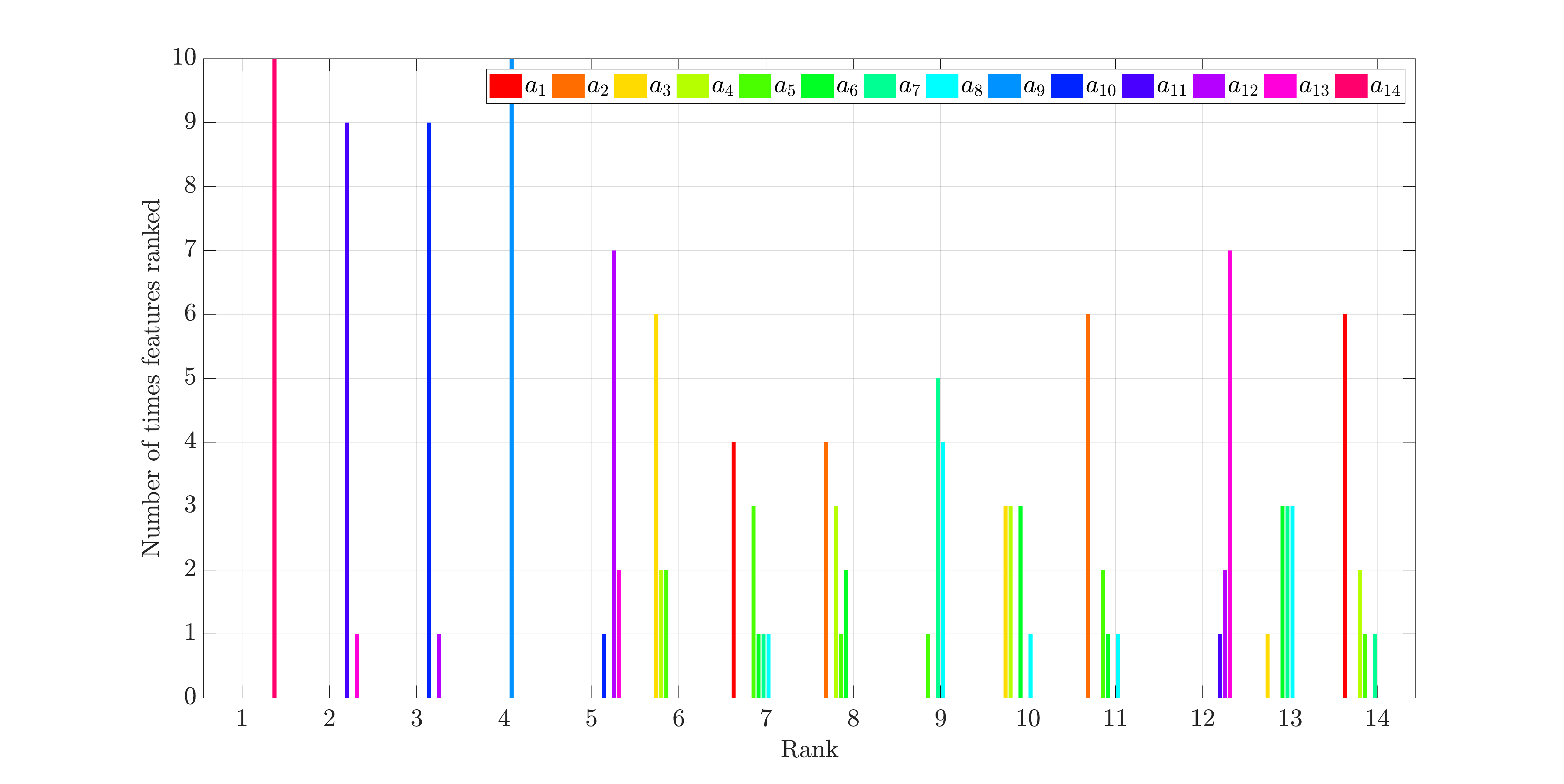}
\caption{Bar plot for feature ranking obtained with SVM-RFE for 6.35 cm (2.5 inch) stickout case at level 4 WPT.}
\label{fig:feature_ranking_2p5}
\end{figure}
%------------------------------

%------------------------------
\begin{figure}[H]
\centering
\includegraphics[width=\textwidth,height=.95\textheight,keepaspectratio]{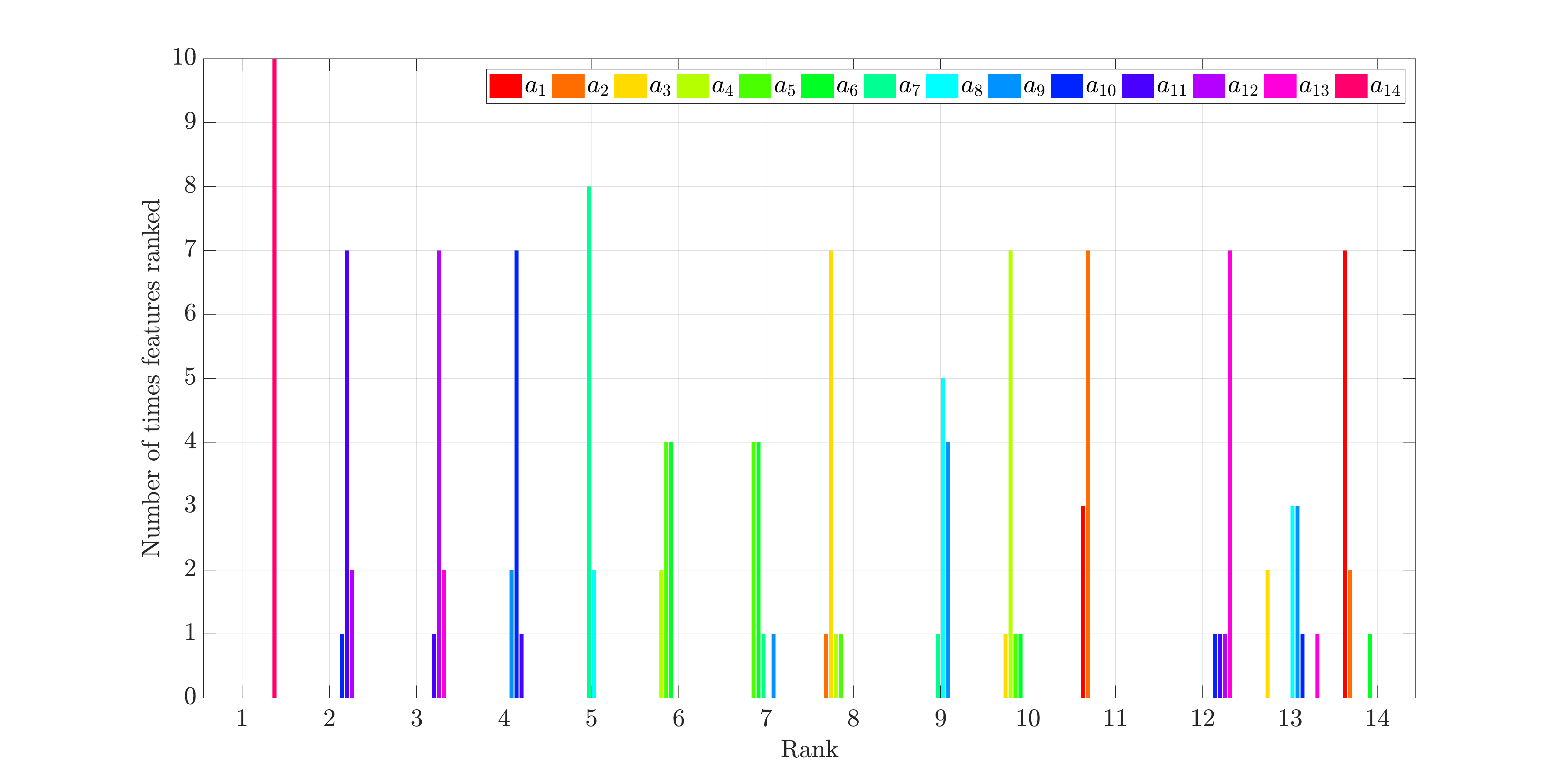}
\caption{Bar plot for feature ranking obtained with SVM-RFE for 8.89 cm (3.5 inch) stickout case at level 4 WPT.}
\label{fig:feature_ranking_3p5}
\end{figure}
%------------------------------

%------------------------------
\begin{figure}[H]
\centering
\includegraphics[width=\textwidth,height=.95\textheight,keepaspectratio]{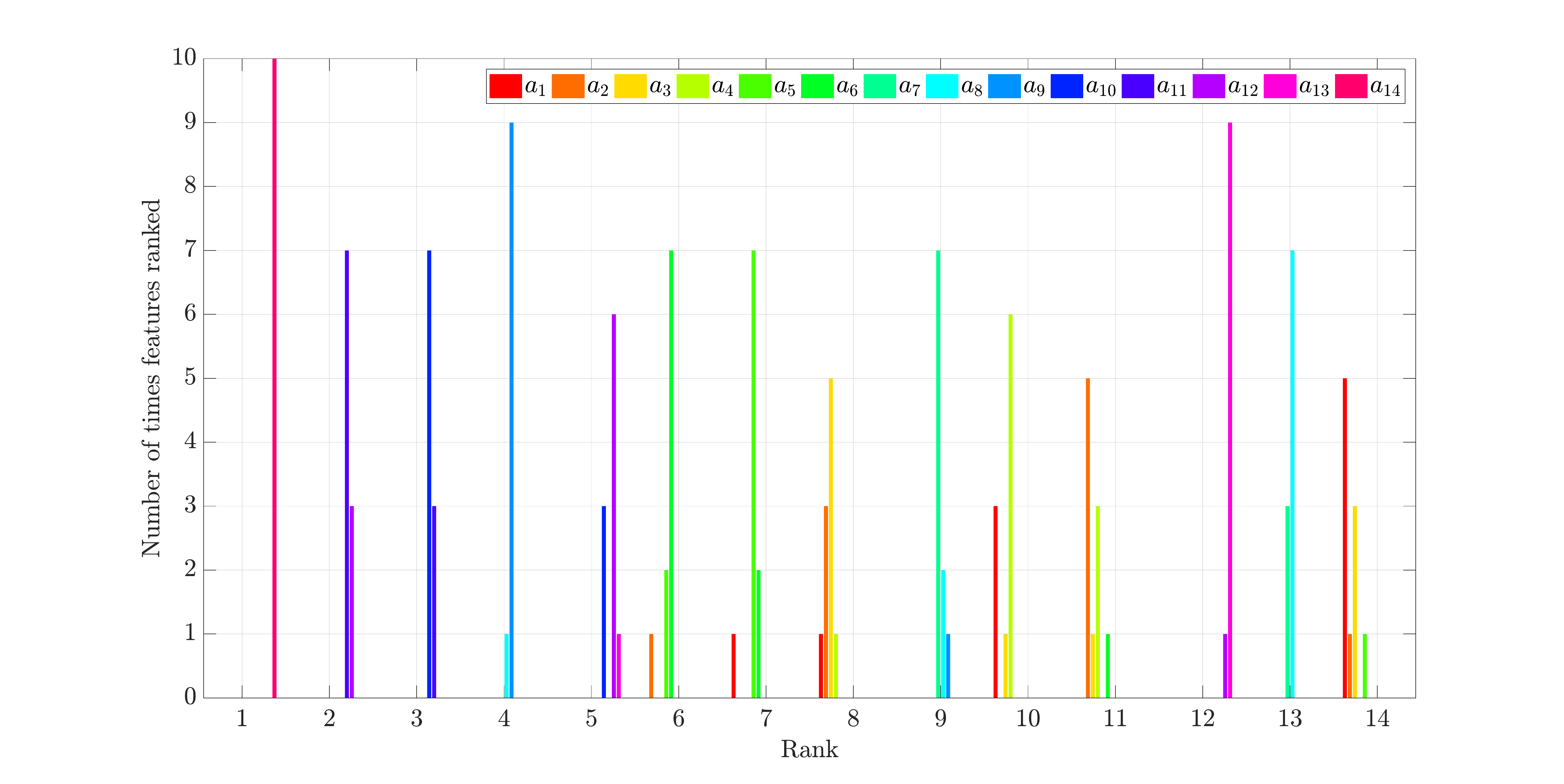}
\caption{Bar plot for feature ranking obtained with SVM-RFE for 11.43 cm (4.5 inch) stickout case at level 4 WPT.}
\label{fig:feature_ranking_4p5}
\end{figure}
%------------------------------

%------------------------------
\begin{figure}[H]
\centering
\includegraphics[width=\textwidth,height=.8\textheight,keepaspectratio]{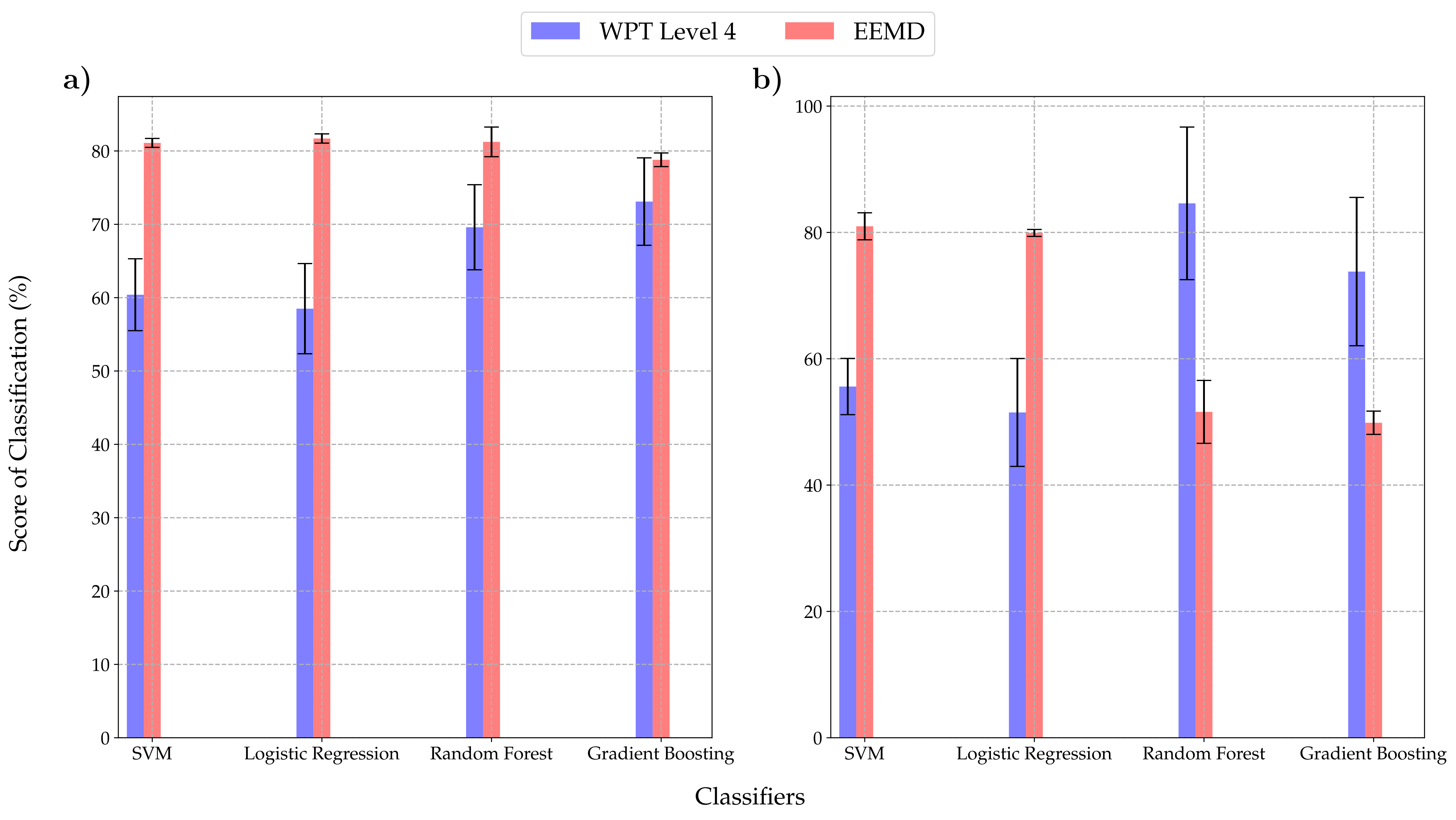}
\caption{Best accuracies obtained with four different classifier for transfer learning application where {\bf{a)}} one classifier is trained with 5.08 cm (2 inch) and 6.35 cm (2.5 inch) data sets features and tested on 8.89 cm (3.5 inch) and 11.43 cm (4.5 inch) data sets features (left) {\bf{b)}} where another classifier is trained with 8.89 cm (3.5 inch) and 11.43 cm (4.5 inch) data sets features and tested on 5.08 cm (2 inch) and 6.35 cm (2.5 inch) data sets features (right).}
\label{fig:transfer_learning_trained_with2case_tested_on2case}
\end{figure}
%------------------------------

% \clearpage
%\appendix

\end{document}